\input harvmac.tex      
\input psfig.sty

%
%
%

\def\lb{\langle}
\def\rb{\rangle}
\def\tilde{\widetilde}

\def\hat{\widehat}
\def\*{\star}
\def\[{\left[}
\def\]{\right]}
\def\({\left(}		
\def\){\right)}

%
%

\def\frac#1#2{{#1 \over #2}}

\def\2pi{\hbox{$2\pi i$}}

\def\dsl{\raise.15ex\hbox{/}\kern-.57em\partial}
\def\Dsl{\,\raise.15ex\hbox{/}\mkern-.13.5mu D}
%
%
\def\th{\theta}		
		
\def\be{\beta}
\def\al{\alpha}
\def\ep{\varepsilon}
\def\la{\lambda}	
\def\de{\delta}		
\def\om{\omega}		
\def\sig{\sigma}	

%
%

	\def\CH{{\cal H}}	
		
		\def\CO{{\cal O}}

\def\2pi{\hbox{$2\pi i$}}

\def\dsl{\raise.15ex\hbox{/}\kern-.57em\partial}
\def\Dsl{\,\raise.15ex\hbox{/}\mkern-.13.5mu D}
%
%
%
\font\numbers=cmss12
\font\upright=cmu10 scaled\magstep1
\def\stroke{\vrule height8pt width0.4pt depth-0.1pt}
\def\topfleck{\vrule height8pt width0.5pt depth-5.9pt}
\def\botfleck{\vrule height2pt width0.5pt depth0.1pt}
\def\Zmath{\vcenter{\hbox{\numbers\rlap{\rlap{Z}\kern
0.8pt\topfleck}\kern
2.2pt
                   \rlap Z\kern 6pt\botfleck\kern 1pt}}}
\def\Qmath{\vcenter{\hbox{\upright\rlap{\rlap{Q}\kern
                   3.8pt\stroke}\phantom{Q}}}}
\def\Nmath{\vcenter{\hbox{\upright\rlap{I}\kern 1.7pt N}}}
\def\Cmath{\vcenter{\hbox{\upright\rlap{\rlap{C}\kern
                   3.8pt\stroke}\phantom{C}}}}
\def\Rmath{\vcenter{\hbox{\upright\rlap{I}\kern 1.7pt R}}}
\def\Z{\ifmmode\Zmath\else$\Zmath$\fi}
\def\Q{\ifmmode\Qmath\else$\Qmath$\fi}
\def\N{\ifmmode\Nmath\else$\Nmath$\fi}
\def\C{\ifmmode\Cmath\else$\Cmath$\fi}
\def\R{\ifmmode\Rmath\else$\Rmath$\fi}


\def\del{\partial}
\def\clsd{c_{l\sig}^\dagger}
\def\cls{c_{l\sig}}
\def\cesd{c_{e\sig}^\dagger}
\def\ces{c_{e\sig}}
\def\up{\uparrow}
\def\do{\downarrow}
\def\il{\int^{\tilde{Q}}_Q d\la~}
\def\ilp{\int^{\tilde{Q}}_Q d\la '}
\def\ik{\int^{B}_{-D} dk~}
\def\ila{\int d\la~}
\def\ilpa{\int d\la '}
\def\ika{\int dk~}
\def\tQ{\tilde{Q}}
\def\rh{\rho_{\rm bulk}}
\def\ri{\rho_{\rm imp}}
\def\sh{\sig_{\rm bulk}}
\def\si{\sig_{\rm imp}}
\def\rph{\rho_{p/h}}
\def\sph{\sig_{p/h}}
\def\rp{\rho_{p}}
\def\sp{\sig_{p}}
\def\drph{\delta\rho_{p/h}}
\def\dsph{\delta\sig_{p/h}}
\def\drp{\delta\rho_{p}}
\def\dsp{\delta\sig_{p}}
\def\drh{\delta\rho_{h}}
\def\dsh{\delta\sig_{h}}
\def\enp{\ep^+}
\def\enm{\ep^-}
\def\enpm{\ep^\pm}
\def\enph{\ep^+_{\rm bulk}}
\def\enmh{\ep^-_{\rm bulk}}

\def\enh{\ep_{\rm bulk}}
\def\eni{\ep_{\rm imp}}

\Title{cond-mat/}
{\vbox{\centerline{Transport in Quantum Dots}
\centerline{from the Integrability of the Anderson Model}}}

\centerline{Robert M. Konik}
\centerline{Department of Physics, University of Virginia}
\centerline{Charlottesville, VA 22903}

\medskip
\medskip
\centerline{Hubert Saleur}
\centerline{Department of Physics, University of Southern California}
\centerline{Los Angeles, CA 90089-0484}

\medskip
\medskip
\centerline{Andreas Ludwig}
\centerline{Department of Physics, University of California}
\centerline{Santa Barbara, CA 93106}

\bigskip
\bigskip

In this work
we exploit the integrability of the two-lead
Anderson model to compute transport
properties of a quantum dot, in and out of 
equilibrium.  Our method combines the properties of integrable scattering
together with a Landauer-Buttiker formalism.
Although we use integrability, the nature of the
problem is such that 
our results are not generically {\it exact}, but must only be considered
as excellent approximations which nonetheless are
valid all the way through crossover regimes.  

The key to our approach is to identify the excitations that correspond
to scattering states and then to 
compute their associated scattering amplitudes.
We are able to do so both in and out of equilibrium.  In 
equilibrium and at zero temperature, 
we reproduce the Friedel sum rule for an arbitrary magnetic
field.  From this we compute exactly the behaviour
of the zero temperature linear response conductance as a function of 
both the gate voltage and the field.  We also
study transport quantities requiring knowledge of scattering
states away from the Fermi surface.
We compute the linear response
conductance at finite temperature at the symmetric point of the Anderson
model, and  reproduce Costi et al.'s numerical renormalization
group computation of this quantity.  We then explore the out-of-equilibrium
conductance for a near-symmetric Anderson model, and   
arrive at quantitative expressions for 
the differential conductance,
both in and out of a magnetic field.
We reproduce the expected splitting of the differential conductance
peak into two in a finite magnetic field, $H$.  We determine the
width, height, and position of these peaks.  In particular
we find for $H \gg T_k$, the Kondo temperature, the differential
conductance has maxima of $e^2/h$ occuring for a bias $V$ close to
but smaller than $H$.
The nature of our construction of scattering states suggests
that our results for the differential magneto-conductance are not
merely approximate but become
exact in the large field limit.

\vskip .2in

\Date{08/00}

\newsec{Introduction} 

The Kondo effect is a cynosure of modern condensed matter physics.  
Due to the strongly coupled nature of its IR fixed point,
understanding its low energy behaviour has proven a major theoretical
challenge.
Typically the phenomena refers to the interaction of isolated magnetic
impurities in a bulk metal.  However in the last several years the
experimental study of single magnetic impurities has moved to a new
arena, that of quantum dots connected to external leads
\ref\gold{D. Goldhaber-Gordon, J. G\"{o}res, M. Kastner, H. Shtrikman, 
D. Mahalu, and U. Meirav, cond-mat/9807233.}
\ref\cron{S. Cronenwett, T. Oosterkamp, and L. Kouwenhoven, cond-mat/9804211.}
\ref\golda{D. Goldhaber-Gordon, H. Shtrikman, D. Mahalu,
D. Abusch-Magder, U. Meirav, and M. Kastner, Nature 391 (1998) 156.}
\ref\wiel{W.G. van der Wiel, S. De Franceschi, T. Fujisawa,
J. Elzerman, S. Tarucha, and L. Kouwenhoven, Science 289 (2000) 2105.}
\ref\nygard{J. Nygard, D. Cobden, and P. Lindelof, Nature 408 (2000) 342.}.
In analogy to the traditional realization of the Kondo effect, the leads
serve as the bulk metal and the dot as the magnetic impurity.
The appearance of the Kondo effect in this new setting has brought
a new set of theoretical challenges: how to compute transport quantities
that form the main experimental signatures of these systems.

Quantum dots come in at least two forms.
Semi-conductor quantum dots \gold\cron\golda\wiel~are 
a product of the continuing project of the miniaturization
of solid state devices.
They are fabricated by confining electrons in a two dimensional
electron gas (2DEG) within a GaAs/AlGaAs structure
using a combination of metallic gates.  The region to which the electrons
are confined is small enough that its energy levels may be considered discrete.
The dot is connected to source and drain contacts 
(the two leads).  Schematically
the quantum dot can be pictured as follows:
\vskip .15in
\centerline{\psfig{figure=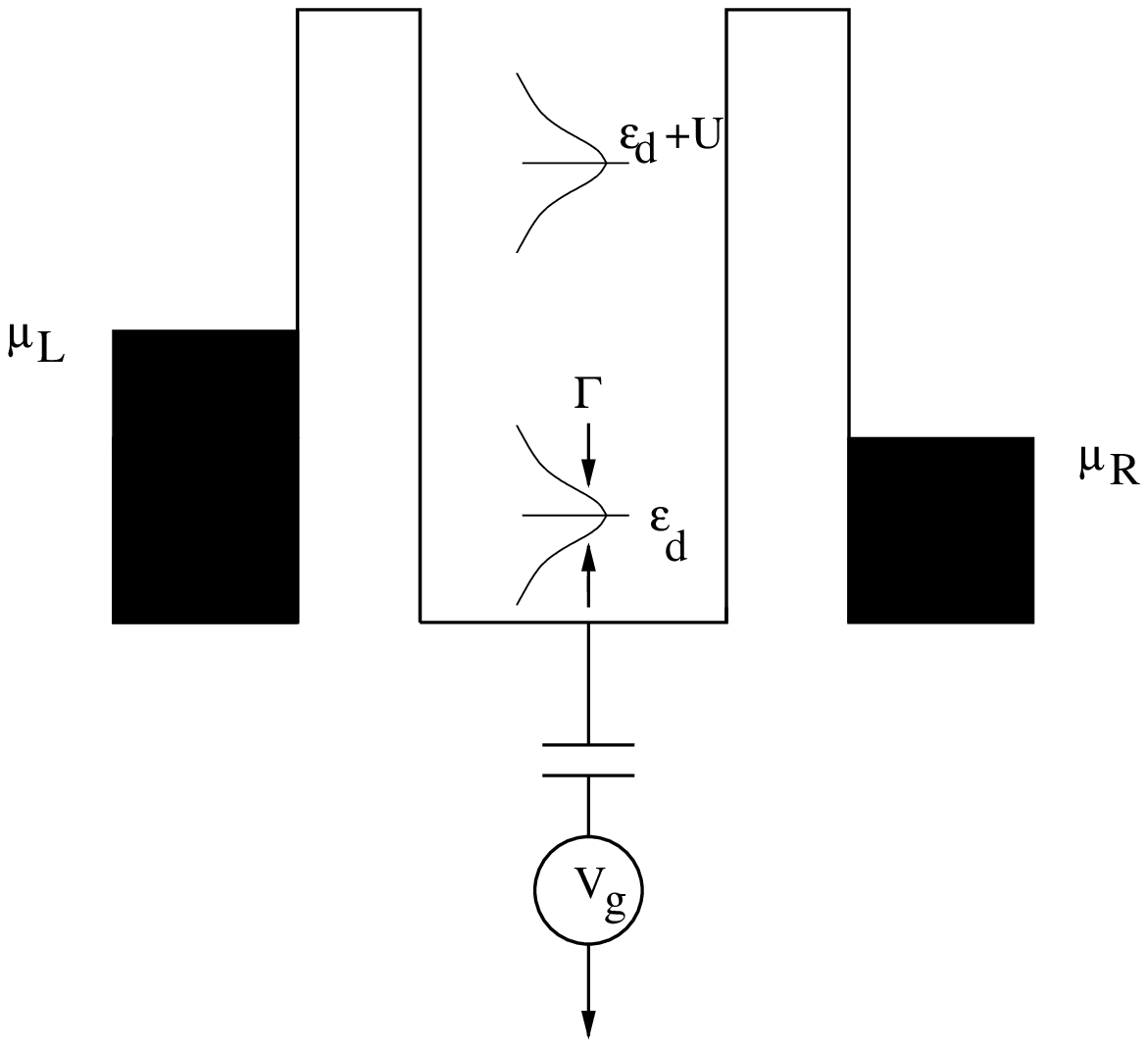,angle=-90,height=3in,width=3in}}
\indent\vbox{\noindent\hsize 5in Figure 1: A schematic of the
quantum dot.}

\noindent The source and drain can be held at any relative voltage, thus 
enabling the study 
of both linear response and out-of-equilibrium transport quantities.  Beyond
the gates that serve to confine the electrons of the 2DEG, additional gates
can be deposited on the GaAs/AlGaAs heterostructure.  Such gates capacitively
couple to the quantum dot through a gate voltage, $V_g$, thus allowing
the chemical potential of the dot to be adjusted.  This has two
important consequences.  By adjusting the gate voltage, one can tune
the number of electrons in the confined region to be odd so that there
is a single unpaired electron.  
The presence of an unpaired electron allows the
appearance of Kondo-like physics.  Moreover by tuning $V_g$, the unpaired
electron's chemical potential can be adjusted thus controlling the
scale, $T_k$, where Kondo physics sets in.  Such systems are thus said to
possess a ``tunable'' Kondo effect \cron .

Quantum dots may also be fabricated from metallic carbon 
nanotubes \nygard .  By depositing metallic leads on top of a 
small section of a carbon
nanotube, an effective
quantum dot is made.  Like their semiconductor counterparts, these
dots are tunable: gates may be added to the semiconductor substrate
upon which the nanotube/leads lay.  Semiconductors dots typically
carry 10-100 electrons.  Nanotube dots, in contrast, have
many thousands of electrons and yet still exhibit Kondo-like physics.

The transport quantities that lie at the focus of the experimental study
of a tunable Kondo effect in quantum dots have been measured under a
variety of conditions.  Conductances of the lead-dot system have been
determined both in and out of equilibrium and both in and
out of the presence of a magnetic field.  Remarkably this wide variety
of experimental phenomena is thought to be described by a conceptually
simple theory: the Anderson model.

The Anderson model is fashioned from a chain of non-interacting spinful
fermions (the leads) connected via hopping to a single site
impurity on which alone Coulomb repulsion is present.
On a lattice, the Hamiltonian reads (with no bias and H=0):
\eqn\eIi{\eqalign{
H &= \sum_{i\alpha} -t (c^\dagger_{i,\alpha} c_{i+1,\alpha} + {\rm h.c.})
+ Un_{d\up}n_{d\do}\cr
& + \sum_{\alpha} \big(V_1 (c^\dagger_{-1,\alpha} d_{\alpha} + h.c) +
V_2 (c^\dagger_{1,\alpha} d_{\alpha} + h.c)\big)
+ \ep_d (n_{d\up} + n_{d\do}).}}
Here $c^\dagger_{\alpha}/c_\alpha$ are the lead electron creation/destruction
operators, $d^\dagger/d$ the dot electron creation/destruction operators,
and $n_d = d^\dagger d$.  The dot is considered to reside at $x = 0$.
The index $\alpha$ indicates spin species.
The interaction on the dot is present in the term $Un_{d\up}n_{d\do}$.
Although deceptively simple, the presence of a non-zero U makes the 
problem many-body with all of its manifold complications.

U is pictured in Figure 1 and represents the charging energy incurred when an
electron is added to the dot.  Roughly it can be estimated as
\eqn\eIii{
U = {e^2 \over 2C} + \Delta \ep ,}
where $C$ is the capacitance of the dot and $\Delta \ep$ is 
the dot's energy level
spacing.  For the experiments at hand, $U \sim  1 {\rm meV}$.  The 
counterpart of 
the gate voltage, $V_g$, in the above Hamiltonian is $\ep_d$, the parameter
that controls the chemical potential of the electrons on the dot.
By adjusting $\ep_d$, the number of electrons on the model dot can be varied
from zero to two.  (Although
there may be a large number of electrons on the actual dot,
the concern here both theoretically and experimentally is of electrons
in the highest occupied energy level.)  
The final pair of parameters, $V_{1,2}$, measure
the height of the tunnel barriers between the leads and the dot.  In general,
they differ between the two leads.  The quantity, $\Gamma$, (see Figure 1)
measuring the width of the level resonance that arises through
the interaction of the leads and the dot is given in terms
of the tunnel barrier heights to be $\Gamma = (V_1^2+V_2^2)/2$.
Typically $\Gamma/U \ll 1$ in the experiments of concern \gold\cron\golda\wiel .

Although the above described experimental results 
have come relatively recently, the theoretical
study of transport through impurities is much older.
Appelbaum and Anderson both studied conductance anomalies 
present in tunnel junctions due to the presence of
magnetic impurities \ref\aa{J. Appelbaum, Phys. Rev. Lett. 17 (1966) 91;
P. Anderson, Phys. Rev. Lett. 17 (1966) 95.}.  
However their efforts were perturbative
in nature and did not describe the strong coupling nature of
the Kondo effect.
More recently, Ng and Lee
\ref\nglee{T. Ng and P. Lee, Phys. Rev. Lett. 61 (1988) 1768.},
studied the linear response conductance both in and out of a
magnetic field using the Friedel sum rule.  The Friedel sum rule
relates the scattering phase of the electrons at the Fermi surface
to the average number of
electrons sitting on the dot.  However the Friedel sum rule is
useful only in determining the linear response conductance, a consequence
of the rapid variation in the scattering phase as one moves away from the
Fermi surface.  In contrast to the linear response conductance where
the Friedel sum rule is an exact relationship, the techniques used to determine
out-of-equilibrium transport are limiting in nature.  
In one approach, a non-crossing approximation (NCA)
\ref\hersh{
M. Hettler, J. Kroha, and S. Hershfield,
Phys. Rev. Lett. 73, 1967 (1994); ibid, Phys. Rev. B 58 (1998) 5649.}
\ref\win{
N. Wingreen and Y. Meir, Phys. Rev. B 49 (1994) 11040.}
was employed.
The NCA approach has drawbacks.  In order to implement the associated
use of slave bosons, one must take $U = \infty$. 
Moreover,
NCA is in some sense a
large N approximation where N is the number of spin degrees of freedom 
of the impurity
(in this case N=2).  It is known to be remarkably accurate in computing
thermodynamics.  However it is less accurate when it comes
to transport quantities ($\sim 15\%$ errors \win ) due 
to its less accurate prediction
of behaviour right at the Fermi surface.  And as such 
these difficulties render it unusable in non-zero magnetic fields \win .
In another approach to computing non-equilibrium properties, 
a clever combination of the analysis of
the equations of motion with perturbation theory was employed
to study the differential magneto-conductance
\ref\meir{Y. Meir, N. Wingreen,
and P. Lee, Phys. Rev. Lett. 70 (1993) 2601.}.
However the truncation of the equations of motion necessary
to perform the analysis in this work is in some sense an uncontrolled
approximation.  The authors of \meir\ indicate
that their methodology underestimates the
magnitude of the differential conductance.
Another set of approaches 
have relied upon perturbation theory
\ref\sivan{N. Sivan and N. Wingreen, Phys. Rev. B 54 (1996) 11622.}
\ref\hersh1{S. Hershfield, J. Davies,
and J. Wilkins, Phys. Rev. B 46 (1992) 7046.}
\ref\kaminski{A. Kaminski, Y. Nazarov, and L. Glazman, cond-mat/0003353.}.
As with the results of \aa ,
perturbation theory requires relatively small
U (Coulomb repulsion) or alternatively, temperatures
far in excess of the Kondo temperature, and so presumably
can access, at best, qualitative, not quantitative,
features of the strongly coupled
physics found in the Kondo regime of quantum dots at low temperatures.

These inherent difficulties with the out-equilibrium Anderson
were circumvented in the study of a non-equilibrium Kondo impurity
at its Toulouse point \ref\shl{A. Schiller and S. Hershfield,
Phys. Rev. B 51 (1995) 12 896; ibid 58 (1998) 14978.}.
At this point, the model can be mapped
to a system of non-interacting fermions, thus permitting an
exact solution.  It is unclear, however, how the Toulouse limit
affects the underlying physics.  Although the ordinary Kondo
model shares the same IR fixed point as its Toulouse counterpart,
we are interested in part in physics for large applied field,
bias, and temperature, that is, in physics far away from this fixed point.

Given the limitations of these methods, one cannot help but
notice that the Anderson model is exactly solvable.  Indeed this integrability
has already been exploited through Bethe ansatz solutions to compute
thermodynamic quantities
\ref\kao{N. Kawakami and A. Okiji, Phys. Lett. A 86 (1982) 483;
ibid. J. Phys. Soc. Japan 51 (1982) 1143; ibid Solid St. Commun. 
43 (1982) 365; Okiji, A. and Kawakami, N., 
J. Phys. Soc. Japan 51 (1982) 3192.}
\ref\wie{P. Wiegmann, V. Filyov, and A. Tsvelick, Soviet Phys. JETP Lett. 
35 (1982) 77; 
P. Wiegmann and A. Tsvelick, Soviet Phys. JETP Lett. 35 (1982) 100;
ibid., J. Phys. C 16 (1982) 2281;
A. Tsvelick and P. Wiegmann, Phys. Lett. A 89 (1982) 368;
ibid., J. Phys. C 16 (1983) 2281;
V. Filyov, A. Tsvelick, and P. Wiegmann, Phys. Lett. A 89 (1982) 157.}
such as the specific heat and magnetic susceptibility.  But what of transport
quantities?  A limited attempt 
to deduce information about transport properties
from the Bethe ansatz solution of the {\it Kondo} model 
was made recently
\ref\mwen{J. Moore and X. Wen, cond-mat/9911068.}.
There the equilibrium impurity density
of states that arises from the Bethe ansatz was studied.  
In general, the impurity density of states coming from the Bethe ansatz
is unrelated to the spectral density of states arising from the
dot correlator, ${\rm Im}\langle dd^\dagger\rangle$.
Indeed at zero temperature and zero field, it is clear the
two are much different quantities (the heights of 
zero energy peaks in both quantities are controlled by far
different energy scales).  But in methodology of \mwen , it is this latter
quantity, ${\rm Im}\langle dd^\dagger\rangle$, that is 
directly related to transport \hersh\win\meir .
Moreover, 
the context of their computation, as determined in \hersh\win\meir , demands that
the {\it non-equilibrium} properties of ${\rm Im}\langle dd^\dagger\rangle$
be computed.
Given the general unavailability from integrability of information 
about correlators 
such as ${\rm Im}\langle dd^\dagger\rangle$, a different approach
is needed to extract transport properties from the exact solvability
of the model.  Here we advocate a Landauer-B\"uttiker approach
to transport,
and so are instead faced with the task of identifying
scattering states in the context of integrability.

The key feature of an integrable system is the exact knowledge of
eigenfunctions of the fully interacting Hamiltonian.  In turn there is a
well-defined notion of elementary excitations.  
In particular these excitations have an infinite lifetime:
integrability forbids any decay processes from occurring.  This arises
from the infinite series of non-trivial conservation laws in the model.
In some sense an integrable system is a superior version of a Fermi liquid.

In the Anderson model, there is such a set of excitations, as detailed
in Sections 2 and 4.  They are not
on the face of it, however, particularly electronic.
And if we are to understand the transport of the
sea of electrons in the attached leads, we necessarily need
scattering states which carry the quantum numbers of an electron.
Rather the excitations divide 
into separate spin and charge sectors.  The closest they come to being
electronic is in bound states between excitations which can be thought
of as bound states of electrons.
This is not so unnatural.  If one were
to bosonize the Anderson model, one would find that the degrees of freedom
separate into spin and charge bosons.
But this is only one problem with the excitations arising from integrability.
These excitations, as explained in Section 2, are a combination
of degrees of freedom in {\it both} of the leads connected to the dot.  
And it is the case that this entanglement cannot always be simply reversed.

And so there is the difficulty.  The scattering states are not necessarily
electronic in
nature 
and not confined to a single lead.  Only if one can understand 
electronic
excitations in an individual lead can one hope to make sense of scattering
amplitudes off the dot.  It is these two facts that have prevented the 
integrability of the Anderson model from being applied to transport quantities
up to now.

However we have managed to circumvent these problems in a number
of cases.  In particular we have successfully described both scattering
states at the Fermi surface for generic values of $U$, $\ep_d$, and $\Gamma$,
and scattering at finite energies at the
symmetric point of the Anderson model, $U = -\ep_d/2$.
There we argue that
by correctly
gluing together a spin and charge excitation, we are able to form an
excitation that is electronic in nature.  
Moreover the excitations are such
that one can understand them in terms of the individual leads and so
compute reflection and transmission amplitudes of the excitation
off the dot.  We do so in an argument akin to that used by N. Andrei
\ref\andrei{N. Andrei, Phys. Lett. A. 87 (1982) 299.} in computing
the magnetoresistance in the Kondo model.  There he argues that the scattering
phase of an excitation can be identified with its impurity momentum.  
In turn this momentum is related to the impurity density of states
as it appears in the Bethe ansatz and so can be directly computed.

We now turn to how we use these excitations to compute transport
quantities.  All such quantities could be expressed in terms
of the scattering of asymptotically free electrons (i.e. electrons
in the attached leads) off the quantum dot.  
However such scattering, in general, is not particular simple.
In general away from the Fermi surface such scattering will be
inelastic and involve particle-hole production.
We, however, recast the density of states of
asymptotically free fermions in terms of the
integrable excitations we have identified.  Because of their
integrability, their scattering is simple: their character does
not change in scattering through the dot and their transport
can be described individually: they scatter one-by-one through
the dot.

It is however unlikely that 
the integrable electronic 
excitations we use in computing transport properties provide
exact results in all cases  - the issues involved here are subtle and will
be discussed in detail in the next sections.  
In particular, it is unlikely the high energy limit of the
excitations we construct are entirely confined to a single lead.
However this methodology successfully passes
a number of tests.
The first test of our method comes in proving the Friedel sum rule.  The
Friedel sum rule relates the occupancy of spin $\up/\do$ electrons on 
the quantum dot, $n_{d\up/\do}$,
to the scattering
phase of an electron of the same spin, $\delta_{e\up/\do}$, 
at the Fermi surface:
\eqn\eIiii{
\delta_{e\up/\do} = 2\pi n_{\up/\do}.}
It thus relates a dynamic 
quantity to a thermodynamic quantity.  With this 
in hand, previous works have
computed the linear response conductance from the knowledge of this occupancy,
at least at $H=0$
\ref\vdelft{J. von Delft, 
U. Gerland, T. Costi, and Y. Oreg, cond-mat/9909401.}
\ref\meyer{D. Meyer, 
T. Wegner, M. Potthoff, and W. Nolting, cond-mat/9905089.}.
However such works do not make any attempt to explicitly identify the
excitations that scatter according to the Friedel sum rule.  Here
we do so.  We show that the scattering phase of the excitations we have
identified to be the same as that predicted by the Friedel sum rule both
in and out of a magnetic field.  As we have 
reproduced the Friedel sum rule, we can say that
the excitations we have identified coincide exactly with the free fermions
{\it at the Fermi surface}.

Now while the 
Friedel sum rule only deals with excitations at the Fermi surface,
our methods goes beyond excitations directly at the Fermi surface,
at least near the symmetric point of the Anderson model.
To determine
whether the excitations we have identified together with their
associated scattering amplitudes provide a complete solution of the problem,
we compute the linear response
conductance at finite temperature.
At finite temperature, the excitations
needed for the linear response conductance using a Landauer-B\"uttiker
formalism exist over a range of
energies. 
In particular, we compute the linear response conductance at the 
symmetric point ($-U/2 = \ep_d)$ of the Anderson model as a function 
of temperature, T, and
compare it to Costi et al.'s \ref\costi{T. Costi, A. Hewson, and
V. Vlatic, Journal of
Physics: Cond. Mat. 6 (1994) 2519; T. Costi, cond-mat/0004302} 
numerical renormalization group (NRG)
computation of this quantity and find excellent agreement.
We thus are able
to conclude that using our excitations away from the Fermi
surface is a valid procedure.  

Although our finite temperature computation suggests we have correctly
identified the low (but finite) energy excitations at the symmetric point, 
we do not claim that our
result is {\it exact}. 
Again our inability to make this claim hinges on the question
of whether the integrable excitations we construct are entirely confined
to a single lead and so make appropriate scattering states.
Moreover we know that our prescription for scattering
fails once we leave the 
Kondo regime where approximately one electron sits on the dot
and enter the mixed valence regime of the Anderson model.
This again suggests that near the symmetric point, our methods
are merely highly accurate.
The situation here
is not dissimilar
to form-factors computations of correlation and response functions, where
integrable techniques can provide, if not the exact result 
(which would involve resumming an infinite number of contributions), 
controlled approximations
of excellent accuracy, from the lowest energies through crossover regimes
\ref\mus{J. Cardy and G. Mussardo, Nucl. Phys. B 410 (1993) 451.;
G. Delfino and G. Mussardo, Nucl. Phys B 455 (1995) 724.;
G. Delfino and J. Cardy, hep-th/9712111.;
J. Cardy and G. Mussardo, Nucl. Phys. B 410 (1993) 451.}
\ref\usc{F. Lesage, H. Saleur and S. Skorik, Phys. Rev. Lett. 76
(1996) 3388; Nucl. Phys. B474 (1996) 602.}.

The physical origin of the accurate reproduction of scattering
at non-zero energies relative to the Fermi surface at the Anderson
model's symmetric point lies in a separation of scales.  At this point
there are two relevant scales in the problem: one is the Kondo temperature,
$T_k$, while the other is $\sqrt{U\Gamma}$, a function of the
Coulomb repulsion, $U$, and the resonance width of the dot level, $\Gamma$.
At the symmetric point, $T_k \ll \sqrt{U\Gamma}$.  We exploit this
fact to make our identification of integrable scattering states.
But in turn this means that we expect errors 
in transport quantities involving scattering away
from the Fermi surface of  $\CO (T_k/\sqrt{U\Gamma})$.

Bootstrapping from our success with the finite temperature
linear response conductance, we look at the non-equilibrium conductance 
near the symmetric point both in and
out of a magnetic field at zero temperature.  In order to compute this 
conductance we again 
employ a Landauer-B\"uttiker formalism akin to that employed in
computing the out-of-equilibrium conductance of interacting quantum Hall 
edges
\ref\FLS{P. Fendley, A.W.W. Ludwig, and H. Saleur,
Phys. Rev. Lett. 74 (1995) 3005; ibid. 75 (1995) 2196;
Phys. Rev. B52 (1995) 8934.}.
We imagine placing each lead at two differing chemical potentials, 
$\mu_1$ and $\mu_2$.  
These differing voltages induce different
populations of free electrons in the leads.  As with the finite temperature
linear response problem, we recast these electrons in terms of
our integrable scattering states.  We then compute the (equilibrium) 
scattering amplitudes of scattering states in the leads.
These scattering amplitudes then provide the probability for a state
to tunnel from one lead to the other.  Although the system is
interacting, its integrability again implies the states scatter
one-by-one.  It is important to understand that this means
of computation introduces no additional error into the calculation.
The sole source of uncertainty is found in whether the scattering
states that we construct are entirely confined to a single lead.
But because of the
excellent agreement of the finite T linear response conductance with the
previous NRG results, we expect this error to be similarly insignificant
in our non-equilibrium computations.

It is important to understand that this approach to the non-equilibrium
physics differs from that used in \hersh\win\meir\ in a fundamental way.
There the non-equilibrium conductance is expressed in terms of the 
non-equilibrium density of states of the impurity as determined
from the correlator, ${\rm Im}\langle dd^\dagger\rangle$.
Here we have nothing 
direct to say about the 
non-equilibrium (or indeed, the equilibrium) behaviour of this quantity.
 
We must stress this as the reader may be confused by the fact
we do use the impurity momentum (which is in turn related to 
the Bethe ansatz impurity density of states as explained in Section 2) 
to compute the scattering amplitudes of
excitations off the dot.  This confusion may be heightened in that
we employ the equilibrium Bethe ansatz
density of states in computing the scattering
matrices.  It would thus seem legitimate to ask why we do not need to use 
a non-equilibrium impurity density of states in computing the 
out of equilibrium conductance \ref\skorik{S. Skorik, cond-mat/9708163}.

The answer lies in correctly understanding the basis of excitations
by which we
compute the conductance
\ref\reply{P. Fendley, A. Ludwig, and H. Saleur, cond-mat/9710205.}.
We are able to use the equilibrium scattering
phases as we employ the basis that is naturally present when
the system is equilibrated.  However because of the integrability
of the system, we can continue to employ this basis when we move
the system out of equilibrium.  These particles continue to scatter
as they do in equilibrium.  We note that if one were to compute
out-of-equilibrium scattering matrices, one would find
that they differ from their 
in-equilibrium counterparts
by an overall phase alone.  As transport quantities depend upon the absolute
value of the scattering, this overall phase would then have no effect \reply .
While the application of finite voltage 
does not effect the scattering of the excitations,
it does change their distribution in the leads.  And indeed we must and do take
this into account.  

The rational behind this understanding has been tested
beyond the various checks of these ideas 
applied by \FLS~in their computations of
conductances of quantum Hall edges.  
Generically 
the thermodynamics of an integrable field theory can be computed using
thermodynamic Bethe ansatz which employs zero temperature S-matrices
(and not finite temperature S-matrices as might again be n\"aively expected --
the finite temperature distribution might be thought
to {\it necessarily} dress the scattering).
With the thermodynamics, one can determine the finite temperature
scaling behaviour of a field theory.  If
in the UV or IR limit, the theory flows to a conformal field theory,
the finite scaling behaviour in these limits is independently
determined by the central charge, c, of the theory.  That the two computations
always agree provides strong evidence we are handling the problem
correctly.

Turning to our non-equilibrium results, we find that they 
reproduce the expected gross features of the 
experimental differential
conductance.  When $H=0$ and we are near the symmetric point,
the differential conductance is sharply peaked about its linear response
value.  The peak width is controlled by the scale, $T_k$, the Kondo
temperature.  We find that the peak is roughly symmetrical
about $V=0$  in accordance with experiment \gold .

In a non-zero field, Meir and Wingreen \meir\ predicted that the differential
conductance would peak at $eV = \pm H$.  We find such peaks with
our techniques although our peaks are found shifted to values of $e|V|$
notably smaller than $|H|$.  Even in the limit of fields much larger than
$T_k$, we do not find the peaks at $|H|$.
This is again consistent with
experiment \golda .  
We should not necessarily expect the peaks to occur
exactly at $eV=H$ as Meir and Wingreen's prediction is predicated in part 
upon a second order perturbative result.  Moreover our construction of
the scattering states suggests 
that in the particular case of the differential magneto-conductance, our
results become exact in the limit of large applied fields.

A portion of the results of this paper have been reported in
\ref\short{R. M. Konik, H. Saleur, A. W. W. Ludwig, cond-mat/0010270.}.
Here in this work we provide far greater detail on the nature of
our computations.
The paper is so organized as follows.  In Section 2 we introduce the continuum
version of the two lead Anderson model.  The two lead Anderson model
is integrable as is.  However we first map it onto a one lead problem.
If we were to explicitly solve the two lead problem, we would find nevertheless
that we would be implicitly
implementing the map to the one lead case.  
Having done this we review the Bethe ansatz for
the one lead Anderson model together with the excitations necessary
to form the ground state at zero temperature.  The remaining portion of 
Section 2 is devoted to identifying the excitations 
(both at and away from the
Fermi surface) that can be identified with scattering states and
then computing their scattering amplitudes.  We provide
further details of the approximate nature of the scattering states
so identified away from the Fermi surface.  In the course of this
discussion we demonstrate the Friedel sum rule.  

In Section 3 we explore the behaviour of the $T=0$ linear response conductance
both in and out of a magnetic field.
Because Wiegmann and Tsvelick 
\ref\rev{A. Tsvelick and
P. Wiegmann, Adv. in Phys. 32 (1983) 453.}
computed expressions for the occupancy of the dot, $n_{d\up/\do}$,
as a function of the gate
voltage, $\ep_d$, and H in a variety of regimes, and the Friedel sum rule
relates the electron scattering phase to this occupancy, we can derive
closed form
expressions for the linear response conductance in these same regimes.
Outside these regimes we compute the occupancy numerically.
Using these results, we show how the linear response conductance behaves as
function of $\ep_d$.  

We plot the linear response conductance as a function of the
gate voltage, $\ep_d$, in zero field in Figure 4.
However the most interesting results of this section are found in
our computations of the linear response conductance at finite H.
In Figure 5 we plot the linear response conductance 
as function of $\ep_d$ for a variety of values of $H$.
As $H$ is increased from zero we
see that 
the linear response conductance is suppressed as a result of the
destruction of the Kondo effect in a finite field.  We also
see the structure of the conductance peak evolve from its zero field
value to that of free fermions.  With increasing
H, the full width at half maximum becomes narrower, decreasing from 
its zero field value of $U$ to its free-fermion value of $2\Gamma$.
The peak height, in turn, decreases from its maximal value of $2e^2/h$ 
to $e^2/h$.  And finally the location of the peak shifts from $\ep_d=-U/2$
to $H/2$ (we scale H here and throughout this paper such that $g\mu_B = 1$), 
appropriate to a spin up free-fermion with a field-induced,
shifted chemical potential.  These results are encoded in Figure 6.

At the symmetric point $U/2=-\ep_d$, we provide a simple closed
form expression for the conductance (see Section 4.2).  We then find
that the conductance deviates from its unitary maximum for small fields
via 
\eqn\eIiv{
G = 2{e^2\over h}(1-{\pi^2 \over 16}{H^2\over T_k^2}+\CO (H^4/T_k)).
}
This deviation from the maximal conductance is quadratic,
appropriate for the controlling $H=0$ strongly coupled Fermi liquid fixed
point.
Here $T_k$, the Kondo temperature,
is given in (3.16).

In Section 4 we compute the finite temperature linear response conductance
at the symmetric point, $U/2= -\ep_d$.  
This requires recomputing the scattering
of Section 2.  At finite temperature one must consider the thermal bath
of all possible excitations.  
This highly non-trivial bath modifies the scattering.
However doing so leads us to a highly pleasing result.  We are able to
reproduce Costi et al.'s NRG result for G as a function of T 
(see Figure 9).  This is 
convincing evidence that we have correctly identified the scattering states
(at least at the symmetric point of the dot).
More specifically, we know the linear response conductance will
again have a Fermi liquid form:
\eqn\eIv{
G(T) = 2{e^2\over h}(1-c{T^2\over T_k^2}+\CO (T^4/T_k)).
}
Costi et al. demonstrated in perturbation theory
that the constant, $c$, takes the value,
$c = \pi^4/16 = 6.088$.  We compute in comparison
$c = 6.05 \pm .1$.  Beyond the low temperature
Fermi liquid regime, we emphasize we are able to describe accurately
the conductance in the crossover regime $T \sim T_k$.

We also compare the scaling curve for the finite temperature
linear response conductance with the experimental data found
in \gold .  The comparison is plotted in Figure 10.
We find the experimental measurements, even though taken
away from the symmetric point of the dot (although
still in its Kondo regime), agree well with our scaling curve.
This suggests the Kondo regime of the dot exists over a 
wide range of gate voltages (i.e. $\ep_d$).

In Section 5 we move on to compute the non-equilibrium conductance at
zero temperature.  Again the scattering amplitudes need to be recomputed
to take into account the change in the distributions of electrons in
the leads induced by the finite bias.  We also discuss subtleties with 
understanding how to think of a finite biased system in its one lead
formulation.  We then present results of the differential conductance both
in and out of a magnetic field as discussed above.

We are able to derive a number of simple closed form results for the
out-of-equilibrium conductance.  The differential conductance in zero
field at the symmetric point is computed to be
\eqn\eIvi{
G(\mu_1,\mu_2) = 2{e^2\over h} {1\over (1+{\pi^2(\mu_1-\mu_2)^2\over 4T_k})},
}
a remarkably simple result. 
We are also able to characterize
the peak in the differential conductance that develops in the presence
of a magnetic field.  The quantity most discussed in the literature, the
bias at which the peak occurs, is given by the expression
\eqn\eIvii{\eqalign{
eV_{\rm max} &= -H(1- {1\over 2\pi}\tan^{-1}{1\over I^{-1} -b} +\cdots ),\cr
I^{-1} - b &= {1\over \pi} \log ({H\over 2T_k}{\sqrt{\pi e\over2}}),
}}
valid for $H \gg T_k$ and $H \ll \sqrt{U\Gamma}$.  Because of the
logarithmic dependence of $b$ upon $H$, the position of the peak
approaches $eV=|H|$ extremely slowly.
In addition to the location
of the peak, we are also able to describe both the peak width and
the height of the peak.  The peak width is given by
\eqn\eIviii{
e\Delta V = {H\over 2\pi}\bigg(\tan^{-1}{1 \over I^{-1}-1/2-b} 
- \tan^{-1}{1 \over I^{-1}+1/2-b}\bigg),
}
while the peak height is equal to
\eqn\eIix{
G_{\rm max} = {e^2\over h}\big({3\over 2} - {(I^{-1} - b)\over
\sqrt{4(I^{-1}-b)+1}}\big).
}
We see that the height of the conductance peak
approaches $e^2/h$, one half the unitary limit,
in the asymptotic limit of
large $H$.  
Interestingly, the non-equilibrium Kondo model in the Toulouse limit 
\shl\ also predicts a
zero temperature differential magneto-conductance peak at $eV=H$
which has a peak of $e^2/h$,
one-half the unitary limit.

\newsec{Basic Formalism}
\subsec{Description of the System}

Pictured in Figure 2 is a sketch of the quantum dot connected to two leads.
The Hamiltonian for this model in the continuum limit is given by 
\eqn\eIIi{\eqalign{
\CH &= \sum_{l\sig} \int^\infty_{-\infty} dx \{ -i\clsd (x) \del_x \cls (x)
+ V_l \delta (x) [ \clsd (x) d_\sig + d^\dagger_\sig \cls (x)]\} +\cr
& \hskip 1in \ep_d \sum_\sig n_\sig + U n_\up n_\do ,}}
where $n_\sig = d^\dagger_\sig d_\sig$.  Here $\sum_l$ is a sum over the
two leads ($l = 1,2$).  We have allowed for the possibility that the
hopping matrix element, $V_l$, differs between the leads as is
typical in any experimental realization.  
Rather than treating the leads as half-lines with both left and right
moving fermions, we represent the leads as `unfolded' with fermions
that are solely right-moving.  
Fermions in either lead that are incident upon the dot are considered to lie
in the region, $x<0$, while those traveling away from
the dot in either lead are found with $x>0$.
We represent this in Figure 2 by drawing the leads as elongated arcs.
\vskip .4in
\centerline{\psfig{figure=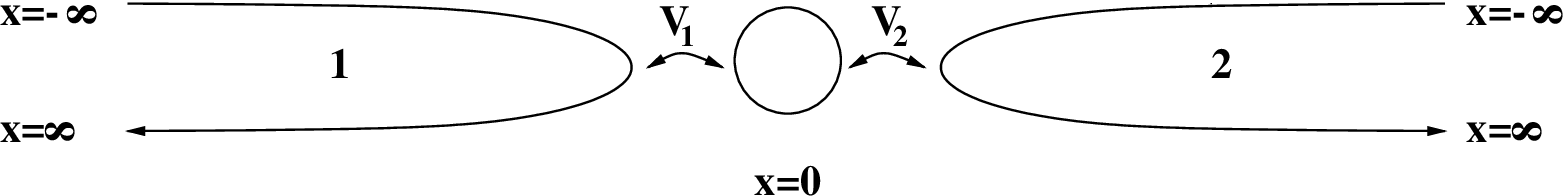,height=.75in,width=5in}}
\indent\vbox{\noindent\hsize 5in Figure 2: A sketch of two leads
attached to a quantum dot.}
\vskip .4in 
\noindent We stress that there are no interactions in the leads (as opposed to 
the calculations in \FLS\ for instance): 
the non-trivial physics of the problem 
arises solely from the strongly interacting dynamics of the dot.

It will be advantageous to reformulate this problem as a one-lead
Anderson model (i.e. a single lead model).  To do so, we introduce
even/odd electrons:
\eqn\eIIii{
c_{e/o} = {1 \over \sqrt{V_1^2 + V_2^2}} ( V_{1/2} c_1 \pm V_{2/1} c_2 ).}
Recasting $\CH$ in this new basis, the odd electron, $c_o$, decouples and we
are left with
\eqn\eIIiii{\eqalign{
\CH &= \sum_{\sig} \int dx \{ -i\cesd (x) \del_x \ces (x)
+ (V_1^2+V_2^2)^{1/2} \delta (x) 
\[ \cesd (x) d_\sig + d^\dagger_\sig \ces (x)\]\} + \cr
& \hskip 1in \ep_d \sum_\sig n_\sig + U n_\up n_\do .}}
We have thus reduced the problem to that solved using Bethe
ansatz in a series of papers by Kawakami and Okiji \kao~and
Filyov, Wiegmann and Tsvelick \wie .

With this reformulation of the model, we have to address the question
of computing scattering amplitudes of electronic excitations in the
original two lead problem.  Na\"{\i}vely it would seem we can do this.
Let 
$T(\ep)$ and $R(\ep)$ be defined by:
\eqn\eIIiiia{\eqalign{
& T(\ep ) : {\rm transmission~amplitude~for~an~electronic~
excitation~of~energy,~\ep} ,\cr
& \hskip .37in {\rm to~scatter~from~lead~1~to~lead~2~(or~2~to~1),}\cr
& R(\ep ) : {\rm reflection~amplitude~for~an~
electronic~excitation~of~energy,~\ep} ,\cr
& \hskip .37in {\rm to~scatter~from~lead~1~to~lead~1~(or~2~to~2).}\cr}}
Assuming the amplitudes behave linearly under the even/odd map \eIIii ,
we have (for $V_1=V_2$)
\eqn\eIIiiib{\eqalign{
e^{i\delta_e(\ep )} &= R(\ep ) + T(\ep ) ;\cr
e^{i\delta_o(\ep )} &= 1 =  R(\ep ) - T(\ep ) .}}
We then conclude that,
\eqn\eIIiiic{\eqalign{
T(\ep) &= {e^{i\delta_e(\ep)} - 1 \over 2};\cr
R(\ep) &= {e^{i\delta_e(\ep)} + 1 \over 2},}}
govern scattering in the two lead problem.
However not all electronic excitations behave linearly under the map
\eIIii\ and so things are not always this simple.  We will consider
this in more detail in Section 2.4 and in particular Section 2.5.

As noted, in writing the above two equations, we have assumed $V_1=V_2$.
If $V_1\neq V_2$, the transmission amplitude is scaled by the factor,
$$
{2V_1V_2 \over V_1^2 + V_2^2}.
$$
As this is a constant factor, it always possible to rescale results
to take into account an asymmetry in the dot-lead couplings.  As such,
we will assume throughout the paper that $V_1=V_2$.

\subsec{Bethe Ansatz Solution of the One-Lead Anderson Model}

In order to understand the scattering between the two leads we will
rely upon  
aspects of the one-lead Bethe ansatz solution found in \kao~and
\wie .  As such we summarize briefly the results of this work.  
Applying the Bethe ansatz yields a set
of quantization conditions describing a finite number of 
bare excitations in the system:
\eqn\eIIiv{\eqalign{
e^{ik_j L + i \delta (k_j)} &= 
\prod^M_{\alpha = 1} { g(k_j) - \lambda_\alpha + i/2 \over  
g(k_j) - \lambda_\alpha - i/2};\cr
\prod^N_{j = 1} {\lambda_\alpha - g(k_j) + i/2 \over  
\lambda_\alpha - g(k_j) - i/2} &= - \prod^M_{\beta=1}
{\lambda_\alpha - \lambda_\beta + i \over 
\lambda_\alpha - \lambda_\beta - i},}}
where
\eqn\eIIv{\eqalign{
\delta (k) &= - 2 \tan^{-1} ({\Gamma \over (k-\ep_d)});\cr
g(k) &= {(k-\ep_d-U/2)^2 \over 2 U \Gamma};\cr
\Gamma &= (V_1^2 + V_2^2).}}
As in all problems with an SU(2) symmetry, there are two types of
excitations: charge (with rapidities, k) and spin (with rapidities,
$\lambda$).  Here $N$ is the total number of particles in the system,
and $M$ marks out the spin projection of the system, $2 S_z = N - 2M$
(in zero magnetic field, M = N/2).

When $\ep_d > -U/2$, the 
ground state of the system consists of the following set of
excitations:

\eqn\eIIvi{\eqalign{
&{\rm i) ~N-2M~real~k_j's} ;\cr
& {\rm ii) ~M~ real~\lambda_\alpha's};\cr
& {\rm iii) ~associated~with~each~of~the~M~\lambda_\alpha 's~are~}\cr
& {\rm two~complex~k's,~k^\alpha_\pm ,~described~by~}\cr
& \hskip .2in g(k^\alpha_\pm) = g(x(\lambda_\alpha) \mp iy(\lambda_\alpha))
= \lambda_\alpha \pm i/2;\cr
& \hskip .2in x(\lambda ) = U/2 + \ep_d - 
\sqrt{U\Gamma}(\lambda + (\lambda^2+1/4)^{1/2})^{1/2};\cr
& \hskip .2in y(\lambda ) 
= \sqrt{U\Gamma}(-\lambda + (\lambda^2+1/4)^{1/2})^{1/2}.}}
Although only valid for $\ep_d > -U/2$, we can also understand the
case, $\ep_d < -U/2$, through a particle hole transformation.
If we take the continuum limit of \eIIvi , we no longer deal with
discrete values of $\lambda$ and $k$, but rather go over to smooth 
distributions, $\rho (k)$ for the real $k_j$'s, and $\sig (\la)$
for the $\lambda_\alpha$'s and their associated $k^\al_\pm$'s.
To derive these distributions, we first take the log of \eIIiv :
\eqn\eIIvii{\eqalign{
k_j L + \de (k_j) &= 2\pi N_j - \sum^M_{\beta = 1} 
\th_1 (g(k_j)-\lambda_\beta);\cr
2\pi J_\alpha + \sum^M_{\beta=1} \th_2 (\la_\al-\la_\be) &
+  \sum^{N-2M}_{j=1} \th_1 (\la_\al-g(k_j)) \cr
&= -2Lx(\la_\al) -2{\rm Re}\de(x(\la_\al ) + iy(\la_\al));\cr
\th_n (x ) &= 2\tan^{-1} ({2\over n}x) + \pi .}}
$N_j$ and $J_\alpha$ are the quantum numbers of the charge and spin
excitations respectively.
Taking the thermodynamic limit (i.e. $N,M,L \rightarrow \infty$
with $N/L$ and $M/L$ finite),
followed by derivatives of the above, gives
\eqn\eIIviii{\eqalign{
\rho (k) &= {1\over 2\pi} + {\Delta (k) \over L} + 
g'(k) \il a_1(g(k)-\la) \sig (\la); \cr
\sig (\la ) &= - {x'(\la)\over\pi} + {\tilde{\Delta}(\la)\over L}
- \ilp a_2(\la '-\la)\sig (\la ') - \ik a_1(\la - g(k))\rho (k),}}
where
\eqn\eIIix{\eqalign{
\Delta (k) &= {1\over 2\pi} \del_k \delta (k);\cr
\tilde{\Delta} (\la ) &= -{1\over \pi} \del_\la 
{\rm Re}\delta (x(\la)+iy(\la));\cr
a_n(x) &= {1\over 2\pi}
\partial_x \th_n (x) =  {2n\over\pi} {1\over (n^2 + 4x^2)}.}}
Various limits appear in the above equations for the distributions.
$-D$ marks the lower allowed limit of the $k$'s while $\tQ$ marks out
the bandwidth of the $\la$'s.  As each $\la$ has a pair of complex 
$k$'s, its associated energy is $2 x(\la )$.  We thus determine
$\tQ$ by 
\eqn\eIIx{
x(\tQ ) = -D.}
Often it will be possible to replace $\tQ$ and $-D$ by $\infty$
and $-\infty$.  $B$ and $Q$ on the other hand give the spin and
charge Fermi surfaces.  They are determined by the constraints
\eqn\eIIxi{\eqalign{
{N-2M \over L} &= \ik \rho (k) ;\cr
{M \over L} &= \il \sigma (\la ). }}

\subsec{Determination of the Scattering Phase at the Fermi Surface: The Friedel 
Sum Rule}

In this section we examine the relationship between the scattering
phase of  electrons 
$\delta_e(\epsilon)$ at the Fermi surface and the number of electrons
on the dot and so verify the Friedel sum rule.  

To determine $\delta_e (\ep )$, 
we employ an energetics argument of
the sort used by N. Andrei in the computation of the magnetoresistance
in the Kondo model \andrei.
Imagine adding an electron to the system.  
Through periodic boundary conditions,
its momentum is quantized, $p = 2\pi n / L$.  If the dot was absent,
the quantization condition would be determined solely by the conditions
in the bulk of the system and we would write, $p_{\rm bulk} = 2\pi n/L$.
Upon including the dot, this bulk momentum is shifted by a term scaling
as $1/L$.  The quantization condition is then rewritten as
\eqn\eIIxii{
p = {2\pi n \over L} = p_{\rm bulk} + {\delta_e(\ep ) \over L},}
where $L$ is the system's length.  
The coefficient of the $1 \over L$ term is 
identified with the scattering phase of the electron off the dot. 

As we are interested in expressing $\delta_e$ in
terms of the number of electrons on the dot,
it is useful to separate out from $\rho (k)$ and 
$\sig (\la)$ the impurity contribution to the density of states.
We thus write
\eqn\eIIxiii{\eqalign{
\rho (k) &= \rh (k) + {1 \over L} \ri (k) ;\cr
\sig (\la) &= \sh (\la )  + {1\over L} \si (\la ) .}}
$\rh /\sh $ represent the bulk contribution to the densities while
$\ri /\si$ determine the number of electrons of definite spin,
$n_{d\up}/n_{d\do}$, sitting on the
dot.  From \eIIxi\ we have
\eqn\eIIxiv{\eqalign{
n_{d\up} &= \il\si (\la ) + \ik\ri (k) ;\cr
n_{d\do} &= \il\si (\la ) .}}
These relations will be key in verifying the Friedel sum rule.
Substituting \eIIxiii\ into \eIIviii , we obtain separate equations
for $\rh / \ri$ and $\sh / \si$:
\eqn\eIIxv{\eqalign{
\rh (k) &= {1\over 2\pi} + 
g'(k) \il a_1(g(k)-\la) \sh (\la); \cr
\sh (\la ) &= - {x'(\la)\over\pi} 
- \ilp a_2(\la '-\la)\sh (\la ') - \ik a_1(\la-g(k))\rh (k),}}
and
\eqn\eIIxvi{\eqalign{
\ri (k) &= \Delta (k) + 
g'(k) \il a_1(g(k)-\la) \si (\la); \cr
\si (\la ) &= \tilde{\Delta}(\la)
- \ilp a_2(\la '-\la)\si (\la ') - \ik a_1(\la-g(k))\ri (k).}}
In Appendix A we give alternative forms to the above equations governing
the density functionals.  These alternatives are far more amenable
to numerical analysis and in practice, the ones used in solving for the
densities.

Having obtained the equations governing the impurity densities of state,
we now focus on the scattering phase itself.  From \eIIvii\ we can
read off the bulk momentum of a charge/spin excitation with quantum number
N/J to be
\eqn\eIIxvii{\eqalign{
p(k) &= {2\pi N \over L} = k + \il \sigma_{\rm bulk} (\la ) 
\theta_1(g(k) - \la );\cr
p(\la ) &= -{2\pi J \over L} = 2 x(\la ) 
+ \ilp \sigma_{\rm bulk} (\la ') \theta_2 (\la - \la ') 
+ \ik \rho_{\rm bulk} (k) \theta_1(\la - g(k)).
}}
We assume here $\tan^{-1}$ in $\theta_{1,2}$ varies from $\pi/2$ to $\pi/2$
so ensuring a simple relationship between the momentum and the energy
functionals to be derived in the next subsection.

The impurity contribution to the momentum for each type
of excitation can be similarly determined to be
\eqn\eIIxviii{\eqalign{
p_{\rm imp}(k) &= \delta (k) + \il \si (\la ) 
(\theta_1(g(k) - \la)-2\pi) ;\cr
p_{\rm imp}(\la ) &=  2 Re \delta ( x(\la) +iy(\la))
+ \ilp \si (\la ') (\theta_2 (\la - \la ')-2\pi) \cr
& \hskip .2in + \ik \ri (k) (\theta_1 (\la - g(k))-2\pi) .
}}
Here we have chosen a different range for $\tan^{-1}$ in $\theta_{1,2}$
for describing the 
impurity momentum.  Shifting back to the original range leads then to
the appearance of the $2\pi$'s.
This choice is governed by our ultimate desire to
give the scattering phases in terms of the impurity momentum.
In particular we want 
$p_{\rm imp}(k\rightarrow -\infty) = p_{\rm imp}(\la\rightarrow \infty)
= 0$.

According to \eIIxii , we identify $p_{\rm imp} (k)$ with the scattering
phase of a charge excitation and $p_{\rm imp} (\la )$ with the scattering
phase of a spin excitation.
By differentiating these expressions and comparing to \eIIxvi , we obtain
the relations
\eqn\eIIxix{\eqalign{
\partial_k p_{\rm imp}(k) &= 2\pi \ri (k);\cr
\partial_\la p_{\rm imp}(\la ) &= -2\pi \si (\la ) .
}}
Again we have relations crucial to verifying the Friedel sum rule.

In order to determine the scattering phase of an electron (as opposed to
a spin or charge excitation), we must
must specify how to glue together a spin and a charge excitation
to form the electron.  The situation is analogous to adding
a single particle excitation in the attractive Hubbard model
\ref\hubbard{N. Andrei, cond-mat/9408101.}.
Adding a single spin $\up$ electron to the system demands that we add
a real $k>B$ (charge) excitation.  In doing so
we create a hole at
$\la > Q$ in the spin distribution as the number of the
available slots in the spin distribution is determined by
the number of electrons in the system.  Adding an electron to the
system thus opens up an additional slot in the $\la$-distribution.

The electron scattering
phase off the impurity is then the difference of the right-moving k-impurity
momentum, $p_{\rm imp} (k)$, and the 
left-moving $\la$-hole impurity momentum
$-p_{\rm imp} (\la )$:
\eqn\eIIxx{\eqalign{
\delta^\up_e &= p^\up_{\rm imp} = p_{\rm imp} (k) + p_{\rm imp}(\la );\cr
& = 2\pi \int^{k}_{-D} dk' \ri (k') 
+ 2\pi\int^{\tilde{Q}}_\la d\la '\si (\la '),
}}
where we have used \eIIxix\ in writing the last line.
If the excitations are 
added/removed at the Fermi surfaces, i.e. $k=B$, $\la = Q$,
we obtain the Friedel sum rule for spin up electrons,
\eqn\eIIxxi{
\delta^\up_e = 
2\pi \int^{B}_{-D} dk \ri (k) + 2\pi\int^{\tilde{Q}}_Q d\la \si (\la ) 
= 2\pi n_{d\up},
}
where \eIIxiv\ has been used.  The total energy of this excitation
is $\ep (k=B) + \ep (\la =Q ) = 0$, as it should be.

To determine the scattering of a spin down electron we employ
particle-hole symmetry.  A particle-hole transformation is 
implemented via
\eqn\eIIxxii{\eqalign{
c^\dagger_\up (k) &\rightarrow c_\do (-k) ;\cr
c^\dagger_\do (k) &\rightarrow c_\up (-k) ;\cr
d^\dagger_\up &\rightarrow d_\do ;\cr
d^\dagger_\do &\rightarrow d_\up ;\cr
\ep_d & \rightarrow -U - \ep_d .}}
Consequently the scattering phase of a spin $\do$ hole is related
to that of a spin $\up$ electron via
\eqn\eIIxxiii{
\delta^\do_{ho}(-U-\ep_d) = \delta^\up_e (\ep_d).}
The phase of this excitation
is then
\eqn\eIIxxiv{
\delta^\do_{ho} (-U-\ep_d)= 
2\pi\int^{\tilde{Q}}_\la d\la ' \si (\la ' ) 
+ 2\pi\int^{k}_{-D} dk' \rho_{\rm imp} (k') 
= 2\pi n_{d\up}(\ep_d).
}
where the last equality holds 
if we take the hole to be at the Fermi surface, $\la =Q$ and $k=B$.
As $n_{d\up} (\ep_d) = 1 - n_{d\do} (-U-\ep_d )$, we have
\eqn\eIIxxv{
\delta^\do_{ho} (-U -\ep_d )~{\rm mod}~2\pi = -n_{d\do} (-U-\ep_d ).}
At the Fermi surface, hole and electron scattering are identical 
(up to a sign)
and so we verify the Friedel sum rule for spin down electrons.

The reader may be puzzled why we rely on a particle-hole transformation
in computing the scattering amplitude for spin-down electrons.
Although it would be desirable to do this computation directly,
it does not seem to be possible.  To construct
a spin $\do$ electron at the Fermi surface, it is natural to
remove a $k = B$ excitation while
adding a $\lambda = Q$ excitation.  The corresponding 
scattering phase is then given by
\eqn\eIIxxvi{\eqalign{
\delta^\do_e &= p^\do_{\rm imp} \cr
&= p_{\rm imp} (k) + p_{\rm imp}(\la )\cr
&= 2\pi\int^{B}_{-D} dk \ri (k) +
2\pi\int^{\tilde{Q}}_Q d\la '\si (\la ') \cr
&= 2\pi n_{d\up}.}}
But this is obviously not what we want - a manifest violation
of the Friedel sum rule.  Rather by comparing \eIIxxvi\ with \eIIxxiv ,
the scattering indicates
we have constructed a spin $\do$ electron not at $\ep_d$ 
but at the particle-hole
conjugate point, $-U-\ep_d$.  Why this is so it is not entirely clear. 
However one can notice that the k-excitations are not only
charge excitations, but are in some sense unbound spin $\up$ electrons (the
number of k-excitations is directly proportional to the magnetization
of the system).  So in removing a k-excitation to form the spin $\do$ 
electron, we are in some sense creating a spin $\up$ hole.  And a spin
$\up$ hole at chemical potential, $\ep_d$, will scatter as a spin $\do$
electron at $-U-\ep_d$.

This entire discussion has concerned itself with proving the
Friedel sum rule for the one-lead Anderson model.  However 
we can argue that it applies, appropriately revised, to the two-lead model.
More precisely, we can argue that the excitations at the Fermi surface
behave linearly under the map \eIIii\ and so have two-lead scattering amplitudes
given by \eIIiiic .  This will be detailed in the
following two sections.

In appendix B we give an alternate derivation of the scattering phase
that focuses upon the impurity energy of an excitation as opposed
to its impurity momentum.  In doing so, we elucidate subtleties
not explicitly discussed in \rev .
We also give a third derivation
of the scattering phase in Appendix C by 
directly considering the 
dressing of the bare scattering.

\subsec{Excitations Away From the Fermi Surface in the Kondo Regime}

In the previous section we were mainly concerned with scattering
at the Fermi surface.  However as made clear by taking 
$k \neq B$, $\lambda \neq Q$, we can look at scattering 
above the Fermi surface. 

It is tempting to ask first whether the non interacting 
electrons in the lead can still be described in this formalism (by 
electrons we mean the standard plane wave excitations 
of appropriate spin and charge).  Here it is useful to recall some 
well known results from many-body theory:
Langreth, in verifying the Friedel sum rule for $H = 0$
\ref\lan{D. Langreth, Phys. Rev. 150 (1966) 516.}, computed the ratio
of the elastic inverse lifetime, $\tau^{-1}_{el}$, of a plane wave mode 
to that of its total inverse lifetime, $\tau^{-1}$, finding
\eqn\eIIxxvii{
{\tau^{-1}_{el}(\ep) \over \tau^{-1}(\ep)} = 
{\Gamma \over 2 {\rm Im}\Sigma (\ep)},
}
where $\Sigma (\ep)$ is the self-energy for the dot electron Green's function.
At the Fermi surface,
\eqn\eIIxxviii{
{\rm Im}\Sigma (\ep = 0) = {\Gamma \over 2},}
and there are no inelastic processes.  However away from the Fermi surface
\eqn\eIIxxix{
{\rm Im}\Sigma (\ep) = {\Gamma \over 2} + c \ep^2 , ~~~ c>0,}
and $\tau^{-1}_{el} < \tau^{-1}$, so electrons with energies 
above the Fermi surface do not scatter elastically. 
  
On the other hand, the simple excitations 
we construct within the integrable description by gluing 
spin and charge excitations will necessarily 
scatter elastically: beyond the Fermi surface, they cannot be the free 
electrons one would initially like to describe.  In and of itself,
this does not matter as all
we are interested in at the end is charge transport,
irrespective of what kind of objects actually do carry this charge. 
A similar situation occurs in the fractional quantum Hall effect \FLS, where
the integrability approach uses quasi-particles which are neither electrons
nor Laughlin quasi-particles.   This approach merely
provides a more convenient basis for the space of excitations, chosen
such that scattering at the impurity is as simple as possible. 
However we do have to concern ourselves with 
rewriting the original free electrons in terms of the integrable
scattering basis.  As indicated in the introduction we are not
able to provide an answer to this problem in its entirety.

Our approach will then be to build excitations which are ``electronic'',
that is carry the same quantum numbers as electrons, but scatter simply
(i.e. elastically) 
at the impurity - they will also scatter in a simple, factorized
way among themselves, although their S-matrix is non trivial  
(it is not $S=-1$ anymore).  One can certainly think of these excitations 
as dressed electrons. 

This being understood, another difficulty remains:
the potential parameter space
of electronic excitations, i.e. ($k$,$\la $), is two dimensional
(provided we neglect other solutions of the Bethe ansatz, see Section 2.6),
whereas we naturally want the space to be one dimensional.
For the moment,
we can only make the 
necessary dimensional
reduction when we are in the Kondo regime of the Anderson model.
The first step in doing so is to
determine the energy-momentum of an excitation labelled by
($k$,$\la$).  

We already know the momentum of the excitations from \eIIxvii .
We thus must only compute the energies.
To facilitate the calculation of excitation energies, it
is useful to decompose $\rho (k)$
and $\sig (\la )$ into particle and hole densities:
\eqn\eIIxxx{\eqalign{
\rph (k) &= \theta ( \pm B \mp k)\rho (k);\cr
\sph (\la ) &= \theta (\mp Q \pm \la) \sig (\la).}}
Now imagine varying $\rph$ and $\sph$ and asking what is the corresponding
variation in the energy.  We can
write this variation in two ways: one in terms of the bare energies 
and one in terms of new functions,
$\enpm (k)$ and $\enpm (\la)$, governing the dressed energies:
\eqn\eIIxxxi{\eqalign{
\delta E &= L \ika \{\enp (k) \drp (k) - \enm (k) \drh (k)\}
+ L \ila \{\enp (\la )\dsp (\la ) - \enm (\la ) \dsh (\la) \}\cr
& = L \ika (k-{H\over 2}) \drp (k) +  2L \ila x(\la ) \dsp (\la).}}
The variations on $\drph$ and $\dsph$ are not independent.
From \eIIviii\ we see
\eqn\eIIxxxii{\eqalign{
\drp (k) + \drh (k) &= g'(k) \ila \dsp(\la) a_1(g(k) - \la);\cr
\dsp (\la) + \dsh (\la) &= - \ilpa \dsp(\la') a_2(\la' - \la)
- \ika \drp (k) a_1 (\la - g(k)) .}}
Substituting \eIIxxxii\ into \eIIxxxi , we obtain
\eqn\eIIxxxiii{\eqalign{
\enp (k) + \enm (k) &= k - {H\over 2} - \ila \enm (\la ) a_1(\la - g(k));\cr
\enp (\la ) + \enm (\la) &= 2x(\la ) - \ilpa \enm (\la ' )a_2(\la ' - \la)\cr
&+ \ika g'(k)\enm(k)a_1(g(k)-\la) .}}
$\enpm (\la )$ and $\enpm(k)$ are characterized by
\eqn\eIIxxxiv{\eqalign{
\enp (\la ) &= \theta (Q-\la )(\enp (\la ) + \enm (\la )) > 0;\cr 
\enm (\la ) &= \theta (\la -Q)(\enp (\la ) + \enm (\la )) < 0;\cr 
\enp (k) &= \theta (k-B)(\enp (k) + \enm (k)) > 0;\cr 
\enm (k) &= \theta (B-k)(\enp (k) + \enm (k)) < 0.
}}
The functions, $\ep = \enp + \enm$, are
continuous and monotonic.  $\enpm$ have been defined such that
$\enp (k/\la )$ is the cost of adding an excitation
at $k/\la$ while $-\ep^-(k / \la)$ is the energy needed to create
a hole at $k/\la$.
Again in Appendix A, we give alternative forms to the above equations
governing the energy functionals which are more amenable to numerical
analysis.

Having determined the energy of the excitation, we can easily relate it to
its corresponding momentum.  We consider the case of $H=0$ first.
Comparing \eIIxvii\ and \eIIxxxiii , and using 
$\del_k \ep (k) = 2\pi \rho_{\rm bulk} (k)$
and $\del_\lambda \ep (\la ) = -2\pi \sigma_{\rm bulk} (\la )$, we see
\eqn\eIIxxxv{\eqalign{
p_{\rm bulk}(k) &= \ep (k);\cr
p_{\rm bulk}(\la ) &= \ep (\la),
}}
where $p_{\rm bulk}$ is the portion of momentum not scaling as $1/L$.

\vskip .4in
\centerline{\psfig{figure=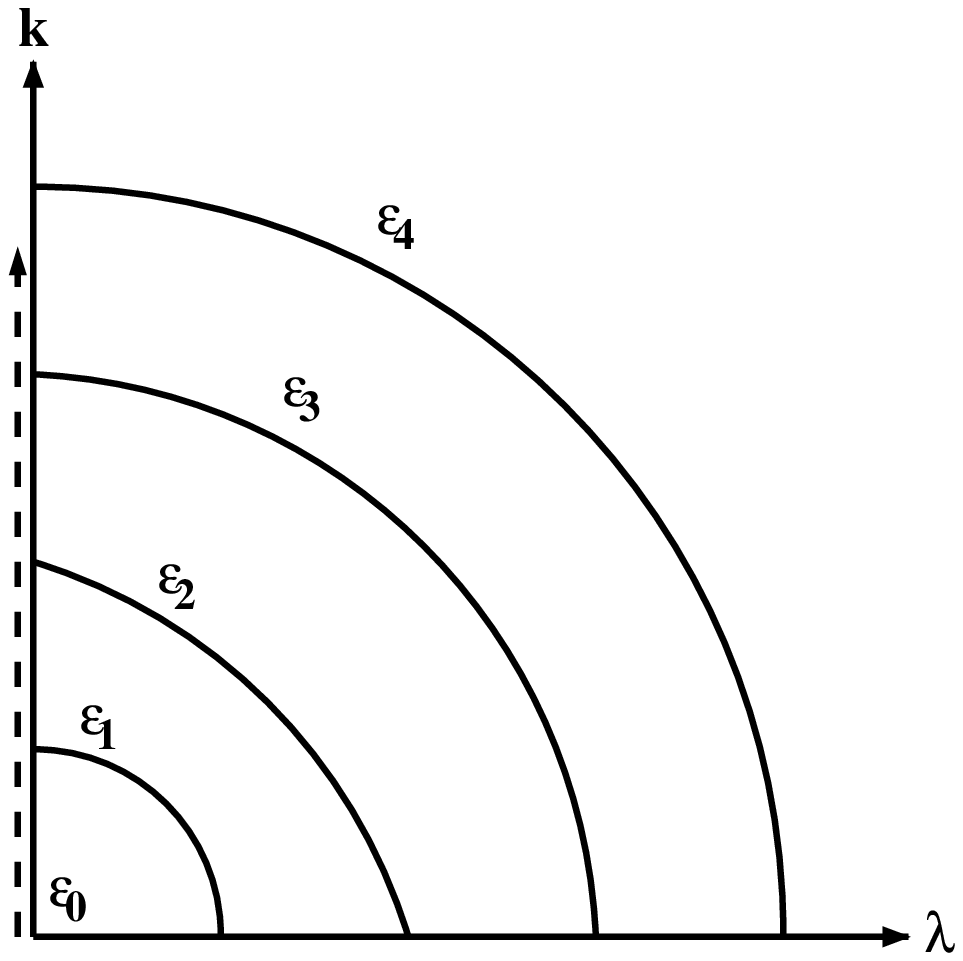,angle=-90,height=3in,width=3in}}
\vskip .1in
\centerline{\vbox{\noindent\hsize 5in Figure 3: 
A cartoon of the parameter space describing electronic excitations.
The drawing supposes that $\ep_0 = 0 < \ep_1 < \ep_2 < \ep_3 < \ep_4$.
Each curve represents a set of the excitations that share
the same energy $\ep$.  Only in the case, $\ep_0 = 0$, i.e.
when we are at the Fermi surface, is the pair $(k,\la )$ uniquely
specified.  The dashed line marks out the ansatz we employ
in the Kondo regime.}}
\vskip .4in 

With this in hand, we can parameterize the scattering phases of electronic
excitations away from the Fermi surface.  
Suppose we want to 
characterize a spin $\up$ electron
with energy, $\ep_{el}$.  (This is sufficiently general 
for $H=0$ as we know spin $\do$
electrons will scatter identically.)
The possible $k$ and $\la$ forming this excitation
must satisfy
\eqn\eIIxxxvi{
\ep(k ) - \ep (\la ) = \ep_{el}.
}
Given \eIIxxxv , this choice automatically satisfies $\ep_{el} = p_{el}$ 
(up to $1/L$ corrections).

This parameterization leaves an unresolved issue.  It 
does not in general uniquely specify a particular
pair $(k,\la )$, crucial if we are to actually compute quantities
involving information away from the Fermi surface.
We have schematically illustrated
the degeneracy of choices in Figure 3: as the energy is increased
the multiplicity of pairs $(k,\la )$ correspondingly increases.

In certain cases, however, the specification is unique.  
At the Fermi surface, the degeneracy of pairs is lifted. 
This already illustrated in Figure 3.  However there is
another case not captured by the cartoon in Figure 3 where the specification
is unique.  We have only a single possible pair $(k,\la )$ for
each energy in the case of
spin $\up$ hole scattering in 
finite magnetic field at the symmetric point of the model.  
The reason behind the
reduction of the parameter space for this
case will be made clear in what follows.

For the other cases where the choice is not unique, the
question becomes on what operative principle do we reduce the
space.  The key that we have identified to reducing the parameter space 
is determining how the excitations behave under the map
\eIIii .  Only if they behave linearly under the map are they
of use for it is only then that we can compute their scattering amplitudes
in the two-lead picture via \eIIiiic .

Although we have refined the question, we cannot in general determine
whether a given excitation unfolds linearly in the case when
there are multiple possible pairs $(k,\la )$. 
We can however make some progress when
we are in the Kondo regime of the Anderson model (i.e. $U+2\epsilon_d \sim 0$).
In this regime we expect the scattering phase to vary on the scale
of the Kondo temperature,
$T_k$.  The electron scattering phase is determined by $\rho_{\rm imp}$
and $\sigma_{\rm imp}$, the two impurity densities.  Of the two, only
$\rho_{\rm imp}$ varies on scales on the order of $T_k$.  In
contrast, $\sigma_{\rm imp}$
is controlled by the much larger scale $\sqrt{U\Gamma}$.  Thus
in computing electronic scattering phases away from the Fermi surface
at zero temperature, it is natural to keep $\la = Q$, its Fermi surface
value, and vary $k$.  Specifically, to describe an electron of
energy, $\ep_{el}$, we chose $(k,\la )$ such that
\eqn\eIIxxxvii{\eqalign{
&k~{\rm particle}, ~~~ \ep (k) = \ep_{el};\cr
&\la~{\rm hole~at}~\la = Q.
}}
With this ansatz, we then have restricted
the two dimensional phase space, $(\la , k)$, of potential
excitations carrying the quantum numbers of an electron to an one dimensional
subspace.
Hence the scattering phases of electrons of energy, $\ep_{el}$, above the
Fermi surface at $H = 0$ are given by
\eqn\eIIxxxviii{
\delta_e^{\up}(\ep_{el} ) = \delta_e^{\do}(\ep_{el} ) = 
2\pi\int^k_{-D} dk' \ri (k') + 2\pi \il \si (\la ) , ~~~\ep (k) = \ep_{el} .
}

When $H\neq 0$, we still have a simple relation between the
energy and momentum, i.e., we have 
\eqn\eIIxxxix{\eqalign{
\ep (k) &= p (k) - {H \over 2};\cr
\ep (\la) &= p (\la ).
}}
Hence with $H\neq 0$ we are still faced with an oversized parameter
space.  But we conjecture similar relations to those in \eIIxxxvii\ hold in
constructing the electronic spin $\up$ excitations:
\eqn\eIIxl{\eqalign{
\up~{\rm electron:}~~&k~{\rm particle}, ~~~ \ep (k) = \ep_{el};\cr
&\la~{\rm hole~at}~\la = Q.}}
The scattering phase of this
excitation is accordingly
\eqn\eIIxli{
\delta_e^{\up}(\ep_{el} ) = 2\pi\int^k_{-D} dk' \ri (k') 
+ 2\pi \il \si (\la ) ,
~~~\ep (k) = \ep_{el}.} 
With $H\neq 0$ and consequently $\ep^-(k)$ not identically zero,
we can construct spin $\up$ hole states by 
removing a $k$-state 
and a $\la$-hole.
The scattering phase of spin $\up$ holes 
is then equal to
\eqn\eIIxlii{\eqalign{
\delta_{ho}^{\up}(\ep_{ho}>0 ) &= p_{\rm imp} (k) + 
p_{\rm imp} (\la );\cr
& = 2\pi \int^{k}_{-D} dk' \ri (k') + 2\pi\int^{\tilde{Q}}_{Q} 
d\la '\si (\la '), ~~\ep(k) = -\ep_{ho} .}}
For this particular excitation we do not need the scattering ansatz.  If
we are to remove a $\la$-hole we must do it for $\la < Q$.
However at the symmetric point $Q=-\infty$ and so the choice is
unique.  This fact will allow us to conclude that in the large field
limit, our computation of the differential magneto-conductance becomes
exact.
We also point out that as $\ep (k)$ is bounded below, i.e. $\ep (k = -D) = -H$,
we are limited in the energy range ($[0,H]$) in which we can construct
spin $\up$ holes.  

With $H\neq 0$, we must compute the scattering of spin $\do$ objects
separately.  To do so we again employ a particle-hole transformation.
We so obtain
\eqn\eIIxliii{\eqalign{
\delta_e^{\do}(\ep_{el},-U-\ep_d) &= 
\delta_{ho}^{\up}(\ep_{ho}=\ep_{el},\ep_d);\cr
\delta_{ho}^{\do}(\ep_{ho},-U-\ep_d) &= 
\delta_e^{\up}(\ep_{el}=\ep_{ho},\ep_d) .}}
Unlike scattering at the Fermi surface where we could infer behaviour
at $\ep_d > -U/2$ from behaviour at $\ep_d < -U/2$, we cannot do
so for scattering away from the Fermi surface.  Thus when we need to
compute explicitly the scattering of spin $\do$ objects, say in computing
the magneto-conductance out of equilibrium, we will be restricted to
the symmetric point $\ep_d=-U/2$.

\subsec{Returning to the Two-Lead Problem}

In this section we explore in more depth the map 
between the one and
two lead models and its attendant problems.  
To review the map in more formal terms
call $E_{e,o}$ the integrable
excitations in the even 
and odd leads.
We then describe the factorized scattering by the relations
\eqn\eIIxliv{
\eqalign{E_eD=e^{i\delta_e} DE_e\cr
E_o D=e^{i\delta_o} D E_o ,}}
where $D$ is a formal symbol representing the impurity.
Again the phase, $\delta_e$, is non trivial, while $\delta_o = 0$.
Under the map \eIIii , integrable excitations in the two lead
picture are given by
\eqn\eIIxlv{
E_{1,2} = E_e \pm E_o .
}
Scattering of an excitation in the first lead is then described
by
$$
E_1 D = R D E_1 + T D E_2;
$$
or
\eqn\eIIxlvi{
(E_e+E_o)D = R D (E_e+E_o)  +T D (E_e-E_o),
}
where consistency with \eIIxliv\ demands that 
the transmission and reflection amplitudes, $R$ and $T$, satisfy
$R+T=e^{i\delta_e}$ and $R-T=e^{i\delta_o}$. 
An implicit assumption in this determination of the scattering in the
two lead picture is that 
the
superposition $E_o+E_e$ of an electronic excitation in the even
sector and an electronic excitation in the odd sector 
carries unit charge in lead 1 and no charge in lead 2. 
For an arbitrary fermionic excitation in the even and odd leads this
will not be the case.  For example,
imagine a electronic excitation in the even lead, that if decomposed
into a plane wave basis of free electrons, consists in part of
particle-hole excitations:
\eqn\eIIxlvii{
E_e = \sum_{k} a_k c^\dagger_{ek} |{\rm Fermi~sea}\rb + \sum_{k,k_p,k_h}
a_{kk_pk_h} c^\dagger_{ek}c^\dagger_{ek_p}c_{ek_h}|{\rm Fermi~sea}\rb
+ \cdots ,
}
where here $c_{ek}$ is a plane wave electron in the
even lead with wave vector $k$.
The linear combination $E_e+E_o$, where the excitation, $E_o$,
is arrived at from \eIIxlvii\ through $c_e \rightarrow c_o$, does
then not carry unit charge in lead 1.  Rather it carries indefinite
charge in both leads.  In this case the excitation
does not transform between the two pictures
as indicated by \eIIxlv\ and its scattering 
cannot be expected to be described by \eIIxlvi .  If $E_e$
was strictly a linear combination of 
terms of the form $c^\dagger_{ek} |{\rm Fermi~sea}\rb$,
this problem would not surface.

Thus in order to exploit the map between the two pictures,
we must limit ourself to excitations that behave linearly under
the map as described in \eIIxlv .  There are then two questions
to be answered.  Do such excitations in general exist?  And if they
do exist, are they sufficient for our purposes, the description
of transport properties.
We have two arguments that excitations with scattering described by
\eIIxlv\ do exist.  More precisely we have two arguments giving that 
excitations falling along some line
in the two dimensional $(k,\lambda )$ parameter space have such scattering.
Moreover, we argue that the scattering of such excitations is
sufficient to determine transport properties.

The first argument relies upon the transformation properties of
the Fermi field, $c(x)$.  Recall that in order to implement the
map between the one and two lead pictures, it is $c(x)$ that is
transformed, i.e.
$$
c_e (x) \rightarrow {1\over \sqrt{2}}(c_1 (x) \pm c_2 (x)) .
$$
Thus any integrable excitation $E_e$ that has a finite matrix element
with $c_e$, i.e.
$$
\lb c_e | E_e \rb \neq 0 ,
$$
must also behave linearly under the map from one to two leads.
To see this more explicitly imagine making a mode expansion
of the field, $c_e(x),$ in terms of the integrable excitations.  
As $E_e$ couples to $c_e$, $E_e$ must appear in this expansion,
\eqn\eIIxlviii{
c_e (x) = \sum_{r} a_r e^{ip_rx} E_{er} + \cdots ,
}
where $E_{e}$ is one of the $E_{er}$'s.  As $c_e$ and $E_e$ 
are linearly related, they must share the same transformation
properties.  This guarantees that any excitation appearing in
the above mode expansion will have scattering described by
\eIIxlvi .  

But we can say more on the basis of the properties of $c_e (x)$.
Because the underlying model is essentially free, we know the
single particle spectral function of the model will be given by
$$
\lb c_e c_e (0) \rb (E,p) \propto \delta (E - p) .
$$
Thus for any given energy $E=p$, we know that some integrable
excitation with this energy and momentum must appear in
the mode expansion \eIIxlvii .  In terms of the two dimensional
parameter space, $(k,\la )$, this implies that there is at least
one line in this space describing excitations transforming
as \eIIxlv\ and scattering as \eIIxlvi .

The second argument for the existence of this line in
the $(k, \la )$ parameter space relies upon 
combining the properties of the low energy sector of the
theory with the equivalence of the integrable excitation
we have constructed at the Fermi surface with the corresponding
plane wave electron excitation.  Given our reliance on this equivalence,
it deserves further exploration.

This equivalence is, of course, strongly suggested by our ability
to reproduce the Friedel sum rule and the fact Langreth \lan\
demonstrated that plane wave electrons at the Fermi surface scatter
elastically, the hallmark of integrable excitations.  Nevertheless, the statement
that the integrable excitation coincides with a plane wave electron needs
further clarification.  If we denote the wave function of
the integrable excitation as $\psi_{\rm int} (x,x_1,\ldots,x_N)$, where
$x$ is the coordinate of the excitation and the $x_i$ are the
coordinates of the electrons in the Fermi sea, and
$\psi_{\rm free~el.} (x,x_1,\ldots,x_N)$ as the many-body wave function of
the corresponding plane wave electron plus Fermi sea, we know
that the orthogonality catastrophe implies the matrix element,
$\langle {\rm int}|{\rm free~el.}\rangle$, equals
$$
\langle {\rm int}|{\rm free~el.}\rangle  = \int^\infty_{-\infty} dx dx_1\cdots dx_N
\psi^*_{\rm int} (x,x_1,\ldots,x_N)\psi_{\rm free~el.} (x,x_1,\ldots,x_N)
= {\cal O} (1/L),
$$
and so vanishes in the thermodynamic limit.  Thus it would seem that
that in fact that the two excitations do not coincide.

However we are not interested in matrix elements involving full eigenstates
of the Hamiltonian, but matrix elements involving asymptotic scattering
states defined far from the impurity.  These are the states of concern
in applying a Landauer-Buttiker formalism.  With such states, we would
evaluate the above matrix elements by restricting $x,x_i \ll 0$ or $x,x_i \gg 0$,
depending on whether the state is in-going or out-going.  With such a
restriction, the orthogonality catastrophe does not apply and
$$
\langle {\rm int}|{\rm free~el.}\rangle = 1 + {\cal O} (1/L).
$$
In this sense the excitations coincide.

With this equivalence so understood, the two excitations,
the integrable excitation and the plane wave share the same
transformation property \eIIxlv\ under the map.  We now exploit
this fact by combining it with the behaviour of the low energy
sector of the theory.  In this sector we can take a scaling limit
and obtain a relativistic theory invariant under Lorentz transformations.
Under such transformations, we can imagine boosting the integrable
excitations at the Fermi surface, obtaining in the process
an excitation with finite energy and momentum.  However the transformation
properties of the excitation cannot be altered by the boost.  As such
the boosted excitation will still transform via \eIIxlv\ under the map.
This again implies that there is a line in $(k,\la )$ parameter space
describing excitations transforming in the desired fashion.  
In this case moving
along this line amounts to making a Lorentz boost.

Having argued that there do exist excitations transforming as \eIIxlv ,
we now have to address whether the existence of such excitations is
sufficient for our computations of transport properties.
We can answer in the affirmative.
To compute any given transport quantity in the Landauer-Buttiker
approach, we need to sum up 
transmission amplitudes
over some given energy range.  For example, if we were
to compute the zero temperature out-of-equilibrium conductance,
this energy range would be determined by the difference of chemical potentials
in the two leads.  Now imagine looking at
a particular infinitesimal energy interval
within this range.  As we know the density of states of the free electron
in the lead, and we know that the interactions in the problem do
not affect this particular quantity, we know precisely
how much charge lies in this interval.  Now we are able
to construct an integrable state that transforms via \eIIxlv\ with an 
energy in this interval and with the same density
of states as the free particles.  Thus our integrable 
state completely exhausts the
charge lying within this infinitesimal interval.  Given that
we are able to compute its
transmission amplitude \eIIxlvi , we can
compute the contribution of this infinitesimal energy interval to
the transport quantity.

As a corollary to this, the manifold other integrable
states arising from the Bethe ansatz equations 
are then not needed for the computation of transport
properties.  We do not need to account for the $(k, \la )$ states
that do not transform as \eIIxlv .  We also do not need to worry
about states consisting of $(k,\la )$ excitations together with
particle-hole excitations of $k$ and $\la$, or indeed excitations
involving more complicated string solutions of the Bethe
ansatz equations.
The inclusion of such states in the computation of any transport quantity
would amount to a double counting, given that the line of $(k,\la )$
states transforming as $\eIIxlv$ completely exhausts the density
of states of free electrons in the leads.

Given all of this, we still must stress
that our computation of scattering amplitudes 
away from the Fermi surface is in general only approximate.
Although we understand that there exists a line of excitations
in the $(k, \la )$ parameter space for which we understand and can
compute scattering, we do not know which line.  Rather, at the
symmetric point of the model we have only an ansatz of how
this line cuts through parameter space.  However we again stress
that this ansatz is supported by the nature of the two
scales in the Kondo
regime, $T_k$ and $\sqrt{U\Gamma}$.  
Moreover our ansatz appears to be extremely good given its
agreement with Costi et al.'s NRG results.

Fortuitously there is one case where this ansatz is exact:
the description of spin $\up$ holes.  There, the $(k,\la )$ parameter
space is one-dimensional from the start and no ansatz is needed.
We point out that the scattering of such holes provides by far and away
the main contribution to the differential magneto-conductance at
large fields, $H$.  As such we expect the differential magneto-conductance
to be exact at asymptotically large fields.

In order to determine for all cases how the line of
linearly transforming excitations cuts through the $(k,\la )$ 
parameter space 
we would have to have complete control of the change of basis between
the free electrons and the integrable excitations.  There must presumably 
exist a complete set of such excitations providing a proper 
change of basis of the form
\eqn\eIIl{
|E \rb =\sum c_{n,F_1\cdots F_k} |F_1,\ldots,F_k\rangle.}
Here, the notation is highly symbolic: $E$ stands for 
any possible integrable excitation,
while $F_i$ stands for free electrons
with some spin and energy with the sum over the 
number, $k$, the types, and the energies of such excitations. 
Presumably, there are some complex selection rules making it possible 
to match the number of degrees of freedom on the left and on the right of 
\eIIl .  At this point however, an exact understanding of \eIIl\ is
beyond our reach.

\vfill\eject

\newsec{Linear Response Conductance at $T=0$}

In this section we discuss the linear response conductance both in
and out of a magnetic field at zero temperature.  We of course note
that none of Section 2 is necessary to compute the linear response
conductance at $T=0$.  Although we have demonstrated the Friedel sum
rule using integrability, all we need for this quantity is the
occupancy of the dot as a function of various parameters, 
something available from the original Bethe ansatz work on the one-lead
Anderson model.  However, the behaviour of the linear response
conductance as predicted by the Bethe ansatz has never been
adequately explored, particularly in the case with a magnetic
field.  Indeed in the original work of Ng and Lee \nglee\ showing
that the Friedel sum rule could be applied to quantum dots,
they
employ a Hartree-Fock approximation in estimating the dot occupancy
and so obtain some qualitatively incorrect predictions as to the
behaviour of the conductance.

Generally, the linear response conductance
equals
\eqn\eIIIi{
G = {e^2 \over h}(|T_\up|^2 (\ep = 0) + |T_\do|^2 (\ep = 0)) ,
}
where $T_{\up / \do} (\ep = 0)$ is the scattering amplitude
at the Fermi surface:
\eqn\eIIIii{
|T_{\up / \do} (\ep = 0)|^2 = \sin^2(\pi n_{d\up /\do}).}
In the case with $H=0$, the number of electrons on the dot as
a function of $\ep_d$, the gate voltage, can be computed exactly,
as has been done by \rev .  When $H\neq 0$, the equations become
more difficult to analyze and in general only numerical solutions
are available.  However at the symmetric point, it is again possible to
compute in closed form the number of spin $\up /\do$ electrons on the
dot \rev, and so arrive at an analytic expression for $G$.
We first consider the case with $H=0$.

\subsec{$H=0$ Linear Response Conductance}

In this case the Friedel sum rule tells us
\eqn\eIIIiii{
|T_{\up /\do}|^2 = \sin ^2 ({\delta_{\up / \do} (\ep =0) \over 2}),
}
where the phase, $\delta_{\up /\do}$, is equal to
\eqn\eIIIiv{
\delta_{\up / \do} = 2\pi n_{d\up /\do} .
}
The number of electrons, $n_{d\up /\do}$, on the dot when $H=0$
simplifies to
\eqn\eIIIv{
n_{d\up /\do} = \int^{\tilde{Q}}_Q d\la \si (\la ).
}
$\si (\la )$ in turn is given by \eIIxvi\ with the charge Fermi
surface, $B$, set to the bottom of the band, $-D$,
\eqn\eIIIvi{
\si (\la ) = \tilde{\Delta}(\la)
- \ilp a_2(\la '-\la)\si (\la ') .
}
The Fermi surface, $Q$, of the spin excitations is determined by
the equations
\eqn\eIIIvii{\eqalign{
{N\over L} &= \il \sigma (\la ) ;\cr
\sig (\la ) &= - {x'(\la)\over\pi} - \ilp a_2(\la '-\la)\sig (\la ') .
}}
These equations are solved explicitly over most of
the relevant parameter range in \rev~using a 
Wiener-Hopf technique.  

The solution breaks down into three cases 
according to the value of $Q$ describing
the Fermi surface:

\vskip .2in
\noindent{\bf case i:}
If we are close to the symmetric point ($U/2+\ep_d \ll \sqrt (U\Gamma)$)
then $Q \ll 0$ (at the symmetric point, $Q = -\infty$) and
we have
\eqn\eIIIviii{
n_{d\up /\do} = {1\over 2} - {1 \over \pi \sqrt{2}}\sum^\infty_{n=0}
{(-1)^n \over (2n+1)} G_+ (i\pi (2n+1)) 
\int^\infty_{-\infty} dk \Delta (k) e^{-(2n+1)\pi (g(k)-Q)},
}
where $G_+$ arises in factoring the kernel of the integral equation,
\eIIIvi :
\eqn\eIIIix{
G_+ (\omega ) = {\sqrt{2\pi} \over \Gamma ({1\over 2} - {i\omega \over 2\pi})}
({-i\om + \ep \over 2\pi e})^{-i\om \over 2\pi}.}
We include above an extra factor of $e$ omitted from \rev\ 
through a typo.
$Q$ is determined implicitly by the equation
\eqn\eIIIx{
{2 \ep_d + U \over \sqrt{2U\Gamma}} = {\sqrt{2} \over \pi} 
\sum^\infty_{n=0} {(-1)^n \over (2n+1)^{3/2}} e^{\pi Q (2n+1)} 
G_+(i\pi (2n+1)).}
This differs from \rev\ by a factor of 2.  This same factor of 2 is
missing from eqn. 8.2.38 of \rev\ which should, we believe, read
\eqn\eIIIxa{
{1\over\pi}(\ep_d + U/2) = \int^Q_{-\infty} d\la \sigma (\la ) .
}
\vskip .2in
\noindent{\bf case ii:}
In the next case the location of the Fermi surface satisfies the constraint
\eqn\eIIIxi{
0 < Q < I^{-1} \equiv {U \over 8\Gamma} - {\Gamma \over 2 U}.}
In this case $n_{d\up / \do}$ is computed to be
\eqn\eIIIxii{
n_{d\up/\do} = 2 - \sqrt{2} + {\pi \over 6\sqrt{2}}(I^{-1} - Q) - 
{\pi^2 \sqrt{2} \over (24)^2}(I^{-1} - Q)^2 +  {\cal O}((I^{-1} - Q)^3).}
The first two terms are found in \rev\ although we
disagree by a factor of 2 in the term of $\CO (I^{-1}-Q)$
while the remaining term was computed by the authors alone.

\vskip .2in
\noindent{\bf case iii:}
In the final case we are far from the symmetric point
such that $(U/2 + \ep_d) \gg \sqrt{U\Gamma}$ and 
$Q > I^{-1} \equiv {-1(U/8\Gamma - \Gamma/(2U))}$.
We then have instead
\eqn\eIIIxiii{
n_{d\up /\do} = {1\over 2\pi^{3/2}}\int^\infty_0 {dw \over w} \Gamma (1/2+w)
e^{2\pi w({1\over I} - Q)} \sin (2\pi w)
({w\over e})^{-w},}
with $Q$ in this case determined by
\eqn\eIIIxiv{\eqalign{
Q &= q^* + {1\over 2\pi}{\log (2\pi e q^*)},\cr
\sqrt{q^*} &= {\ep_d + U/2 \over \sqrt{2U\Gamma}}.
}}

\vskip .4in
\centerline{\hskip -1in 
\psfig{figure=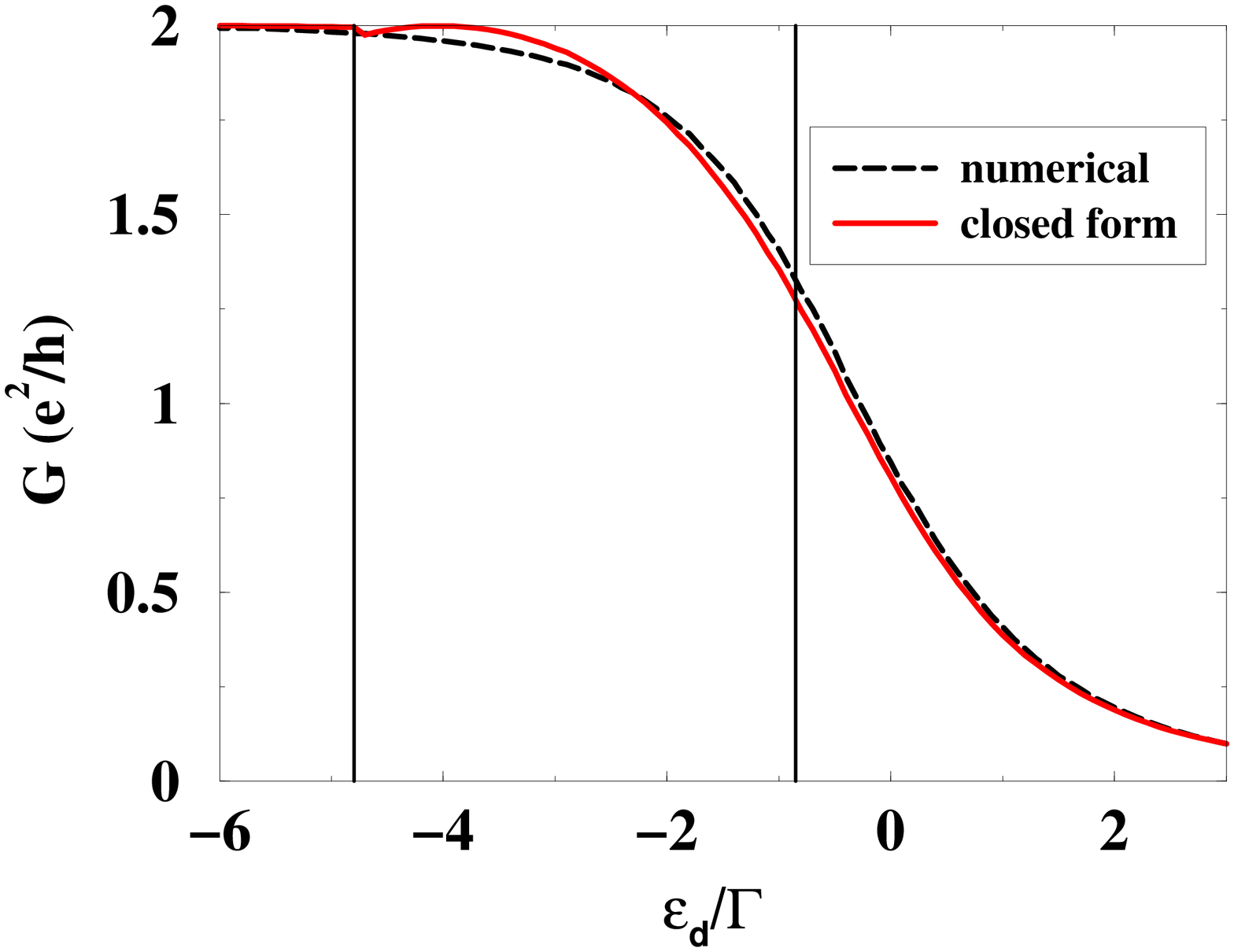,angle=-90,height=3.75in,width=4in}}
\vbox{\noindent\hsize 5in Figure 4: A plot of the linear
response conductance at zero temperature in zero magnetic field.  
The parameters used
are U = .75 and $\Gamma$ = U/12.  The dashed line marks out the conductivity 
derived from a numerical
solution of $n_d$ while the solid line represents the closed form
solution described in this section.}
\vskip .4in 

In Figure 4 is plotted the linear response conductance as a function of the dot
chemical potential, $\ep_d > -U/2$ (for $\ep_d < -U/2$ particle-hole
symmetry tells us the plot is a mirror image about the $\ep_d=-U/2$
axis), according to this closed form solution.
For the purposes of comparison, 
we also present the conductance derived from a numerical
evaluation of the equations determining $n_d$.  The vertical lines 
divide the plot
according to the three cases of the closed form solution.  We see that 
this solution best matches the numerical solution in cases i and iii.  
We also see that the
solution makes a discontinuous transition from case i 
to case ii, a consequence
of the approximate nature of the solution in case ii.

As expected the linear response conductance rises smoothly 
from zero at large, positive
values of $\ep_d$ to its maximum possible value, $2 e^2/h$, at the 
symmetric point of
the model, $U/2 = -\ep_d$.  The ratio of the values of $U$ and $\Gamma$ 
chosen for this plot 
correspond to that of the experimental realization of a quantum dot 
discussed in \gold  \foot{Note that our definition of $\Gamma$ is related
to that of \gold~by $\Gamma = \Gamma_{\gold}/2$.}.

\subsec{$H\neq 0$ Linear Response Conductance}

\vskip .4in
\centerline{\hskip -1in 
\psfig{figure=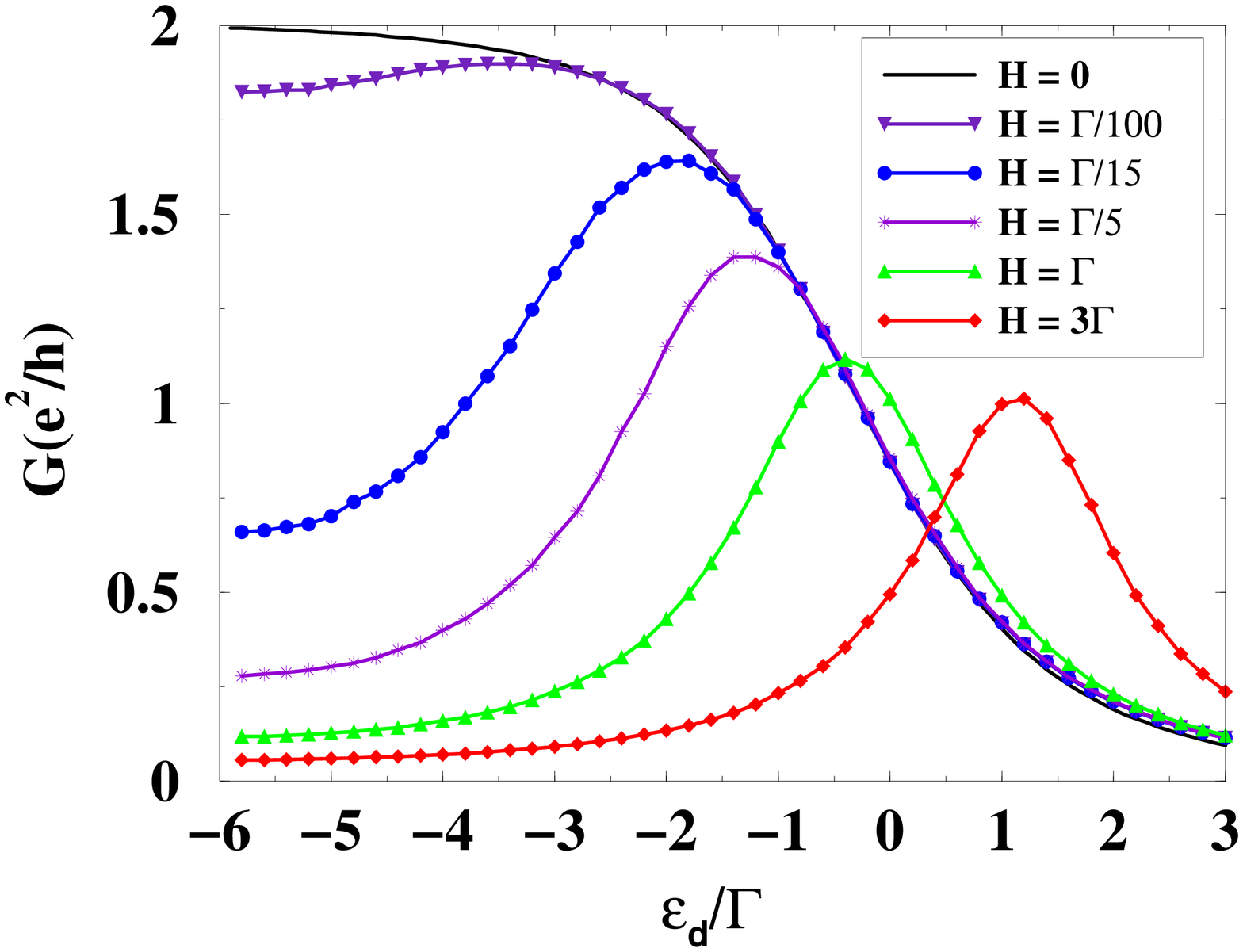,angle=-90,height=3.5in,width=4in}}
\vbox{\noindent\hsize 5in Figure 5: A plot of the linear
response conductance at zero temperature for various values of
the  magnetic field.  By particle-hole symmetry, the conductance
for values of $\ep_d < -U/2$ is obtained by taking the plot's mirror
image about the axis $\ep_d = -U/2$.
The parameters used
are $U = .75/\pi D$ (D being the bandwidth) and $\Gamma$ = U/12. 
For these parameters, the Kondo temperature at the symmetric
point is $T_k = .02508\Gamma$. }

\vskip .25in
\noindent{\it Solution Away from the Symmetric Point ($\ep_d > -U/2$):}
\vskip .25in

Again the transmission amplitude of the electrons is given by \eIIIiii\ and
\eIIIiv .  But in this case
\eqn\eIIIxv{\eqalign{
n_{d\up} &= \int^B_{-D} dk \rho_{\rm imp}(k) + \int^{\tilde{Q}}_Q d\lambda 
\sigma_{\rm imp} (\la)\cr
n_{d\do} &= \int^{\tilde{Q}}_Q d\lambda \sigma_{\rm imp} (\la) ,}}
where $\rho_{\rm imp}$ and $\sigma_{\rm imp}$ are given by \eIIxvi .

In general the equations for $\rho_{\rm imp}$ and $\sigma_{\rm imp}$ cannot be
solved analytically.  Therefore we resort to numerical solutions.
In Figure 5 we plot the result.  Presented there is the linear response
conductance as a function of $\ep_d$ for a variety of magnetics fields ranging
from $H = 0$ to $H = 3\Gamma$.

As $H$ is increased from zero, we see two effects:
the value of $\ep_d$ marking the conductance peak shifts away
from the symmetric point, $\ep_d = -U/2 (= -6\Gamma )$
while the magnitude of the peak decreases.
This is as expected.
The Kondo temperature for the model is given 
by \ref\haldane{D. Haldane, Phys. Rev. Lett. 40 (1978) 416.} \rev :
\eqn\eIIIxvi{
T_k = \sqrt{U\Gamma \over 2}
e^{\pi (\ep_d (\ep_d + U)-\Gamma^2)/(2\Gamma U)},
}
and so varies strongly as a function of the dot chemical potential.
When $H > T_k$ we expect the Kondo effect to be suppressed and any consequent
enhancement in G to disappear.  For values of $\ep_d$ away from the 
symmetric point,
$T_k$ is relatively large and thus strong fields are needed to suppress
the conductance.  Closer to the symmetric point, $T_k$ is exponentially 
suppressed
and weak fields are sufficient to destroy the Kondo effect.

When $H=0$, a conductance maximum of $2e^2/h$ occurs at the symmetric
point.  At the symmetric point, $n_{d\up /\do} = 1/2$,
and so each spin species makes a corresponding $e^2/h$ 
contribution to $G$.  As H is increased to large positive values, 
the gate voltage, $\ep_d$,
at which $n_{d\up /\do} = 1/2$ splits leading to a corresponding
split in the conductance
resonance.  For example for large H the resonance associated with
the spin $\up$ electrons is approximately $e^2/h$ and
occurs at $H/2$.  

We see for example in Figure 5 that when $H = \Gamma/100$ the
Kondo temperature is never exceeded regardless of the value
of the gate voltage and so we see little consequent suppression
of the conductance.  However for the next largest value of
H, $H=\Gamma/15$, the Kondo temperature is exceeded in the Kondo
regime and we see a corresponding depression in the conductance
in this regime.  For the largest value of $H = 3\Gamma$, we
see as expected that the peak value is approximately $e^2/h$
and that it occurs roughly at $\ep_d = H/2$.

We note that the linear response conductance curves are symmetric 
about their peak value.  This differs from the prediction 
based upon a Hartree-Fock computation of Ng and 
Lee \nglee~.  But it is in agreement
with Meir and Wingreen \win .

The conclusions in the above discussion are reiterated
in Figure 6.  There we plot the behaviour of the conductance peak
as the magnetic field is increased from zero.
In the top panel of Figure 6 we see that the location of the
peak rapidly moves away from $\ep_d = -U/2$ towards the
large field value of $H/2$.  The straight line in this panel indicates
the behaviour of the peak if interactions were absent.
Similarly we see the peak height
in the middle panel of Figure 6
change from its maximal value of $2e^2/h$ at $H=0$ to $e^2/h$
at large fields corresponding to a contribution to
the conductance of a single spin species.  And finally
in the bottommost panel of Figure 6 we examine the width of
the peak.  At $H=0$ the width of the peak is approximately $12\Gamma$.
However in the large field limit this settles down to $2\Gamma$
appropriate to the conductance being governed by the Breit-Wigner
formula
\eqn\eIIIxvii{
G = {e^2 \over h} {\Gamma^2 \over \Gamma^2 + (\ep_d - H/2)^2 },}
appropriate to a single non-interacting
electron species.

\vskip .25in
\noindent{\it Solution at the Symmetric Point, ($\ep_d = -U/2$):}
\vskip .25in

Although we cannot in general express the
magneto-conductance in closed form, we can
do so at the symmetric point.
At the symmetric point, the Fermi surface of the spin excitations, $Q$,
goes to $\infty$.  The density equations then simplify to
\eqn\eIIIxviii{\eqalign{
\rh (k) &= {1\over 2\pi} + 
g'(k) \int^\infty_{-\infty} d\la a_1(g(k)-\la) \sh (\la); \cr
\sh (\la ) &= - {x'(\la)\over\pi} 
- \int^\infty_{-\infty} a_2(\la '-\la)\sh (\la ') - \ik a_1(\la-g(k))\rh (k),}}
and
\eqn\eIIIxix{\eqalign{
\ri (k) &= \Delta (k) + 
g'(k) \int^\infty_{-\infty} d\la  a_1(g(k)-\la) \si (\la); \cr
\si (\la ) &= \tilde{\Delta}(\la)
- \int^\infty_{-\infty} d\la 
a_2(\la '-\la)\si (\la ') - \ik a_1(\la-g(k))\ri (k).}}

\vskip .4in
\centerline{\hskip -1in 
\psfig{figure=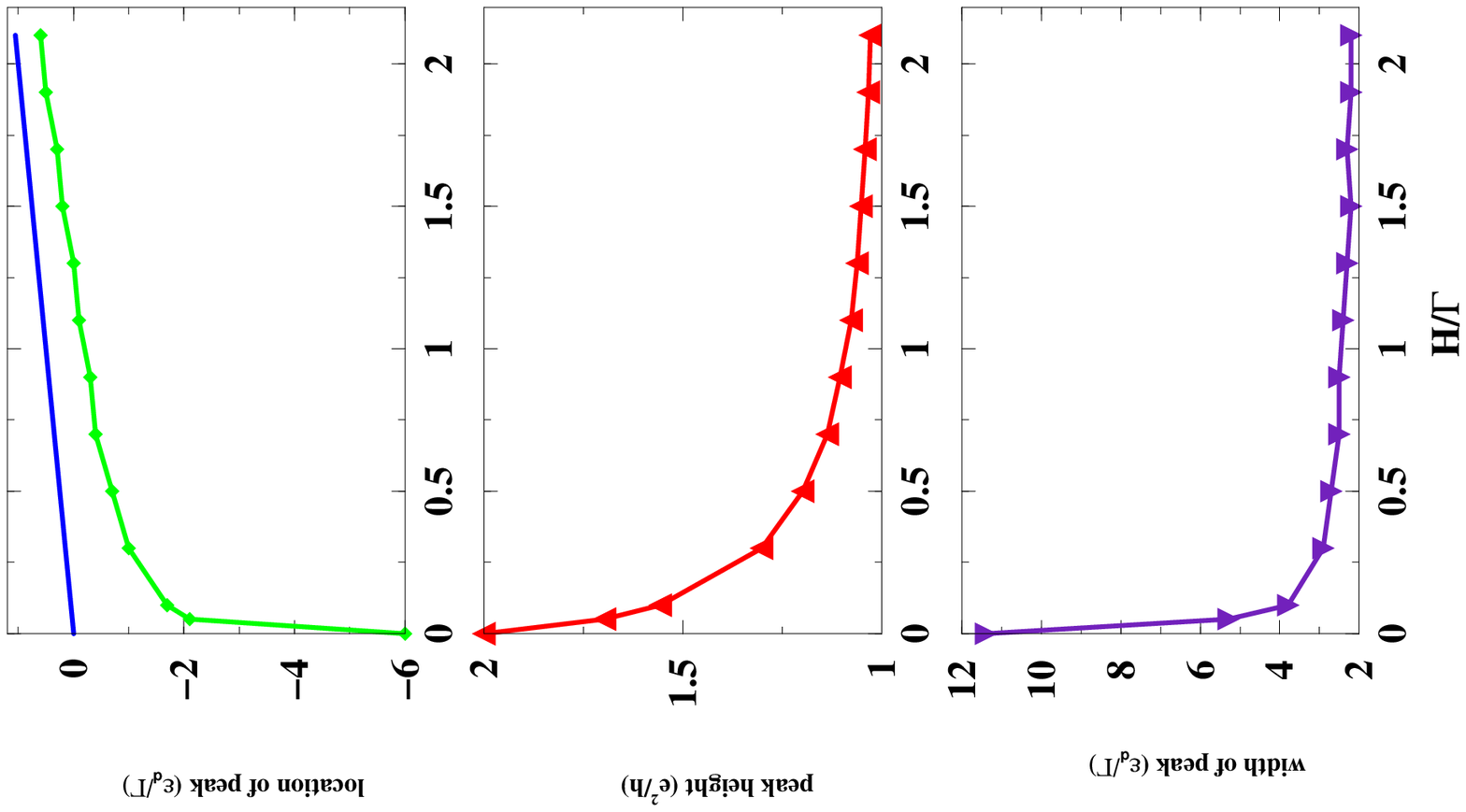,angle=-90,height=6.0in,width=2.5in}}
\vbox{\noindent\hsize 5in Figure 6: Plots of how the 
conductance peak evolves with increasing magnetic field.
In the top panel is a plot of the location of the peak 
while the middle panel records the 
peak height and the
bottom panel gives the peak width.
The parameters used
are $U = .75/\pi D$ (D being the bandwidth) and $\Gamma$ = U/12. }
\vskip .4in

\noindent Here the limit $B$ is determined by
\eqn\eIIIxx{
{2 S_z\over L} = {H \over 2\pi} = \int^{B}_{-D} dk \rho (k).}
As the electrons in the leads are non-interacting, the first equality
is a result of Pauli-paramagnetism.

\vskip .4in
\centerline{\hskip -1in
\psfig{figure=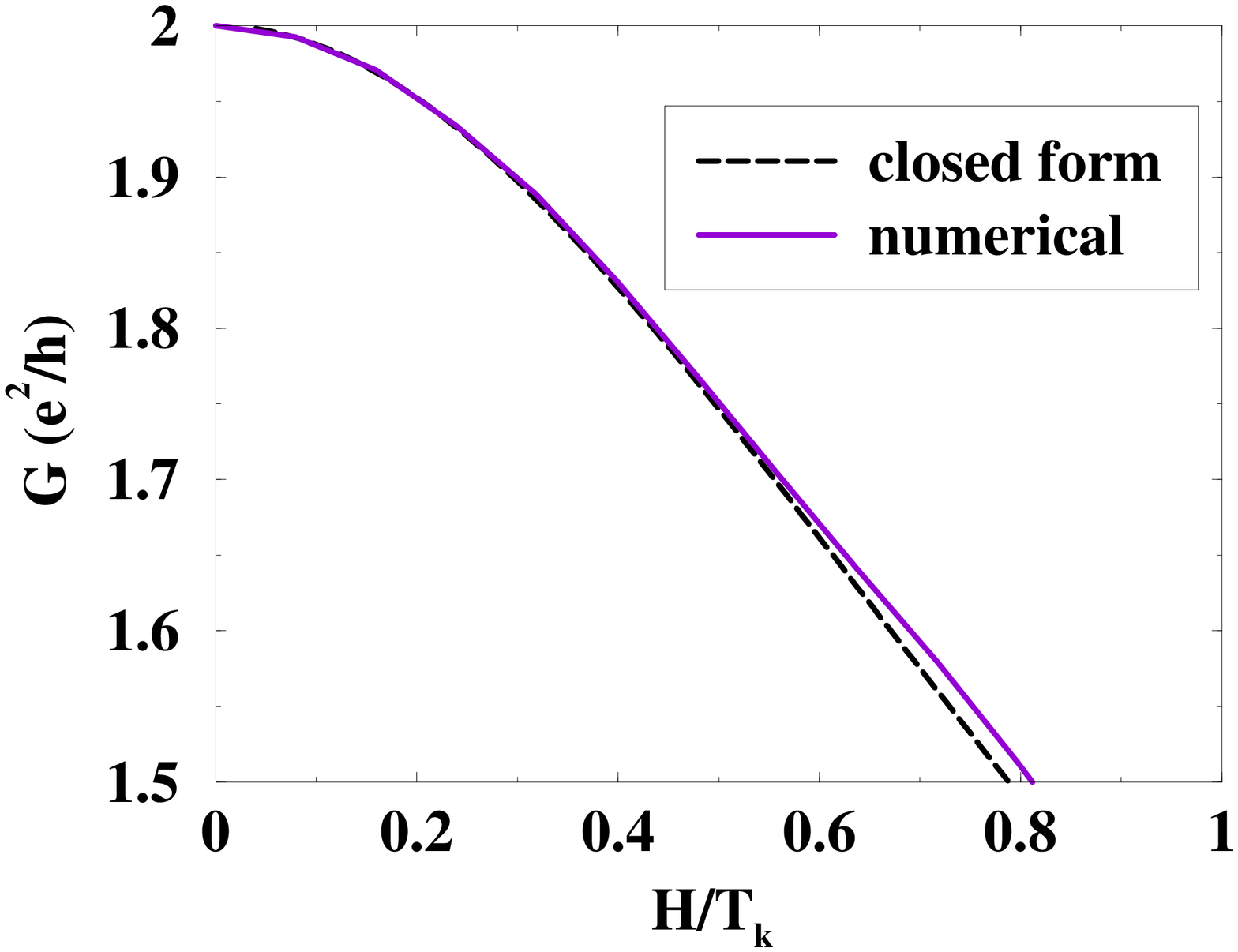,angle=-90,height=3.5in,width=4in}}
\vbox{\noindent\hsize 5in Figure 7: A plot of the linear
response conductance at zero temperature at the symmetric point 
($\ep_d = -U/2$)
as a function of magnetic field.  The parameters used
are U = .75 and $\Gamma$ = U/12.  The dashed line marks out the conductivity 
derived from a numerical
solution of $n_d$ while the solid line represents the closed form solution
described in this section.}
\vskip .4in 

The phase shifts are given by
\eqn\eIIIxxi{\eqalign{
\delta_{e\up} = 2\pi-\delta_{e\do} &= 
2\pi \int^{\tilde{Q}}_Q d\la \si (\la) +
2\pi\int^B_{-\infty} dk \rho_{\rm imp} (k)\cr
&= \pi + \pi \int^B_{-\infty} dk \rho_{\rm imp} (k) ,}}
which follows as
\eqn\eIIIxxii{
\int^{\infty}_{-\infty} d\la \sigma_{\rm imp} (\la ) 
= {1\over 2} - {1\over 2}\int^B_{-\infty} dk \rho_{\rm imp} (k).}
This is established by integrating, $\int^\infty_{-\infty}$, \eIIIxix .
We can thus focus solely upon the k-distribution.

In order to evaluate the phase shift in the equation above, we
are thus interested in computing the integral, 
\eqn\eIIIxxiii{
\int^B_{-\infty} dk \rho_{\rm imp} (k) = 2 M_i ,}
which, as indicated, is directly related to the 
impurity magnetization, $M_i$.
Using the same Wiener-Hopf technique, \rev~determined this integral
in the case $T_k > H$, to be
\eqn\eIIIxxiv{\eqalign{
2 M_i  &= {\sqrt{2} \over \pi} \sum^\infty_{n=0} 
{G_+(i\pi (2n+1))\over (2n+1)} (-1)^n e^{-\pi (2n+1)(b-1/I)};\cr
b &= {1\over\pi}\log ({2\over H}\sqrt{U\Gamma\over\pi e}).}}
Combining this with the expression for the Kondo temperature in
\eIIIxvi ,
we have for the scattering phases at leading order in $H/T_k$
\eqn\eIIIxxv{
\delta_{e\up} = 2\pi - \delta_{e\do} = \pi (1 + {H\over 2T_k}),
}
which in turn gives the magneto-conductance as 
\eqn\eIIIxxvi{
G(H) = 2{e^2\over h}(1 - {\pi^2\over 16} ({H\over T_k})^2 + 
\CO ({H\over T_k})^4).
}
The quadratic deviation from the maximal conductance 
has the expected Fermi liquid form.

In Figure 7 is plotted how the magnitude of the linear response conductance
at the symmetric point changes as a function of $H/T_k$ 
(for small H) according to
this solution.  We plot it against the numerical solution 
and we obtain
agreement at worst of $1.5\%$.  
The disagreement becomes larger as H increases as the
closed form solution is only valid at $H < T_k \propto e^{-1/I}$.  
Beyond $H >T_k$,
we rely entirely on a numerical solution to determine the magneto-conductance.
We plot the behaviour of $G(H)$ in Figure 8 up to $H=\Gamma$.

\vskip .4in
\centerline{\hskip -1in 
\psfig{figure=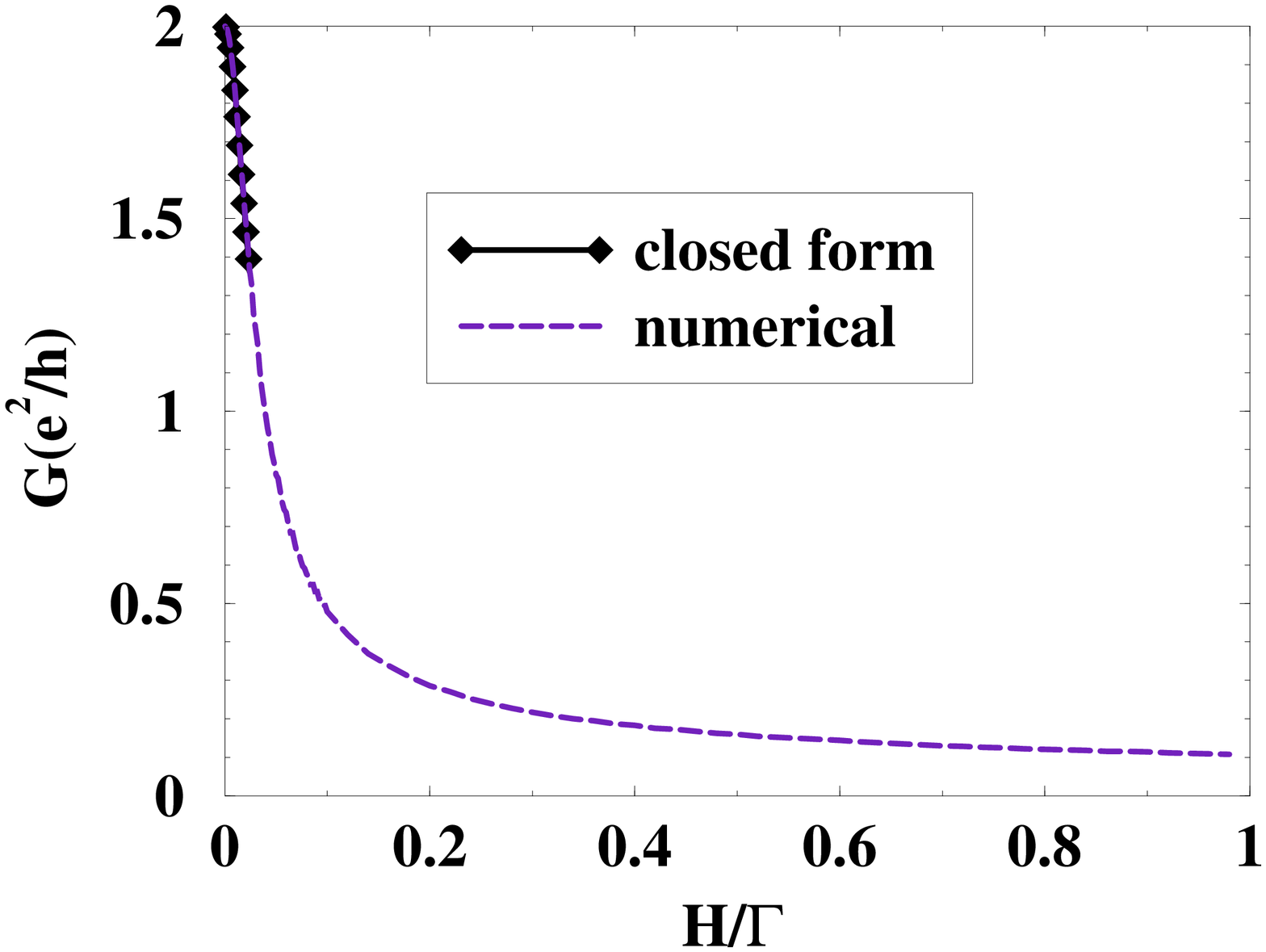,angle=-90,height=3.5in,width=4in}}
\vbox{\noindent\hsize 5in Figure 8: A plot of the linear
response conductance at 
zero temperature at the symmetric point ($\ep_d = -U/2$)
as a function of magnetic field.  The parameters used
are U = .75 and $\Gamma$ = U/12.  The dashed line marks out the conductivity 
derived from a numerical
solution of $n_d$ while the solid line represents the closed form solution
described in this section.}
\vskip .4in

\newsec{Linear Response Conductance at Finite Temperature}

In the previous section we focused upon scattering at the Fermi
surface.
In this section, we discuss a problem that requires
us to understand scattering at finite energy:
the linear response conductance as a
function of temperature.  We compare it to Costi et al.'s
NRG results and find excellent agreement.  This is important as
it indicates we have an essentially correct description of 
the low energy scattering
states, 
i.e. an excellent approximation to the right hand side of \eIIl .

Computing the linear response conductance at finite
T is a complicated matter.  Even though we only consider 
glued charge and spin excitations as explained in Section 2,
we now have to compute their  scattering matrix - the $1/L$ 
correction of their associated momenta - in the presence
of a ``thermalized ground state''.  This ground state is no longer composed
of merely real k states and 2-string bound states of spin and charge as it was at $T=0$.  Rather
all the possible solutions of the Bethe ansatz equations of the
model make an appearance.  Thus to
begin the computation of the scattering amplitudes at finite T,
we give the complete list of excitations in the model.

It is useful to understand how the 
following calculation differs from the exact computation
of the conductances in the fractional quantum Hall problem \FLS .
The logic of \FLS\ would first require the identification of all the 
excitations above the zero temperature ground state.
We then would
need to compute both the bulk and impurity scattering matrices of 
these excitations.  Having done this, the second step 
would be to turn on the
temperature.  The thermalized ground state would them consist
of seas of all these excitations with their
distributions determined by
a thermodynamic Bethe ansatz, employing the zero temperature S-matrices.
Finally, the conductance would
be determined by the zero temperature impurity
scattering matrices as in \FLS.  A potential difficulty
in this approach would come in understanding the gluing necessary
to form excitations carrying electronic quantum numbers as it only such
excitations that 
can be mapped back to the two-lead picture.

Whatever the likelihood for success, one can see that this way of proceeding
requires considerable technical expenditure.  It turns
out to be easier to determine the thermalized ground state directly
from the Bethe ansatz, i.e., by dealing with 
bare excitations rather than first determining the zero temperature
ground state, classifying its ``physical'' or ``dressed''
excitations, and expressing the thermalized
ground state in terms of these excitations.  On the other hand, 
with the quantum Hall edges, it was
the natural way to proceed as the initial
data one had at hand (with no calculations whatsoever)
is precisely the excitations about the zero temperature ground state
together with their S-matrices, both bulk and impurity.

To proceed then, the first step is to identify all solutions of the
Bethe ansatz equations.
These are of three types: real k-states, spin complexes associated
with complex k's (both of which we have seen as the ground state at zero
temperature is composed of such excitations) and spin complexes not associated
with complex k's.  Below we give a more specific description together
with the quantum numbers carried by each excitation.  The spin
quantum numbers are measured relative to a vacuum carrying spin $N/2$
where N is the number of particles in the system.
\vskip .2in
\noindent i) real k's: These appear in the ground state at T=0 in
the presence of a magnetic field.  They carry charge $e$ and no spin.
\vskip .2in
\noindent ii) n-spin complex with no associated k's: An n-complex involves
n $\lambda's$ organized as
\eqn\eIVi{
\lambda^{nj} = \lambda^n + i({n+1\over 2} - j), ~~~~ j=1,\cdots , n.}
Here $\lambda^n$ is a real rapidity and is known as the centre of the
complex.  The n-spin complex carries spin $-n/2$. 
\vskip .2in
\noindent iii) n-spin complex with 2n associated complex k's: The n-complex
is organized as before
\eqn\eIVii{
\lambda^{nj} = \lambda^n + i({n+1\over 2} - j), ~~~~ j=1,\cdots , n.}
but now there are two k's associated with each $\lambda^{nj}$
\eqn\eIViii{\eqalign{
g(k^{+nj}) &= \lambda^n + i({n\over 2} + 1 - j);\cr
&\hskip 1.35in j = 1,\cdots ,n;\cr
g(k^{-nj}) &= \lambda^n + i({n\over 2} - j).}}
These excitations carry charge $2ne$ and spin $-n/2$.
The simplest of these excitations (n=1) appear in the ground state 
at zero temperature.  
Now the claim is that these are solutions to the 
Bethe ansatz equations
\eIIiv\ and \eIIv\ and indeed they are in the thermodynamic limit
(the object of our concern) \rev .

We can derive equations constraining the particle/hole densities of these
various excitations in the same fashion we arrive at \eIIviii\ 
from \eIIiv\ and \eIIv .  The result is
\eqn\eIViv{\eqalign{
\rho_p (k) + \rho_h (k) &= {1 \over 2\pi} + {\Delta (k) \over L} 
+ g'(k) \sum^{\infty}_{n=1} \int^\infty_{-\infty} d\lambda a_n (g(k)-\la )
(\sigma_{pn}(\la ) + \sigma'_{pn} (\la ));\cr
\sigma_{hn} (\la ) &= -{x_n'(\la ) \over \pi}  +
{\tilde{\Delta_n}(\la ) \over L} - 
\int^\infty_{-\infty} a_n(\la - g(k))\rho_p(k)\cr
&- \sum^\infty_{m=1} 
\int^\infty_{-\infty}d\la ' A_{nm}(\la - \la')\sigma_{pm}(\la ');\cr
\sigma'_{hn} (\la ) &= \int^\infty_{-\infty} a_n(\la - g(k))\rho_p(k)
- \sum^\infty_{m=1} 
\int^\infty_{-\infty}d\la ' A_{nm}(\la - \la')\sigma'_{pm}(\la ');\cr
x_n(\la ) &= 
\sqrt{2U\Gamma}{\rm Re}~(\la + i{n\over 2})^{1/2} + n({U\over 2} +\ep_d)\cr
\tilde{\Delta}_n(\la ) &= -{1\over\pi}\partial_\la \delta_n (\la )\equiv
-{1\over\pi} \partial_\la {\rm Re}~
\delta (-\sqrt{2U\Gamma}\,(\la +i{n\over 2})^{1/2}+U/2+\ep_d)\cr
&\hskip .5in
-{1\over 2\pi}\del_\lambda \sum^{n-1}_{k=1} 
\bigg\{\delta(-\sqrt{2U\Gamma}\,(\la+{i\over 2}(n-2k))^{1/2}+U/2+\ep_d)\cr
&\hskip 1.3in 
+ \delta(\sqrt{2U\Gamma}\,(\la+{i\over 2}(n-2k))^{1/2}+U/2+\ep_d))\bigg\}
.}}
The kernels in the density equations are given by
\eqn\eIVv{\eqalign{
a_n(\la ) & = {2n\over\pi} {1\over (n^2+4\la^2)};\cr
A_{nm}(\la ) &= \delta_{nm}\delta(\la ) + a_{|n-m|}(\la) + 
2\sum^{{\rm min}(n,m)-1}_{k=1} a_{|n-m|+2k} (\la ) + a_{n+m}(\la ).}}
Here $\rho_{p/h}$ is as before while $\sigma_{p/h n}$ denotes the particle/hole
densities of n-strings associated with complex k's (in Section 2 we denoted
$\sigma_{p/h 1}$ by $\sigma_{p/h}$) and $\sigma'_{p/h n}$ denotes particle/hole
densities of n-strings not so associated.

As stated in the introduction of this section, we construct the electronic 
excitations in the same fashion as at zero temperature,
the only difference being that the excitations 
are now over the thermal ground state,
not the $T=0$ ground state.
In order to describe
the scattering amplitudes we thus need to specify the 
impurity momentum of
the $\rho (k)$ and $\sigma_1 (\la )$ excitations.  The finite temperature
momenta for such excitations are as follows (compare \eIIxvii ):
\eqn\eIVvi{\eqalign{
p^{\rm imp} (k) &= \delta (k) + \sum^\infty_{n=1} \int^\infty_{-\infty}
(\theta_n (g(k) - \la )-2\pi) (\sigma_{pn}^{\rm imp}(\la ) + 
\sigma_{pn}^{'\rm imp} (\la ));\cr
p_1^{\rm imp} (\la ) &= 2\delta_1 (\la ) 
+\int^\infty_{-\infty} dk \rho_p^{\rm imp} (k)(\theta_1 (\la - g(k))-2\pi)\cr
& + \sum^\infty_{m=1} \int d\la ' 
(\Sigma_{1m} (\la - \la '))
\sigma_{pm}^{\rm imp}(\la '),}}
with $\Sigma_{nm}$ given by
\eqn\eIVvii{\eqalign{
\Sigma_{nm}(\la ) &= (\theta_{|n-m|}(\la)-2\pi) + 
2\sum^{{\rm min}(n,m)-1}_{k=1}(\theta_{|n-m|+2k} (\la )-2\pi) + 
(\theta_{n+m}(\la )-2\pi).}}
We again can read off the $1\over L$ contributions to the momenta and
the densities and arrive at the all important relations:
\eqn\eIVviii{\eqalign{
\partial_k p^{\rm imp}(k) &= 2\pi \rho^{\rm imp} (k);\cr
\partial_\la p^{\rm imp}_1(\la ) &= -2\pi \sigma^{\rm imp} (\la ),
}}
still valid at finite temperatures.  The scattering phases, as they
are given (as in Section 2)
from the impurity momenta, $p^{\rm imp} (k)$ and $p^{\rm imp}_1(\la)$,
can be computed from a knowledge of $\rho^{\rm imp}$
and $\sigma^{\rm imp}_1$.  For example, a spin up excitation constructed
from a charge excitation, k, and a $n=1$ spin-charge complex, $\la$, has
a scattering phase of the form
\eqn\eIVix{\eqalign{
\delta^\up_e &= p^{\rm imp} (k) + p^{\rm imp}_1(\la );\cr
& = 2\pi \int^{k}_{-D} dk \rho^{\rm imp} (k) + 
2\pi\int^{\tilde{Q}}_\la d\la '\sigma^{\rm imp}_1 (\la '),}}
identical to \eIIxx .

Another piece of the prescription of computing the scattering amplitudes
at finite temperature is the energy associated with the charge and spin 
excitations.  The energies of the various excitations can be derived as in
Section 2 with the result
\eqn\eIVx{\eqalign{
\ep (k) &= k + T\sum^{\infty}_{n=1} \int^\infty_{-\infty}
d\la \log ({f(-\ep'_n (\la)) \over f(-\ep_n (\la))}) a_n(\la - g(k));\cr
\log (f(\ep_n(\la ))) &= - {2\over T} x_n(\la ) - \int^\infty_{-\infty}
dk g'(k) \log (f(-\ep (k))) a_n(g(k) -\la)\cr
& + \sum^{\infty}_{m=1} \int d\la' A_{nm}(\la -\la ')\log (f(-\ep_m(\la ')));\cr
\log (f(\ep'_n(\la ))) &= - \int^\infty_{-\infty}
dk g'(k) \log (f(-\ep (k))) a_n(g(k) -\la)\cr
& + \sum^{\infty}_{m=1} \int d\la' A_{nm}(\la -\la ')\log (f(-\ep'_m(\la '))),}}
where $f(\ep) = (1+\exp (\ep /T))^{-1}$ is the Fermi distribution.
These equations are arrived at by relating the densities to the energies
via
\eqn\eIVxi{\eqalign{
\exp (\ep (k)/T) &= \rho_h (k)/\rho_p (k);\cr
\exp (\ep_n (\la)/T) &= \sigma_{nh} (\la)/\sigma_{np} (\la );\cr
\exp (\ep'_n (\la)/T) &= \sigma'_{nh} (\la)/\sigma'_{np} (\la ).}}
This relation is chosen so that the energies determine the particle/hole
distributions in the same fashion that they do in the case of non-interacting
fermionic particles, i.e.,
\eqn\eIVxii{
\rho_p (k) = (\rho_p (k) + \rho_h (k) )f(\ep (k)),}
and likewise for $\sigma_{np/h}$ and $\sigma'_{np/h}$.  This definition
is completely consistent with that at zero temperature.  Taking $T\rightarrow 0$
in the above recovers \eIIxxxiii .  This is a general feature of energy
functionals in a thermodynamic Bethe ansatz analysis.  However here
the energies are related to the densities in an additional way indicative
that the bulk of the system (i.e. the leads) is indeed non-interacting:
\eqn\eIVxiii{\eqalign{
\rho_{p/h} (k) &= {1\over 2\pi}\partial_k\ep (k) f(\pm \ep (k));\cr
\sigma_{np/h} (\la ) &= 
-{1\over 2\pi}\partial_\la\ep_n (\la ) f(\pm \ep_n (\la));\cr
\sigma'_{np/h} (\la ) &= 
{1\over 2\pi}\partial_\la\ep'_n (\la ) f(\pm \ep'_n (\la)).}}

Having specified the energy functionals, we can now determine the particular
$k$ and $\la$ we choose in creating an electron.  As $H=0$, we will simply
focus on spin $\up$ electrons.  In forming a spin $\up$ electron we add a 
k-excitation in $\rho (k)$ and a $\lambda $ hole in $\sigma_1 (\la )$.
The energy of the electron is then 
\eqn\eIVxiv{\eqalign{
\ep_{el} &= \ep (k) - \ep (\la).}}
We again will only allow $k$ to vary in varying $\ep_{el}$
while fixing $\la$ to some $\la_o$.
While at $T=0$ we fixed $\la$ to be $Q$, its value at the Fermi surface,
this is not
appropriate at finite temperature as the Fermi 
surface has become blurred.  However we have another way to characterize
the correct choice for $\la$, at least at the symmetric point, which
we give in the following subsection.

We are now ready to specify the final equation governing the finite 
temperature
linear response conductance.  Given that we construct the 
electronic excitations by gluing
together a fixed spin excitation, $\la_o$, and a range of $k$ excitations,
these excitations  are distributed according to the Fermi distribution, as
they must be.  Thus the conductance at finite T is given by
\eqn\eIVxv{\eqalign{
G(T) &= 2{e^2\over h}\int^\infty_{-\infty} d\ep_{el} \, (-\del_{\ep_{el}}
f(\ep_{el}))\, |T(\ep_{el})|^2;\cr
|T(\ep_{el})|^2 &=
\sin^2 ({1\over 2}\delta_{el}(\ep_{el}=\ep(k) - \ep(\la_o), T)).}}
Here, the first formula is the standard Landauer-B\"uttiker 
formula applied to the electronic excitations discussed in Section 2, 
while the second formula follows from the expression of $|T(\ep_{el})|^2$ 
in terms of phase shifts in the even and odd leads, and 
$\delta_{el}$ is given 
by \eIVix\ with $\la =\la_o$.  Finally, we have used the key result 
that the density of states (per unit of energy) for the electronic 
excitations is a constant as follows from \eIVxiii .

\subsec{Computation at the Symmetric Point}

So far we have discussed the computation of $G(T)$ in general terms.  In
this section we specialize to the symmetric point ($\ep_d = -U/2$).  
There are two reasons to do so.  At the symmetric point the
equations become more amenable to analysis.  However more importantly,
it is only at the symmetric point that we are able to compute the conductance,
for it is 
only at this point that we can
compute electron scattering for arbitrary energy,
as required by \eIVxv .

The problem is a technical one.  The energy functional, $\ep (k)$ is
bounded below.  For example, 
as the analysis of 
\ref\des{H. Desgranges and K. Schotte, Phys. Lett. 91 (1982) 240.} 
\ref\mccoy{J. Johnson and B. McCoy, Phys. Rev. A 6 (1972) 1613.} 
and \ref\gaudin{M. Gaudin, Saclay Note CEA-N-1559 (1), unpublished.}
shows, at the symmetric point 
$\ep (k)$ satisfies
\eqn\eIVxvi{
\ep (k) \geq -T \log (3).}
Thus we are unable to compute directly electron scattering phases for
energies below $\ep (k=-D) - \ep (\la = \la_o)$.  Rather for energies below
this, we must compute hole scattering and relate this to particle scattering
via
\eqn\eIVxvii{
\delta^\up_e (\ep_{el}, \ep_d) = \delta^\up_{ho} (\ep_{ho}=\ep_{el},-U-\ep_d ),}
valid when $H=0$.  In order to exploit then this relation, we need
$\ep_d = -U-\ep_d$, i.e. $\ep_d = -U/2$.  To compute $\delta^\up_{ho}$,
we remove a $k$ and a $\la$-hole.  Thus akin to \eIIxlii\ we
have 
\eqn\eIVxviii{\eqalign{
\delta_{ho}^{\up}(\ep_{ho} ) 
&= 2\pi \int^{k}_{-D} dk' \ri (k') +
2\pi\int^{\tilde{Q}}_{\la_o} 
d\la '\si (\la '), ~~~~\ep(k) = -\ep_{ho} + \ep(\la = \la_o) ,}}
with the difference, $\la_o \neq Q$ and $\ep (\la_o) \neq 0$.

With this, we can now consider the simplifications in the structure of the 
equations that arise at the symmetric point.
Following \rev , we can recast the equations defining the energy
functionals in a universal form for energies comparable to $T_k$, the
Kondo temperature.  Define
\eqn\eIVxix{\eqalign{
\phi_n (\la ) &= {1\over T} \ep_n (\la - {1\over \pi} \log ({2 A \over T}));\cr
\phi'_1 (g(k) ) &= -{1\over T} 
\ep (-g(k) + {1\over \pi} \log ({2 A \over T}));\cr
\phi'_{n+1} (\la ) 
&= {1\over T} \ep'_n (-\la + {1\over \pi} \log ({2 A \over T}));\cr
A &= {\sqrt{2U\Gamma} \over 2\pi}.}}
In our definition of $\phi'_1$, we rely upon the fact that in the 
energy range we are interested in, $\ep (k)$ depends
solely upon g(k).  With these definitions, $\phi_n$ and $\phi'_n$ satisfy
the equations
\eqn\eIVxx{
\xi_n (\la ) = -\int^\infty_{-\infty} d\la' s(\la - \la ') 
\log (f(T\xi_{n-1}(\la )) f(T\xi_{n+1}(\la ))) - \delta_{n1} e^{\pi\la },}
where $\xi_n = \phi_n$ or $\phi'_n$ and $s(\la ) = \cosh^{-1} (\pi \la )/2$.
These equations have been analyzed by \des\ \mccoy\ and \gaudin .
In practice they are highly accurate in determining energies
up to scales of 10's of $T_k$'s.  This is a consequence of two scales
existing in the problem, $T_k$ and $\sqrt{U\Gamma}$.  These equations
focus on the first scale while throwing out information on the second.
But because $T_k \ll \sqrt{U\Gamma}$ in the Kondo regime, 
this approximation is extremely good.

With these equations in hand, we can determine the choice of $\la_o$.
At the symmetric point we expect
the scattering phase to be symmetric in energy, i.e.
\eqn\eIVxxi{
\delta_{el} (\ep )  = \delta_{el} (-\ep ),}
regardless of the temperature.
We thus fix $\la =\la_o$ such that \eIVxxi\ is satisfied.

We now derive the specific equations for the impurity densities at the
symmetric point.  These equations have the initial form
\eqn\eIVxxii{\eqalign{
\rho^{\rm imp}(k) 
&= \Delta(k) + g'(k) \sum^\infty_{n=1}\int^\infty_{-\infty}
d\la a_n(g(k)-\la)(\sigma^{\rm imp}_{pn}+\sigma'^{\rm imp}_{pn})(\la );\cr
\sigma^{\rm imp}_{hn} (\la )&= \tilde{\Delta}_n (\la )
- \int^\infty_{-\infty} dk \rho^{\rm imp}_p(k) a_n(\la - g(k))\cr
& - \int^\infty_{-\infty} d\la' \sum^\infty_{m=1} A_{nm}(\la - \la')
\sigma^{\rm imp}_{pm}(\la' );\cr
\sigma'^{\rm imp}_{hn} (\la )&= 
 \int^\infty_{-\infty} dk \rho^{\rm imp}_{p}(k) a_n(\la - g(k))
-\int^\infty_{-\infty} d\la' \sum^\infty_{m=1} A_{nm}(\la - \la')
\sigma'^{\rm imp}_{pm}(\la' ).}}
We will recast these equations in a simpler form.  We use the inverse
of the matrix $A_{nm}$,
\eqn\eIVxxiii{
A^{-1}_{nm} (\la ) = \delta_{nm}\delta(\la ) - s(\la ) 
(\delta_{nm+1}+\delta_{nm-1}),}
together with the equalities
\eqn\eIVxxiv{\eqalign{
\delta_{n1} s(\la - \la'')
&= \int^\infty_{-\infty}d\la' A^{-1}_{nm}(\la - \la ') a_m (\la' - \la''),\cr
\tilde\Delta_n(\la ) &= \int^\infty_{-\infty} dk \Delta (k) a_n(\la -g(k)),
}}
to rewrite \eIVxxii\ as
\eqn\eIVxxv{\eqalign{
\rho^{\rm imp}(k) 
&= \Delta (k) + g'(k) \int^\infty_{-\infty} d\la s(\la - g(k)) 
\tilde{\Delta_1} (\la )\cr
&- g'(k)\int^\infty_{-\infty} d\la s(\la - g(k))
(\sigma^{\rm imp}_{h1}(\la ) + \sigma'^{\rm imp}_{h1}(\la ));\cr
\sigma^{\rm imp}_{pn}(\la) +\sigma^{\rm imp}_{hn}(\la) &=
\int^\infty_{-\infty}d\la '
s(\la - \la')(\sigma^{\rm imp}_{hn+1}(\la ') +\sigma^{\rm imp}_{hn-1}(\la '))\cr
&+ \delta_{n1} \int^\infty_{-\infty} dk \rho^{\rm imp}_h (k)s(\la - g(k));\cr
\sigma'^{\rm imp}_{pn}(\la) +\sigma'^{\rm imp}_{hn}(\la) &=
\int^\infty_{-\infty}d\la '
s(\la - \la')(\sigma'^{\rm imp}_{hn+1}(\la ')
+\sigma'^{\rm imp}_{hn-1}(\la ') )\cr
&+ \delta_{n1} \int^\infty_{-\infty} dk \rho^{\rm imp}_p (k)s(\la - g(k)).}}
We can further simplify these equations.  For energies $\ll \sqrt{U\Gamma}$,
it is an excellent approximation to take
\eqn\eIVxxvi{\eqalign{
-{g'(k) \over 2} {1 \over \cosh (\pi(g(k)-I^{-1}))} &= 
\Delta (k) + g'(k) \int^\infty_{-\infty} s(\la - g(k)) 
\tilde{\Delta}_1 (\la );\cr
I^{-1} &= {U\over 8\Gamma} - {\Gamma \over 2U}.}}
Together with this approximation we can take
\eqn\eIVxxvii{\eqalign{
\sigma^{\rm imp}_{h1} &= 0;\cr
\sigma^{\rm imp}_m &= 0, ~~~ m > 1.}}
These densities are identically
zero at zero temperature and are governed by the energy scale $\sqrt{U\Gamma}$.
Since we work at temperatures far below this scale, they can be safely
approximated as zero.  With this, the density and energy 
equations can be solved numerically
through iteration.

We do so and plot in Figure 9, G as a function of $T/T_k$.  Comparing to
the NRG computation of Costi et al., we find excellent agreement
for energies up to several $T_k$, the regime where one would
expect the NRG, by its very nature, to be most robust.  
We emphasize that this
agreement is achieved with no fitting parameters.  Our definition
of the Kondo temperature, $T_k$, is the same as that used by Costi et al.
Because of the Fermi liquid nature of this problem, we know
the functional form of the conductance at $T \ll T_k$ is
\eqn\eIVxxviii{
G(T/T_k) = {2e^2\over h}\big( 1 - c{T^2\over {T_k}^2} + \cdots\big) .}
Costi et al. \costi , based upon results borrowed from
\ref\noz{J. Nozi\`{e}res, J. Low Temp. Phys. 17 (1974) 31.}
and \ref\yam{K. Yamada, Prog. Theo. Phys. 53 (1975) 970.}, computed
$c$ to be 
\eqn\eIVxxix{
c = {\pi^4 \over 16} = 6.088 ....}
We find numerically
\eqn\eIVxxx{
c = 6.05 \pm .1.}
We have arrived at this value by fitting the plot in the region
$T/T_k < .1$.  The error is systematic in nature, arising from 
the arbitrary nature of deciding the region over which to fit.

\vskip .4in
\centerline{\hskip -1in\psfig{figure=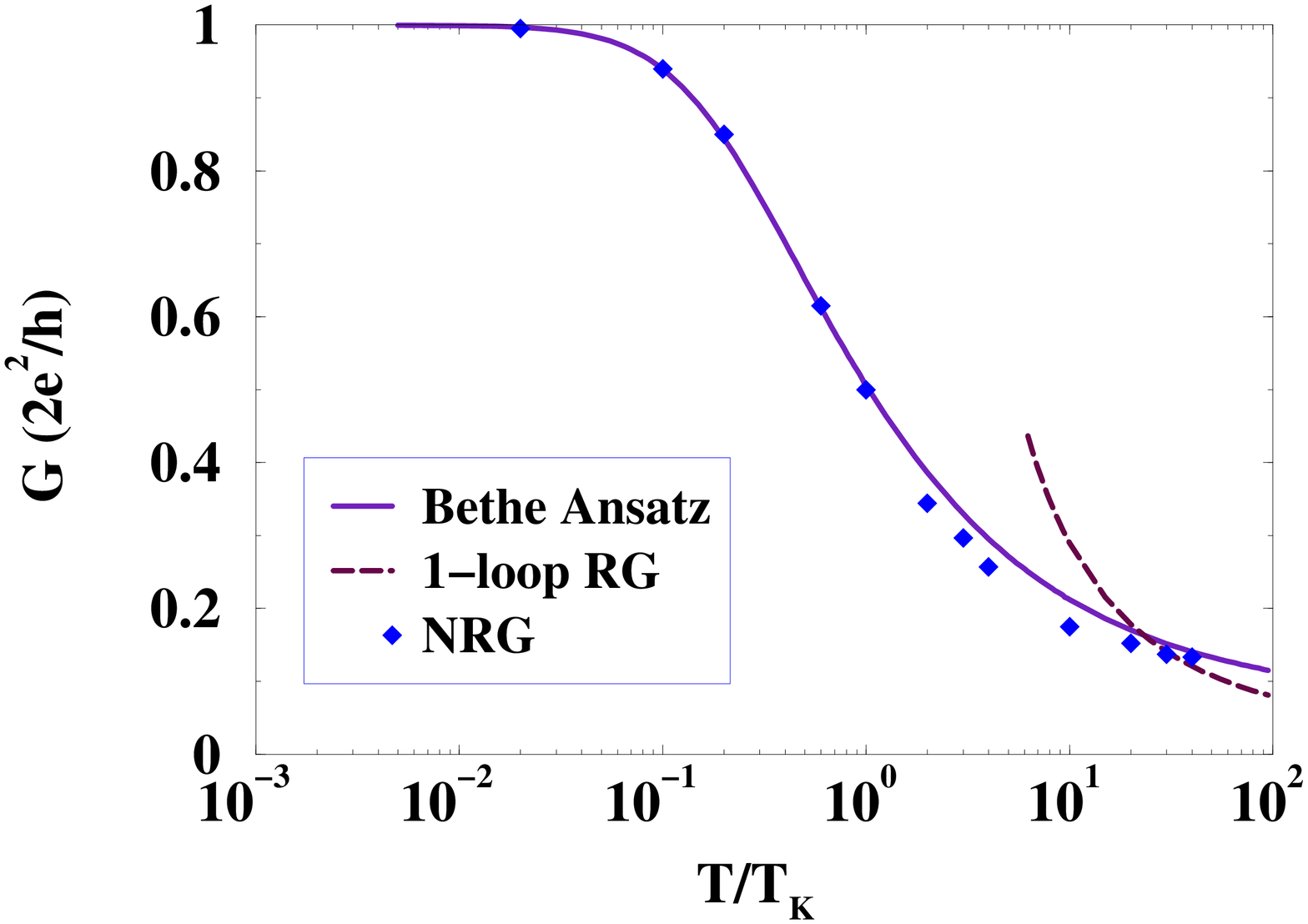,angle=-90,height=4in,width=4in}}
\vskip .15in
\indent\vbox{\noindent\hsize 5in Figure 9: A plot of the scaling
curve for the conductance as a function of $T/T_k$.  $T_k$ is
as defined in Costi et al.'s work and so there are no
free parameters.
Our computation was carried out at the symmetric point in the
Anderson model, $U + 2\ep_d = 0$.}
\vskip .4in 

We also compare our results in Figure 9 with \kaminski .
It appears that the logarithmic dependence in G,
$$
G \sim 1/\log^2(T/T_k),
$$
characteristic of weak coupling and 
arising from a one-loop RG \kaminski , should only be
expected to become qualitatively descriptive
for values of $T/T_k$ in excess of about 20.  This observation
will play a role in our determination of the validity of
our computation of the zero field differential conductance in
the next section.

The quality of the fit is a good indicator of the validity
of our approach in the Kondo regime. 
We expect from arguments given in Section 2 that our methodology
should be characterized by errors of order 
$\CO (T_k/\sqrt{U\Gamma}) << 1$ and as such 
we should see an exact match between our scaling curve and the
NRG results.
We are thus uncertain
whether the slight discrepancy between ours and Costi et al.'s results
at large $T$ is a consequence of the some unguessed
shortcoming in our approach,
some problem with the NRG, or some difficulty with our numerics.
While we cannot speak to the first two, we do note that
our handling of the numerics opens up the possibility for error
at large $T/T_k$; the numerics are fashioned so to more readily reproduce
the low temperature behaviour.

\vskip .4in
\centerline{\hskip -1in
\psfig{figure=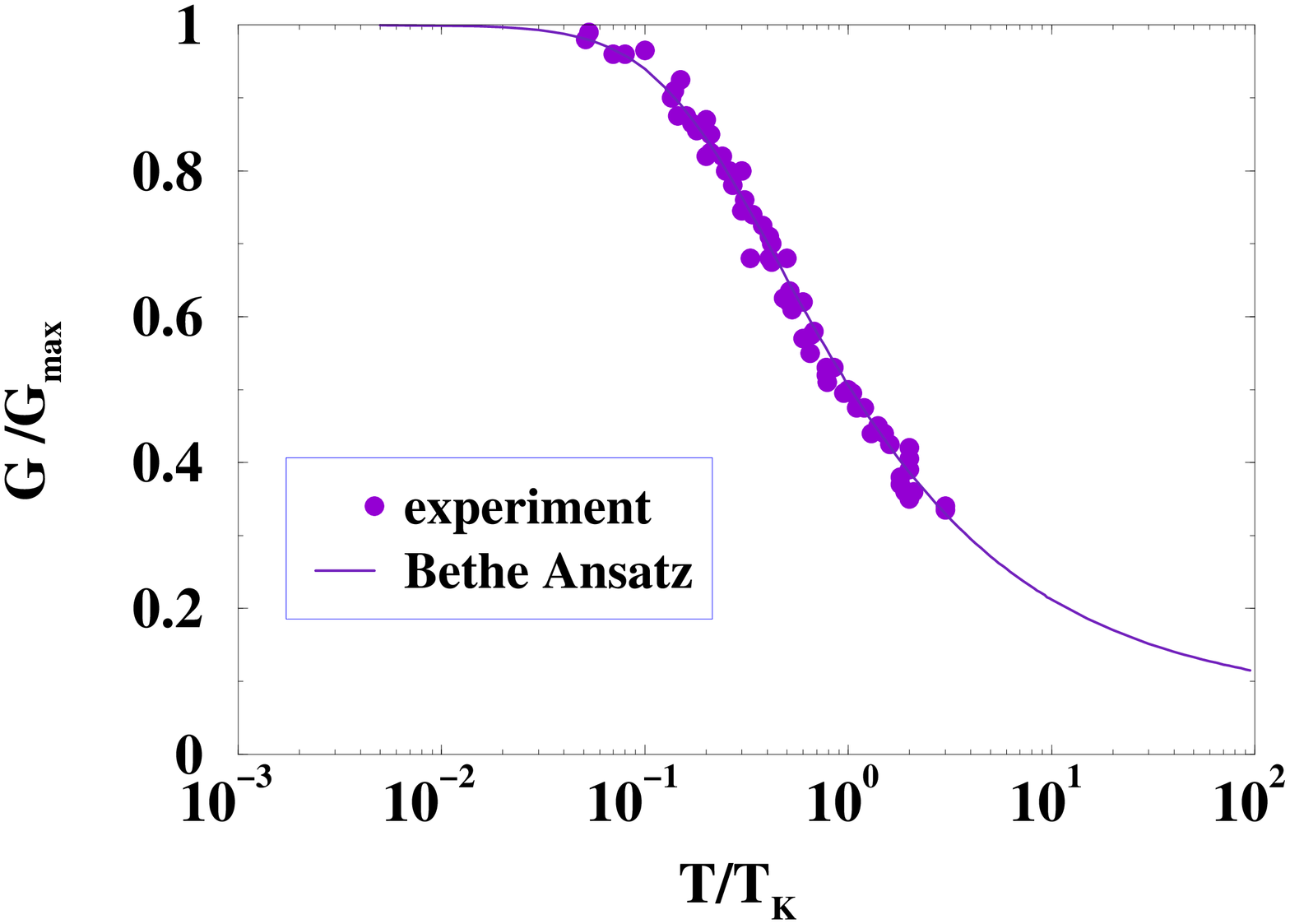,angle=-90,height=4in,width=4in}}
\vskip .15in
\indent\vbox{\noindent\hsize 5in Figure 10: A comparison
of the data from \gold\ with the computed scaling
curve for the conductance.
We now plot on the abscissa the ratio of the conductance
with the maximal possible conductance.  For the experimental
realization in \gold , the dot-lead
couplings, $V_{1,2}$, are asymmetric, and the conductance
does not achieve its unitary maximum, $2e^2/h$.  However
the scaling behaviour of $G/G_{\rm max}$ is expected
to be the same.}
\vskip .4in 

We end this section by comparing in Figure 10 our scaling curve
with the experimental results of \gold .  We see that we find
excellent agreement.  We point however that while we compute the
scaling curve at the symmetric point ($U/2+\ep_d = 0$)
of the Anderson model, the
data in \gold\ was taken away from the symmetric point but still in the 
dot's Kondo regime.  (The Kondo temperature obtains
an exponentially suppressed minimum at the symmetric point and so
is usually below the temperature that can be experimentally realized.
In order to experimentally
see Kondo physics one then must move away from the
symmetric point.)
The continuing applicability of our scaling curve
suggests a certain robustness to the
scaling behaviour.

\newsec{Out-of-Equilibrium Conductance}

\vskip .4in
\centerline{\psfig{figure=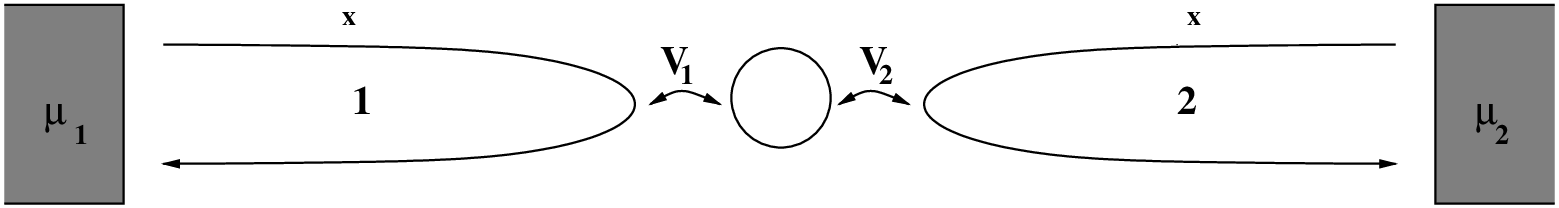,height=.75in,width=5in}}
\vskip .1in
\indent\vbox{\noindent\hsize 5in Figure 11: A sketch of two leads
attached to a quantum dot.  Each lead is, as indicated, at differing
chemical potentials.}
\vskip .4in

\subsec{Basic Formalism}

In order to compute the non-equilibrium conductance we imagine placing the
reservoirs attached to each lead at differing chemical potentials as pictured
in Figure 11.  This
on the face of it poses a problem.  In doing do we add a term 
to the Hamiltonian of the form
\eqn\eVi{
\CH_\mu = \mu_1 \int dx~c^\dagger_1 (x) c_1 (x) + 
\mu_2 \int dx~c^\dagger_2 (x) c_2 (x).
}
This term does not behave well under the map into the even/odd electron
basis in as much as the odd electron no longer decouples.  It would thus seem
it is not possible to employ the results of the previous sections in
analyzing the out-of-equilibrium system.

However we must ask what we need to compute the non-equilibrium conductance.
We need to know the distribution of particles in each of the two reservoirs.
And we need to know the scattering amplitudes of said particles.
For the particle distributions, 
we note that the particles in the two reservoirs
do not interact with one another.  Knowledge of one distribution is not
needed to determine the other.  
Thus to compute the distribution of particles in
reservoir 1 we can set $\mu_2$ to be whatever is convenient and likewise for
the determination of the distribution in reservoir 2.  
This is notably different
from what occurs in the scattering of quasi-particles between quantum Hall
edges.  In the boundary sine-Gordon formulation of this problem, the two
reservoirs, one reservoir of positive solitons and one of negative solitons,
do interact with one another.  The above device would thus not work in
this context.

We emphasize again that the distribution of particles we compute are
not the plane wave modes of an electron but rather are 
``dressed'' electrons.  
But they do share several features with plane wave electrons.
Beyond carrying the same quantum numbers of
electrons, they share the same constant density of states as a function
of energy as the plane wave electron modes.  Moreover their
dispersion relationship is the same as the plane wave modes.

Having dealt with the computation of the distributions we now turn to the
scattering.  Here there is no problem.  As we are using an integrable
basis, we are able to compute the scattering in
the context of the in-equilibrium model.  The scattering of the basis of
integrable excitations is unaffected by the differing chemical
potentials in the leads.  We could, if we wished, adopt a basis of
(dressed) excitations that was aware of the finite voltage.  
Although technically
challenging, we could also compute the corresponding scattering amplitudes.
Such amplitudes, however, would only differ from the original by
an overall phase.
As the conductance
depends upon the absolute value of the scattering matrices,
our answer would go unchanged.
The sole consequence 
of $\mu_1\neq\mu_2$
is then felt in the distributions.  
In this sense then, the problem {\it is} akin to
scattering between quantum Hall edges.

We now proceed with the actual calculation.  Here we keep
$\mu_1$ constant and imagine varying $\mu_2$ alone.  The computation
divides itself into two cases: $\mu_1 > \mu_2$ and $\mu_2 > \mu_1$.
We examine the former first.

\vskip .4in
\centerline{\psfig{figure=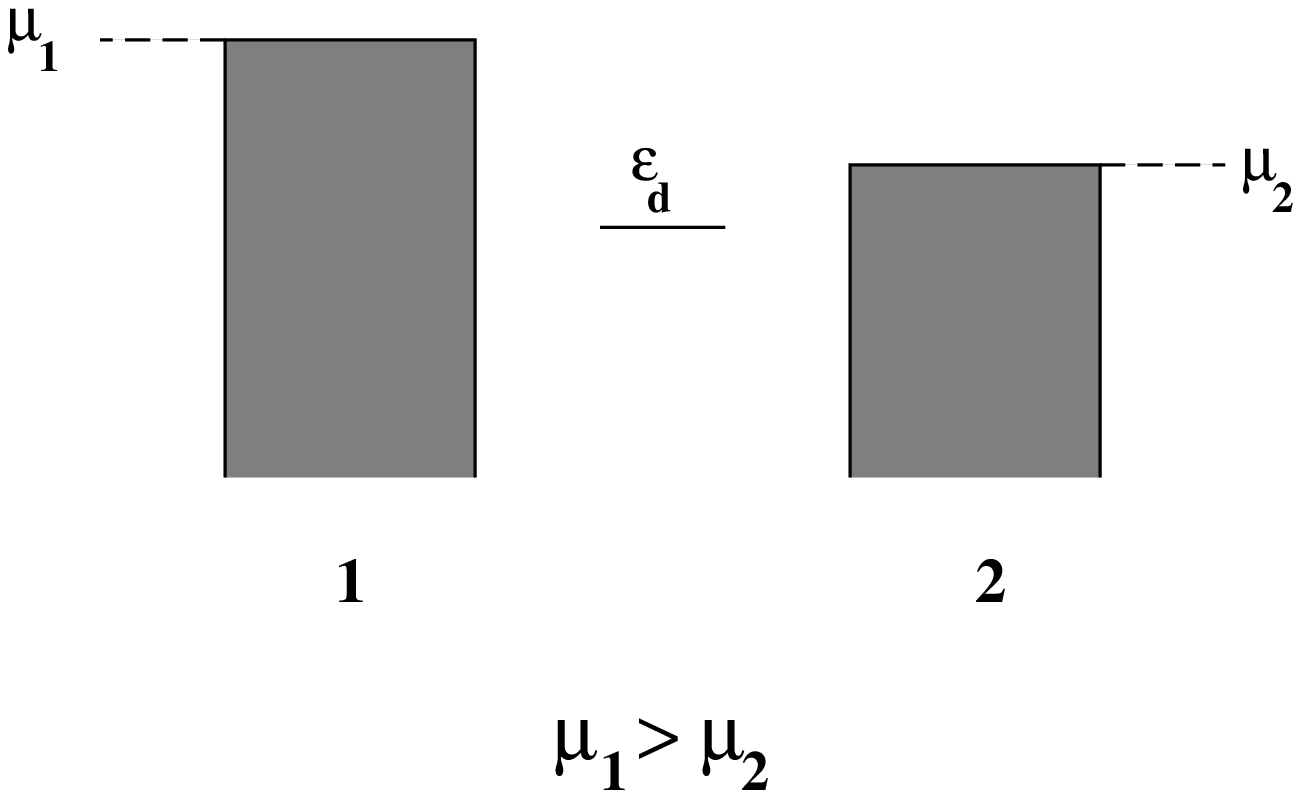,height=2.25in,width=4in}}
\vskip .15in 
\indent\vbox{\noindent\hsize 5in Figure 12: A sketch of the 
distribution of particles in the leads when $\mu_1 > \mu_2$.}
\vskip .4in 

The case $\mu_1	> \mu_2$ is pictured in Figure 12.  
As we are at zero-temperature,
particles will only diffuse from lead 1 to lead 2.  
The current is thus given by
\eqn\eVii{
J(\mu_1 , \mu_2) = {e\over h}\int^{0}_{\mu_2-\mu_1} d\ep 
(|T^{1\rightarrow 2}_\up(\ep ,\mu_1)|^2 
+ |T^{1\rightarrow 2}_\do(\ep ,\mu_1)|^2).
}
We have expressed the current as an integral over all energies ranging
from zero to $\mu_2-\mu_1$.  We note that the above integral reflects
that the density of states as a function of energy is constant.
Our particular choice of limits in the above integral
is a consequence of our conventions: our equations governing the energy
of the excitations (see \eIIxxxiii ) give the Fermi energy of lead 1 as 0.
It is only in this energy range that particles
are available to scatter in lead 1 and that are not Pauli blocked in lead 2.
We choose energies to parameterize the particles as opposed to either of
the parameters $k$ or $\la$.  These latter parameters are not convenient in
determining Pauli blocking in that with $\mu_1\neq\mu_2$
the energy functionals for the leads are also not equal, 
i.e. $\ep_1(k/\la )\neq\ep_2 (k/\la )$.
We note that even though $\mu_1$ and $\mu_2$ are bare chemical potentials, they
are the correct ones to use in determining the current.  They are not
renormalized by interactions, understandably, as there are no interactions in
the bulk.\foot{Technically this may be seen as follows.  Changing the chemical
potential in a lead by $\Delta\mu$ yields a change in the number of particles,
$\Delta N = {\Delta\mu \over \pi}$.  The density of states 
per unit energy of the 
$\la$-particles filling the ground state is $\sigma(\ep) = {1 \over 2\pi}$.
As each $\la$-particle is a bound state of two particles, the shift
in the Fermi energy induced by the change in chemical potential is
precisely $\Delta\mu$.}

It is worthwhile commenting on the dependence of the current
$J$ upon $\mu_1$ and $\mu_2$.  Although the limits governing
the energy range of excitations contributing to transport are a
function of the difference, $\mu_1 - \mu_2$, of chemical potentials,
the current's dependence upon $\mu_1$ and $\mu_2$ is more complicated.
This reflects the dependence of the transmission amplitude,
$T^{1\rightarrow 2}$, upon $\mu_1$ alone.  In particular, in
the Kondo regime of this problem, the Kondo temperature that
appears in the expressions for the current will be a function
of $\mu_1 -\ep_d$ and will not depend at all upon $\mu_2$.

Appearing in \eVii~are the electron
scattering probabilities, $T^{1\rightarrow 2}$, 
from lead 1 to lead 2.  
The amplitude, $T^{1\rightarrow 2}$, 
is given from \eIIiiic :
\eqn\eViii{
|T^{1\rightarrow 2}(\ep ,\mu_1)|^2 = 
\sin^2 ({1\over 2}\delta^1(\ep_{ho}=-\ep,\ep_d-\mu_1)).}
The phase for an electron below the Fermi surface, as indicated above,
is computed by exciting the corresponding hole.
As indicated in the introduction to this section, scattering in this
case is determined solely by the dynamics in lead 1.

To compute $\delta_{ho}$ involves exploiting
particle-hole transformations.  As such it is worthwhile to consider
the cases of zero H and non-zero H separately.
With $H=0$, we can compute the scattering of a spin $\do$ hole by
relating it to a spin $\up$ electron.  According to \eIIxxii\ 
and \eIIxxxviii , we have
\eqn\eViv{\eqalign{
\delta_{ho}^{1\do}(\ep_{ho}>\mu_1,\ep_d-\mu_1) &= 
\delta_{e}^{1\up}(\ep_{el} = \ep_{ho},-U-\ep_d+\mu_1 ) \cr
&= 2\pi\int^k_{-D} dk \rho^1_{\rm imp} (k) 
+2\pi \il \sigma^1_{\rm imp} (\la ) ,~~~\ep^1 (k) = \ep_{ho}-\mu_1 .}}
Here the energies, $\ep_{el}-\mu_1$ and $\ep_{ho}-\mu_1$, are measured 
relative to the Fermi surface in lead 1.
By the SU(2) spin symmetry we then know 
$\delta_{ho}^{1\do}(\ep_{ho}) = \delta_{ho}^{1\up}(\ep_{ho})$.
Because of the behaviour of $\ep_d$ under a particle-hole transformation,
we can only directly compute out-of-equilibrium conductances when
$\ep_d - \mu_1 < -U/2$, unusual in that it is on the other side of the
particle-hole symmetric point.

When $H$ is nonzero the situation is more complicated.  We no longer
can equate spin $\up$ and spin $\do$ scattering.  However we now
can compute spin $\up$ hole scattering directly.
From \eIIxlii\ we have
\eqn\eVv{
\delta_{ho}^{1\up}(\ep_{ho}>\mu_1,\ep_d -\mu_1) = 
2\pi\int^k_{-D} dk' \rho^1_{\rm imp} (k') 
+2\pi \il \sigma^1_{\rm imp} (\la ), ~~~\ep^1 (k) = -(\ep_{ho}-\mu_1).}
Because the bottom bound on $\ep^1 (k)$ is $-H$, we are limited
to computing spin $\up$ hole scattering for energies, $0<\ep <H$.
For spin $\do$ hole scattering we resort to the particle-hole
transformation used above:
\eqn\eVvi{\eqalign{
\delta_{ho}^{1\do}(\ep_{ho}>\mu_1,\ep_d-\mu_1) &= 
\delta_{e}^{1\up}(\ep_{el} = \ep_{ho},-U-\ep_d +\mu_1 ) \cr
&= 2\pi\int^k_{-D} dk' \rho^1_{\rm imp} (k') 
+2\pi \il \sigma^1_{\rm imp} (\la ), ~~~\ep^1 (k) = \ep_{ho}-\mu_1 .}}
We have no similar constraint on the energy range for spin $\do$ scattering.
But we can see another issue arises.  We are able to
compute spin $\up$ hole scattering
for a dot chemical potential, $\ep_d - \mu_1> -U/2$, while for spin $\do$
hole
scattering we can only perform the computation for $\ep_d - \mu_1 < -U/2$.  We
are thus limited in the case of non-zero $H$ to the symmetric point,
$\ep_d - \mu_1 = -U/2$.  But given our belief that our ansatz
for the scattering states is only valid near the symmetric point,
this constraint costs us little.

To compute
the energy functional relating the parameter $k$ to the energy
we employ the equations,
\eqn\eVvii{\eqalign{
\ep^1 (k) &= k - {H\over 2} - \mu_1 - \il \ep^1 (\la ) a_1(\la - g(k));\cr
\ep^1 (\la ) &= 2x(\la ) - 2\mu_1 - \ilp \ep^1 (\la ' )a_2(\la ' - \la)\cr
&+ \ik g'(k)\ep^1(k) a_1(g(k)-\la) .}}
These equations are identical to those of \eIIxxxiii\ but for the presence
of $\mu_1$.  The Fermi surfaces, $Q$ and $B$, are determined as before
by
\eqn\eVviii{\eqalign{
\ep^1 (\la = Q) &= 0;\cr
\ep^1 (k = B) &= 0,}}
that is, the energy functionals 
are defined such that the Fermi energy is always
zero.

Computing the differential 
conductance then amounts to computing -$e\del_{\mu_2} J$:
\eqn\eVix{
G(\mu_1,\mu_2) = -e\del_{\mu_2}J(\mu_1,\mu_2) 
= {e^2\over h}(|T^{1\rightarrow 2}_\up(\ep = \mu_2-\mu_1,\mu_1)|^2 + 
|T^{1\rightarrow 2}_\do(\ep = \mu_2-\mu_1,\mu_1)|^2).
}
As the particle distribution and correspondent
scattering in lead 1 are, as discussed above, only dependent upon $\mu_1$,
G has a particularly simple form: there 
are no terms of the form $\del_{\mu_2}|T|^2 $.

\vskip .25in
\centerline{\psfig{figure=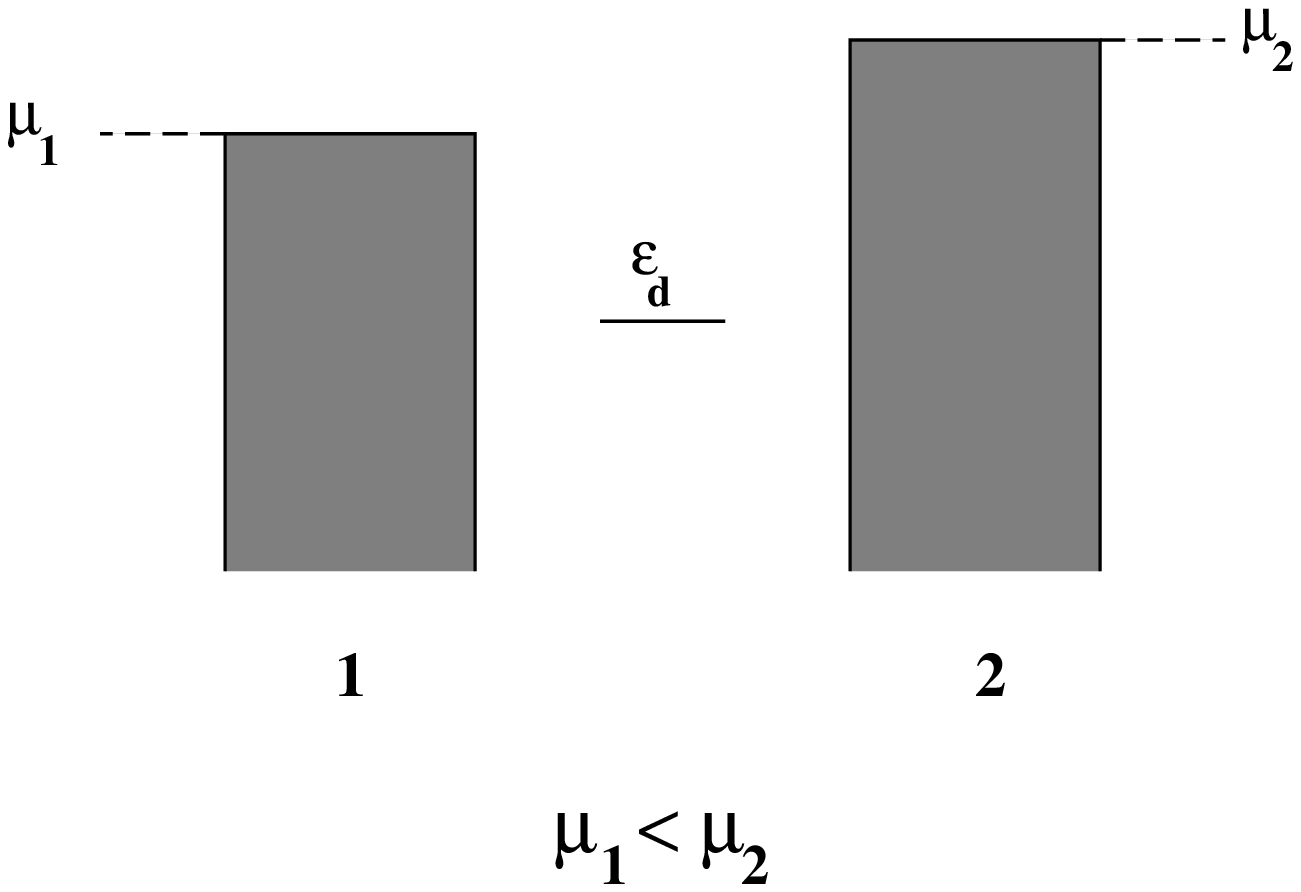,height=2.25in,width=4in}}
\vskip .15in
\indent\vbox{\noindent\hsize 5in Figure 13: A sketch of the 
distribution of particles in the leads when $\mu_1 < \mu_2$.}
\vskip .25in 

In the second case, $\mu_2 > \mu_1$ (pictured in Figure 13), such terms do
come into play.  Here, the current has the form
\eqn\eVx{
J(\mu_1,\mu_2) = -{e\over h}\int^{0}_{\mu_1-\mu_2} d\ep 
(|T^{2\rightarrow 1}_\up(\ep,\mu_2)|^2 
+ |T^{2\rightarrow 1}_\do(\ep,\mu_2)|^2).
}
In this case particles scatter from lead 2 to lead 1.  The choice of limits in
the above integral now reflects that 
the Fermi energy in lead 2 has been taken to be zero.
$T^{2\rightarrow 1}$ can be determined by the same set of equations,
\eViii~to \eVvi , but with the energies and densities defined in lead 2.
The expression for the differential conductance is more complicated
than previously as the scattering matrices are determined on the basis of
distributions in leads 2 and so the scattering varies as $\mu_2$ is varied.
We thus have
\eqn\eVxi{\eqalign{
G = -e\del_{\mu_2} J &= {e^2\over h}
(|T^{2\rightarrow 1}_\up(\ep = \mu_1-\mu_2,\mu_2)|^2 + 
|T^{2\rightarrow 1}_\do(\ep = \mu_1-\mu_2,\mu_2)|^2)\cr
&\hskip .5in + {e^2\over h}\int^0_{\mu_1-\mu_2}d\ep 
(\del_{\mu_2} |T^{2\rightarrow 1}_\up(\ep ,\mu_2)|^2 + \del_{\mu_2}
|T^{2\rightarrow 1}_\do(\ep ,\mu_2)|^2).}}
Here $T^{2\rightarrow 1}$ is given by
\eqn\eVxii{
|T^{2\rightarrow 1}|^2 
= \sin^2({1\over 2}\delta_{ho}(-\ep,\ep_d - \mu_2))
= \sin^2({1\over 2}\delta_{el}(-\ep,-\ep_d + \mu_2 - U)).}
We thus see explicitly $\partial_{\mu_2} |T^{2\rightarrow 1}|^2$ is non-zero.
When $H\neq 0$ recall we can only compute $T_\up$ and $T_\do$ only
at the symmetric point.  Given that we are varying $\mu_2$, and so
varying the effective dot chemical potential, we cannot compute the
differential conductance for non-zero $H$ in the case, $\mu_2 > \mu_1$.  
Moreover
we are restricted to the region where $\ep_d - \mu_2 < -U/2$,
also as discussed previously.

We again comment upon the dependence of the current upon
$\mu_1$ and $\mu_2$.  As with the case $\mu_1 > \mu_2$, the
current is not simply a function of the difference of the two
chemical potentials.  In this case however the scattering
amplitudes depend solely upon $\mu_2$ not $\mu_1$.  In particular
in the Kondo regime, the Kondo temperature is determined by difference
of $\mu_2$ with the dot chemical potential $\ep_d$.

As $\delta_{ho}$ is given by an equation akin to \eVv , we
can compute $\partial_{\mu_2} |T^{2\rightarrow 1}|^2$ to be
\eqn\eVxiii{\eqalign{
\partial_{\mu_2} |T^{2 \rightarrow 1} (\ep ,\mu_2 )|^2
&= {1\over 2} \sin (p_{\rm imp}(Q) + p_{\rm imp} (k))
(\partial_{\mu_2}p_{\rm imp}(Q) + \partial_{\mu_2} p_{\rm imp} (k));\cr
\partial_{\mu_2} p_{\rm imp}(Q) &= 2\pi \int^\infty_Q d\la 
\partial_{\mu_2}\sigma_{\rm imp} (\la, -\ep_d + \mu_2 - U);\cr
\partial_{\mu_2} p_{\rm imp} (k) &= 2\pi \int^k_{-\infty} dk' 
\partial_{\mu_2}\rho_{\rm imp} 
(k', -\ep_d + \mu_2 - U), ~~~~ \ep_2(k) = -\ep .\cr}}
From the density equations in Appendix A (see A.6), we find
that with $H=0$, $\partial_{\mu_2} \sigma_{\rm imp}(\la)$ and
$\partial_{\mu_2} \rho_{\rm imp} (k)$ satisfy
\eqn\eVxiv{
\partial_{\mu_2} \sigma_{\rm imp}(\la ) = 0;}
\eqn\eVxv{\eqalign{
\partial_{\mu_2}\rho_{\rm imp} (k) &= 
\partial_{\mu_2} \Delta(k,-\ep_d+\mu_2-U)
+ {1\over U\Gamma} \int^Q_{-\infty} d\la \sigma_{\rm imp} (\la )
s(\la - g(k))\cr
&- (\partial_k g(k))^2 \int^Q_{-\infty} d\la 
\sigma_{\rm imp} (\la ) s'(\la - g(k))
- {1\over U\Gamma} \int^\infty_{-\infty} dk' \Delta (k')R(g(k)-g(k'))\cr
&- (\partial_k g(k))^2 \int^\infty_{-\infty} dk' \Delta (k')
R'(g(k)-g(k')),}}
where here $g(k) = (k+\ep_d-\mu_2+U/2)^2/(2U\Gamma)$.
In computing $\partial_{\mu_2} T$, we have neglected contributions
from $\partial_{\mu_2} Q$.  We can see from the energy equations
of \eVvii\ and \eVviii\ that 
\eqn\eVxvi{
\partial_{\mu_2} Q \neq 0.}
However, $Q$, determining the Fermi surface relative to the bottom of
the band, is reflective of energy scales on the order of the bandwidth
whereas we consider changes in $\mu_2$ of ${\cal O} (T_k)$.  
Hence $\partial_{\mu_2} Q$
is negligible.

\subsec{Differential Conductance at $H=0$}

\vskip .5cm 

The differential conductance in zero field is expected to
fall off rapidly with a scale, $\sim T_k$,
from its linear response value 
near the symmetric point of $\sim 2e^2/h$.  The characteristics
of this peak are related to the peak, the Kondo resonance, in the spectral
weight of the impurity density of states as determined by the Bethe
ansatz. 
This is similar to the findings of
\win\hersh\meir\ where 
they cast all transport properties in terms of the impurity
density of states though
as determined by ${\rm Im}\langle dd^\dagger\rangle$.  
With the Landauer-Buttiker approach we have adopted, 
all scattering quantites are ultimately related to the
equilibrium density of states; the non-equilibrium density of states
plays not part in the computation marking an important difference
with \win\hersh\meir .
At the symmetric
point we are able to derive in the case of negative bias, $\mu_2 > \mu_1$, 
a closed form expression for the differential conductance.
Away from the symmetric point we must rely upon a numerical solution
of the associated integral equations.

\vskip .5in
\noindent i) Results at the Symmetric Point
\vskip .2in

At the symmetric point (and hence only the case $\mu_1 > \mu_2$ assuming
$\mu_2$ is being varied), we derive a closed form expression
for the current and differential conductance.  
At the symmetric point, $\rho_{\rm imp}$ is given by \rev :
\eqn\eVxvii{\eqalign{
\rho_{\rm imp} (k<0) &= -{g'(k) \over 2} {1 \over \cosh (\pi(g(k)-I^{-1}))}\cr
&-g'(k) \sum^\infty_{n=0} e^{-\pi g(k)(1+2n)} \int dk' e^{-\pi g(k')(1+2n)}
{\rm Re}\Delta (ik').}}
In order to make use of this expression we need to parameterize $k$
in terms of the energy $\ep (k)$.  For energies not far in
excess of $T_k$ we find in solving \eVii\ with $H = 0$,
\eqn\eVxviii{
\ep (k<0) - \mu_1 = \ep^1(k<0) = {\sqrt{2U\Gamma} \over \pi} e^{-\pi g(k)}.}
Hence the scattering phase is given by
\eqn\eVxix{\eqalign{
\delta^1_{ho} (\ep,\ep_d-\mu_1=-U/2) &= \delta^1_{el}(\ep , -U/2)\cr
=& 2\pi \int^\infty_{-\infty} d\la \sigma_{\rm imp}(\la ) 
+2\pi \int^k_{-\infty} dk' \rho_{\rm imp}(k' ) \cr
=& {3\over 2}\pi - \sin^{-1} 
\big({1-(\ep-\mu_1)^2/\tilde{T}_k^2 
\over 1+(\ep-\mu_1)^2/\tilde{T}_k^2}\big)\cr
& + 2\sum^\infty_{n=0} {1\over 1+2n} 
\big({\pi (\ep-\mu_1) \over \sqrt{2U\Gamma}}\big)^{1+2n}
\int dk e^{-\pi g(k)(1+2n)} {\rm Re}[\Delta (ik)],}}
where 
$$
\tilde{T}_k = {2\over \pi}T_k = {2\over\pi}\sqrt{U\Gamma\over 2}
e^{\pi ((\ep_d-\mu_1)(\ep_d-\mu_1+U)-\Gamma^2)/(2\Gamma U)}.
$$
The latter term in the above
is negligible when $(\ep-\mu_1)\sim T_k$ as $T_k << \sqrt{U\Gamma}$.

With this we can compute the current and the differential conductance:
\eqn\eVxx{\eqalign{
J(\mu_1,\mu_2) &= -2{e\over h} \tilde{T}_k
\tan^{-1}({\mu_2-\mu_1\over\tilde{T}_k});\cr
G(\mu_1,\mu_2) &=  -e\del_{\mu_2} J(\mu_2 ) = 2{e^2\over h} 
{1 \over (1+(\mu_2-\mu_1)^2/\tilde{T}_k^2)}.}}
The simplicity of these results is striking.  In our approach,
it is directly related to the simple form of the dressed scattering 
phase \eVxix , 
which only comes about at the end of a complex calculation. 
It is not clear to us whether there is a more direct way to obtain 
the results \eVxx.

We observe 
that no $\log (\mu /T_k)$ terms appear in the above expressions for
the current and conductance whereas we might expect such terms
for large $\mu /T_k$ .  In this regime such terms
appear in weak coupling perturbative computations \kaminski .
However we have already established with our finite temperature
calculation that weak coupling perturbation theory is not even
qualitatively accurate until one exceeds scales of $T/T_k \sim 20$.
We expect the differential conductance to be governed by similar
considerations.  Correspondingly, we would conclude that our scattering ansatz
as applied to the zero field differential conductance is at least good
for energies up to $T/T_k \sim 20$.

Given that we are at the symmetric point, we would expect
to be able to make contact with low energy scattering in the Kondo model
as this model should produce identical results 
to the Anderson model in the low energy regime.
At low energies we have 
\eqn\eVxxa{\eqalign{
\delta^1_{ho} (\ep,\ep_d-\mu_1=-U/2)
&= {3\over 2}\pi - \sin^{-1} \big({1-(\ep-\mu_1)^2/\tilde{T}_k^2 
\over 1+(\ep-\mu_1)^2/\tilde{T}_k^2}\big) ;\cr
&= \pi + 2\tan^{-1}((\ep-\mu_1)/\tilde{T}_k).
}}
This latter form is identical to that found for spin excitations
in the Kondo model \rev .  In the exact solution of the Kondo
model, the role assigned to `charge' and `spin' excitations differs
from that of the Anderson model.  In the Kondo model the charge
excitations are non-interacting and so variations
in the scattering phase occur solely because of changes in the spin
sector.  In this sense it is not surprising that
we find the scattering phase of electronic excitations in the Anderson
model is equal to the scattering phase of spin excitations in
the Kondo model.  If we were
to compute transport properties directly in a two lead Kondo model,
this equivalence suggests how we would have to formulate the scattering
ansatz that governs the gluing together of excitations from the
two sectors (in the case of the Anderson
model, this is discussed in detail in Section 2).   
To compute the finite energy scattering
phase for the case of the Kondo model,
we would leave $k$ at its Fermi surface value while varying 
$\lambda$, the exact opposite of what we find in the Anderson model.

It is also instructive to recast the impurity density of states so
that it is a function of energy:
\eqn\Vxxb{
\rho_{\rm imp}(\ep ) = {1\over \pi \tilde{T_k}} {1\over 1 + \ep^2/\tilde{T_k}^2}.
}
We see then that the impurity density of states is sharply peaked
about zero energy with a peak height proportional to $1/\tilde{T_k}$.
The spectral density of states as determined from
the dot correlator, ${\rm Im}\langle dd^\dagger\rangle$, is also sharply
peaked around zero energy.  In contrast however, its peak height
is proportional to $1/\Gamma$, a wildly different energy scale
than the one governing the Bethe ansatz impurity density of states.
The two quantities then are clearly different thus undermining
an important premise of \mwen .

\vskip .5cm

\noindent ii) Results Away From the Symmetric Point

\vskip .5cm

Pictured below is a plot of the differential conductance 
in zero magnetic field.  We see that the expected
qualitative features appear, namely the differential
conductance sharply varies on energy scales related to the Kondo
temperature, $T_k$.  

Although we are not exactly at the symmetric point, we must remain
close in order to keep true to our methodology of identifying 
scattering states.
If, for example, we were to compute the differential conductance
in the mixed valence regime of the Anderson model, we would find
our results unphysical.  
Our construction of the scattering states was predicated on the
knowledge that in the Kondo regime
the scattering phase varies on the smallest
scale in the problem, the Kondo temperature, 
and that in turn, only the impurity
density of states for the $k$-excitations is governed by this scale.
In the mixed valence regime all of these assumptions
breakdown.  

\vskip .4in
\centerline{\hskip -1in \psfig{figure=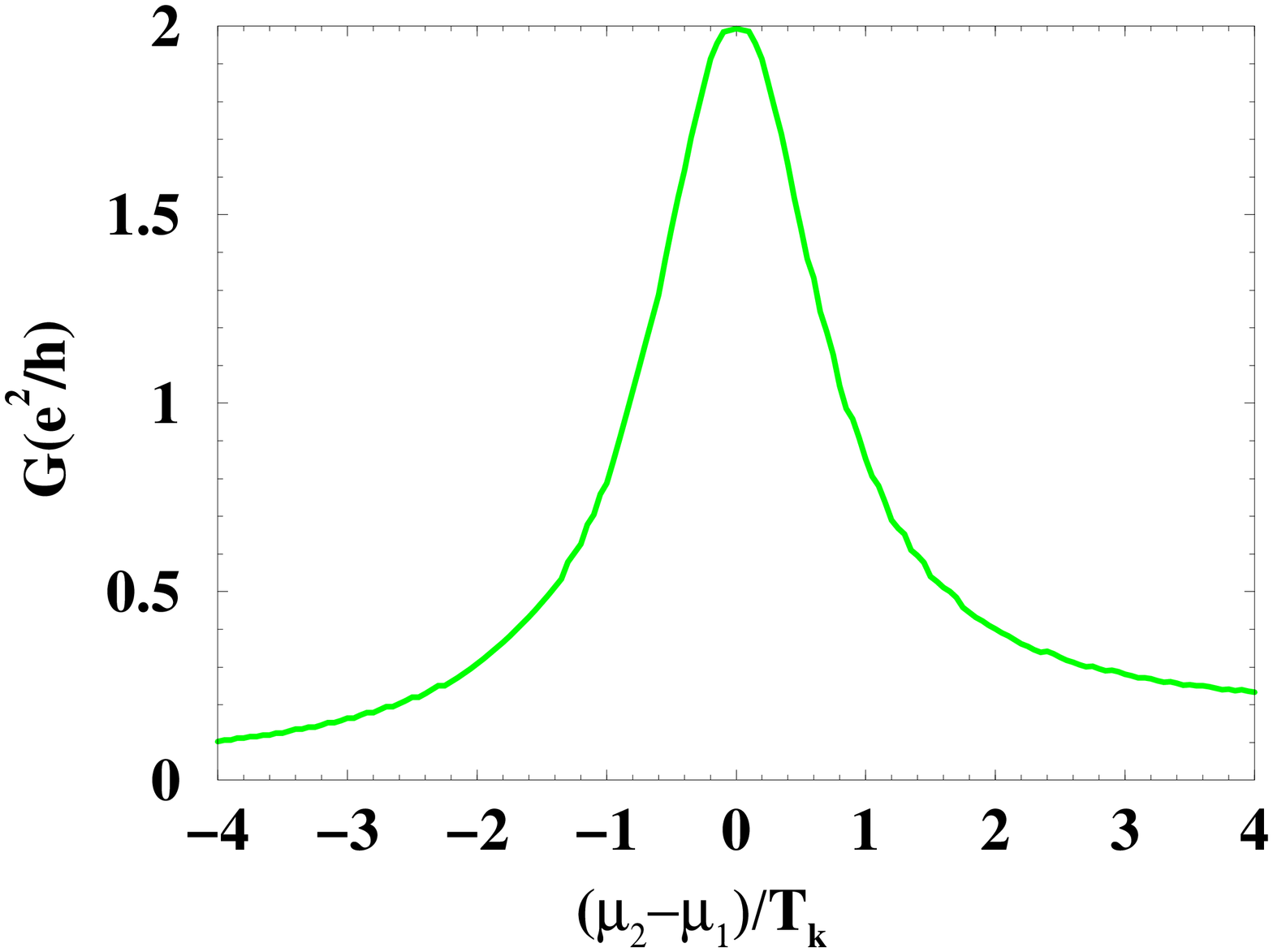,angle=-90,height=4in,width=4in}}
\vskip .15in
\centerline{\vbox{\noindent\hsize 5in Figure 14: A plot of the differential
conductance in zero magnetic field.  The value of the parameters used
in the plot
are $\Gamma = .05$, $U = 10\Gamma$, and $\ep_d = -5.2\Gamma$,
where we set the energy scale in terms of the bandwidth, $D = \pi$. }}
\vskip .2in

We see that the differential conductance curve in Figure
14 is asymmetric about $\mu_2-\mu_1 =0$.  This is a consequence
of the asymmetry introduced by variations in the effective
Kondo temperature.  In the regime $\mu_1 > \mu_2$, $T_k$
does not vary as it is solely a function of $\mu_1 - \ep_d$
and we have assumed only $\mu_2$ is changing.
However in the regime $\mu_2 > \mu_1$,
$T_k$ now depends upon $\mu_2 - \ep_d$ and so changes in $\mu_2$
lead to changes in $T_k$.  Hence the asymmetry in behaviour.

\subsec{Results at the Symmetric Point for $H\neq 0$}

\vskip .1in

At the symmetric point we develop
closed form expressions for the differential conductance.  As
stated previously, the nature of our construction of the
scattering states suggests that our results for the differential
magneto-conductance to become exact as $H/T_k$ becomes large.
Our results are in rough accordance with \meir\ - we find that for fields,
$H > T_k$, the H=0 differential conductance peak at zero bias divides
in two, one peak for each spin species.
Roughly speaking, the origin
of the split in the differential conductance arises from a similar bifurcation
in the impurity density of states.  The spectral weight of
the Kondo resonance present at $\om = 0$ when $H=0$ divides into
two resonances near $eV \sim \pm H$, again one associated with each
spin species.  The peak at negative bias, $\mu_2 < \mu_1$, corresponds
to a spin $\up$ resonance while the peak occurring with $\mu_2 > \mu_1$
is associated with a spin $\do$ resonance.  Given our ability to work
only at $\mu_2 < \mu_1$ at the symmetric point, we explore the former alone.

For fields $H \gg T_k$, this peak is
found at a bias close to $-H$.  
However unlike \meir\ we find the differential conductance peak
does not occur exactly at $eV=-H$.  This is not surprising as this
result was predicated upon a second order perturbative computation.
We find instead that the peak is shifted to values of $e|V|$ smaller than $H$.
For large fields we can develop closed form expressions for the
position, height, and width of the conductance peaks.  A related
computation was done in \mwen\ in the context of the Kondo model.
However there the analysis was restricted to the peaks in the
{\it equilibrium 
impurity density of states} as determined by the Bethe ansatz
and not the conductance per se.  As discussed
previously, the two are not
directly or obviously related, as indeed is clear from the work here.

In order to proceed with the computation, we review the
constituent elements.
The scattering phase for spin $\up$ hole scattering is given by
\eqn\eVxxi{
\delta_{ho}^{\up}(\ep_{ho}>\mu_1) = 
2\pi\int^k_{-D} dk' \rho^1_{\rm imp} (k') 
+2\pi \int^{\tilde{Q}}_{-\infty} 
\sigma^1_{\rm imp} (\la ), ~~~\ep^1 (k) = -(\ep_{ho}-\mu_1).}
Using \eIIIxxii , we can write the phase solely in terms of $\rho_{\rm imp}$:
\eqn\eVxxii{
\delta_{ho}^{\up}(\ep_{ho}>\mu_1) = 
2\pi\int^k_{-D} dk' \rho^1_{\rm imp} (k') 
+\pi(1-\int^B_{-D} dk \rho^1_{\rm imp} (k))   , ~~~\ep^1 (k) = -(\ep_{ho}-\mu_1).}
The scattering phase for spin $\do$ hole scattering is 
found to be
\eqn\eVxxiii{
\delta_{ho}^{\do}(\ep_{ho}>\mu_1) = 
2\pi\int^k_{-D} dk' \rho^1_{\rm imp} (k') 
+\pi(1-\int^B_{-D} dk \rho^1_{\rm imp} (k))   , ~~~\ep^1 (k) = \ep_{ho}-\mu_1,}
through a particle-hole transformation.

For $H$ satisfying $H \ll T_k$, we can arrive at a closed form expression
for the differential magneto-conductance.  With $H \ll T_k$, the impurity
density of states for the k-excitations retains its zero field form,
\eqn\eVxxiv{
\rho_{\rm imp} (k) = -{k\over U\Gamma} {1\over 2\cosh (\pi(g(k)-I^{-1}))}.}
The impurity density is unperturbed by the field
at first approximation as its spectral weight is found at the scale, $T_k$,
while the presence of the field only affects energies far below this
by assumption.
On the other hand the energy is shifted by a constant
from its zero field value (again taking $\mu_1=0$):
\eqn\eVxxv{\eqalign{
\ep^1 (k) &= {\sqrt{2U\Gamma}\over \pi} e^{-\pi g(k)} - H/2, ~~k\gg B;\cr
\ep^1 (k) &= {\sqrt{2U\Gamma}\over \pi} e^{-\pi g(k)} - H, ~~k\ll B.}}
For $k\gg B$ the energy is shifted by the bare energy of a
spin in a magnetic field, $H/2$.  For $k \ll B$, the effect of
the field upon $\ep^1 (k)$ can be determined by
rewriting the energy, $\ep^1 (k) \rightarrow \ep^1(k)-H$ 
and substituting into (A.11).  We then find that the field, $H$, 
disappears from the
the equation, leaving us to conclude that energy is shifted
by a dressed magnetic energy, $H$.

Using these forms for the energy and the impurity density,
the spin $\up$ scattering phase reduces to
\eqn\eVxxvi{
\delta_{ho}^{\up}(\ep_{ho}>\mu_1) = {5\over 4}\pi -
\sin^{-1}\big({1-(\ep_{ho}-\mu_1 - H)^2/{\tilde{T}_k}^2 
\over 1+ (\ep_{ho}-\mu_1 - H)^2/{\tilde{T}_k}^2}\big) +
{1\over 2}\sin^{-1}\big({1-H^2/{\tilde{T}_k}^2 
\over 1+ H^2/{\tilde{T}_k}^2}\big) ,}
while for spin $\do$ scattering, we have
\eqn\eVxxvii{
\delta_{ho}^{\do}(\ep_{ho}>\mu_1) = {5\over 4}\pi - 
\sin^{-1}\big({1-(\ep_{ho}-\mu_1 + H/2)^2/{\tilde{T}_k}^2 
\over 1+ (\ep_{ho}-\mu_1 + H/2)^2/{\tilde{T}_k}^2}\big) +
{1\over 2}\sin^{-1}\big({1-H^2/(4{\tilde{T}_k}^2)
\over 1+ H^2/(4{\tilde{T}_k}^2)}\big) .}
With this we can compute the differential conductance 
\eqn\eVxxviii{\eqalign{
G(\mu_1,\mu_2) &= {e^2\over h} \bigg[1 + {1\over 2} 
{1+(H^2-(\mu_2-\mu_1)^2)/{\tilde{T}_k}^2  
\over (1+H^2/{\tilde{T}_k}^2 )^{1/2}(1+(\mu_2-\mu_1+H)^2/{\tilde{T}_k}^2)}\cr
&+{1\over 2} 
{1+({H^2/4}-(\mu_2-\mu_1)^2)/{\tilde{T}_k}^2  
\over (1+H^2/4{\tilde{T}_k}^2 )^{1/2}
(1+(\mu_2-\mu_1-H/2)^2/{\tilde{T}_k}^2)}\bigg];\cr
\tilde{T}_k &= {2\over\pi}\sqrt{U\Gamma\over 2}
e^{\pi ((\ep_d-\mu_1)(\ep_d-\mu_1+U)-\Gamma^2)/(2\Gamma U)}.}}
Taking $H\rightarrow 0$ recovers \eVxx .

For values of $H > T_k$, we must resort to
a Wiener-Hopf analysis of the scattering phases.  
The details have been relegated to Appendix D
where exact forms of the scattering phase and the
energy, $\ep (k)$, can be found.
Here however we summarize their asymptotic forms.
For $k\ll B$ and $H \gg T_k$, the energy $\ep^1 (k)$ 
as given in (D.8) takes the
form
\eqn\eVxxix{\eqalign{
\ep^1 (k) = -H&\bigg(1 - {1\over 2\pi} \tan^{-1}{1\over g(k)-b}\cr
& ~~- {1\over 4\pi^2}{1\over 1 + (g(k)-b)^2}
\big[ {\psi(1/2) \over \Gamma (1/2)} + 1\
- (g(k)-b)\tan^{-1}({1\over g(k)-b}) \cr
&\hskip 1.5in + {\bf C} + 
 {1\over 2}\log (4\pi^2(1+(g(k)-b)^2))\big]\bigg)\cr
&+ {\sqrt{2\Gamma U} \over \pi^2} \bigg({1\over\sqrt{2e\pi}}
{e^{-b\pi}\over 1 + (g(k)-b)^2} 
+ e^{-\pi g(k)}\tan^{-1}({1\over g(k)-b})\bigg)\cr
&+ \CO ((g(k)-b)^{-3}) ,}}
where $C=.577216\ldots$ is Euler's constant
and $b$ is given by
\eqn\eVxxx{
b = {1\over \pi} \log ({2\over H}\sqrt{U\Gamma \over \pi e}).}
Note that only the first two
terms in the above expansion are at leading order but we
need include the remaining terms in order to 
obtain reasonable estimates of the properties of the conductance
peak.
Under similar conditions for $k$ and $H$ we obtain an expression for
$\int dk \rho_{\rm imp}$:
\eqn\eVxxxi{\eqalign{
2\pi \int^k_{-\infty} dk \rho_{\rm imp} &= 
\pi + 2\tan^{-1}(2(I^{-1}-g(k)));\cr
I^{-1} &= {U\over 8\Gamma} - {\Gamma \over 2U},}}
where again
$I^{-1}$ determines the Kondo temperature, $T_k \sim e^{-\pi I}$.

Combining this analysis with numerical work and the results
in \eVxxviii\ allow us to plot in 
Figure 15 the magneto-conductance for
a variety of values of $H/T_k$.  As explained earlier, we are
able only able to compute the magneto-conductance
for biases satisfying $\mu_2 < \mu_1$.  Nonetheless we expect
the differential conductance to be roughly symmetric about 
the $V=0$ axis. 

We see that for $H\gg T_k$ the
behaviour of the differential conductance is in rough accordance
with the predictions of
Wingreen and Meir \meir , that is, there is a peak approximately
at $\mu_2-\mu_1 \sim H$ for $H\gg T_k$.  However, as we have already noted,
the peak occurs at
a bias noticeably smaller than $H$ while \meir\ finds the peak to
occur exactly at $H$.  As $H$ is decreased this peak moves
to smaller ratios of $(\mu_1-\mu_2)/H$ before disappearing
altogether at $H\sim T_k$.  
For values of $H\ll T_k$, the differential conductance
does not appreciably change from its linear response value for voltage
biases of the same order of magnitude as H.
We also see that as H is increased the width of the peak narrows
and the height of the peak approaches the value of $e^2/h$ indicating
that only a single spin species (in this case spin $\up$) is contributing
to the conductance.

We now analyze the properties of the differential conductance
peak for values of $H \gg T_k$.  As we have already noted in Section 2,
we expect our results to be exact in the regime that
$H/T_k$ becomes asymptotically large.  In this regime the differential
magneto-conductance is determined solely by the spin $\up$ hole
scattering which we are exactly able to determine: no scattering
ansatz is needed here. 

For $H/T_k \gg 1$, we can write the scattering phases as follows:
\eqn\eVxxxii{\eqalign{
\delta_{ho}^{\up}(\ep_{ho}>\mu_1) & =  
2\pi\int^k_{-D} dk' \rho^1_{\rm imp} (k') + 
\pi(1-\int^B_{-D}dk\rho^1_{\rm imp}(k))\cr
&\approx \pi + 2\tan^{-1}(2(I^{-1}-g(k)));\cr
\delta_{ho}^{\do}(\ep_{ho}>\mu_1) &=
2\pi\int^k_{-D} dk' \rho^1_{\rm imp} (k') + 
\pi(1-\int^B_{-D}dk\rho^1_{\rm imp}(k))\cr
&\approx {3\over 2}\pi + \tan^{-1}(2(I^{-1}-b)).}}
In the first of the above equations, the second integral
has been neglected relative to the first, valid for $H\gg T_k$.
We also make the approximation that the spin $\do$ conductance
varies inappreciably from its Fermi surface value as the bias is
varied.  Thus we set $k=B$ in the second
of the equations.
As the conductance of the spin $\do$ species
for large $eV\sim H \gg T_k$
is constant,
the peak maximum occurs when 
\eqn\eVxxxiii{
\delta_{ho}^{\up}(\ep_{ho}>\mu_1) = \pi.}
This is turn implies the condition 
\eqn\eVxxxiv{
g(k) = I^{-1},}
and so from
\eVxxix\ the bias at which the maximum occurs is
\eqn\eVxxxv{\eqalign{
eV_{\rm max} = \ep^1 (k = -\sqrt{2U\Gamma} I^{-1}) 
& = -H\bigg( 1 - {1\over 2\pi} \tan^{-1} {1\over I^{-1} - b}
+ \cdots\bigg),\cr
I^{-1}-b &= {1\over\pi}\log ({H\over 2T_k}\sqrt{\pi e\over 2}),}}
where $\cdots$ indicates we have not written out all the terms
arising from \eVxxix .
The half maxima of the peak occur when 
$\delta_{ho}^{\up}(\ep_{ho}>\mu_1) = {\pi \over 2}/{3\pi \over 2}$, which
in turn implies 
\eqn\eVxxxvi{
g(k) = I^{-1} \pm 1/2 .}
Hence the peak width equals
\eqn\eVxxxvii{\eqalign{
e\Delta V &= \ep^1 (k = -\sqrt{2U\Gamma} (I^{-1}-1/2)) 
- \ep^1 (k = -\sqrt{2U\Gamma}(I^{-1}+1/2));\cr
&= {H\over 2\pi}
\bigg(\tan^{-1} {1\over I^{-1} - {1\over2} - b} 
- \tan^{-1} {1\over I^{-1} + {1\over2}- b}\bigg) + \cdots . }}
Finally we can estimate the peak height.  The peak maximum 
will be characterized by the maximal conductance of the spin $\up$
electrons ($e^2/h$) together with the associated conductance of the spin
$\do$ electrons.  The latter will vary only slightly
from its Fermi surface value as already discussed.
Hence the height of the peak
is given by 
\eqn\eVxxxviii{\eqalign{
G_{\rm max} &= {e^2\over h} 
(1+ \sin^2({1\over 2}\delta_{ho}^{\do}(\ep = \ep_F)))\cr
&= {e^2\over h} ({3\over 2} - {(I^{-1}-b)\over\sqrt{4(I^{-1}-b)^2+1}}).}}

\vskip .15in
\centerline{\hskip -1in
\psfig{figure=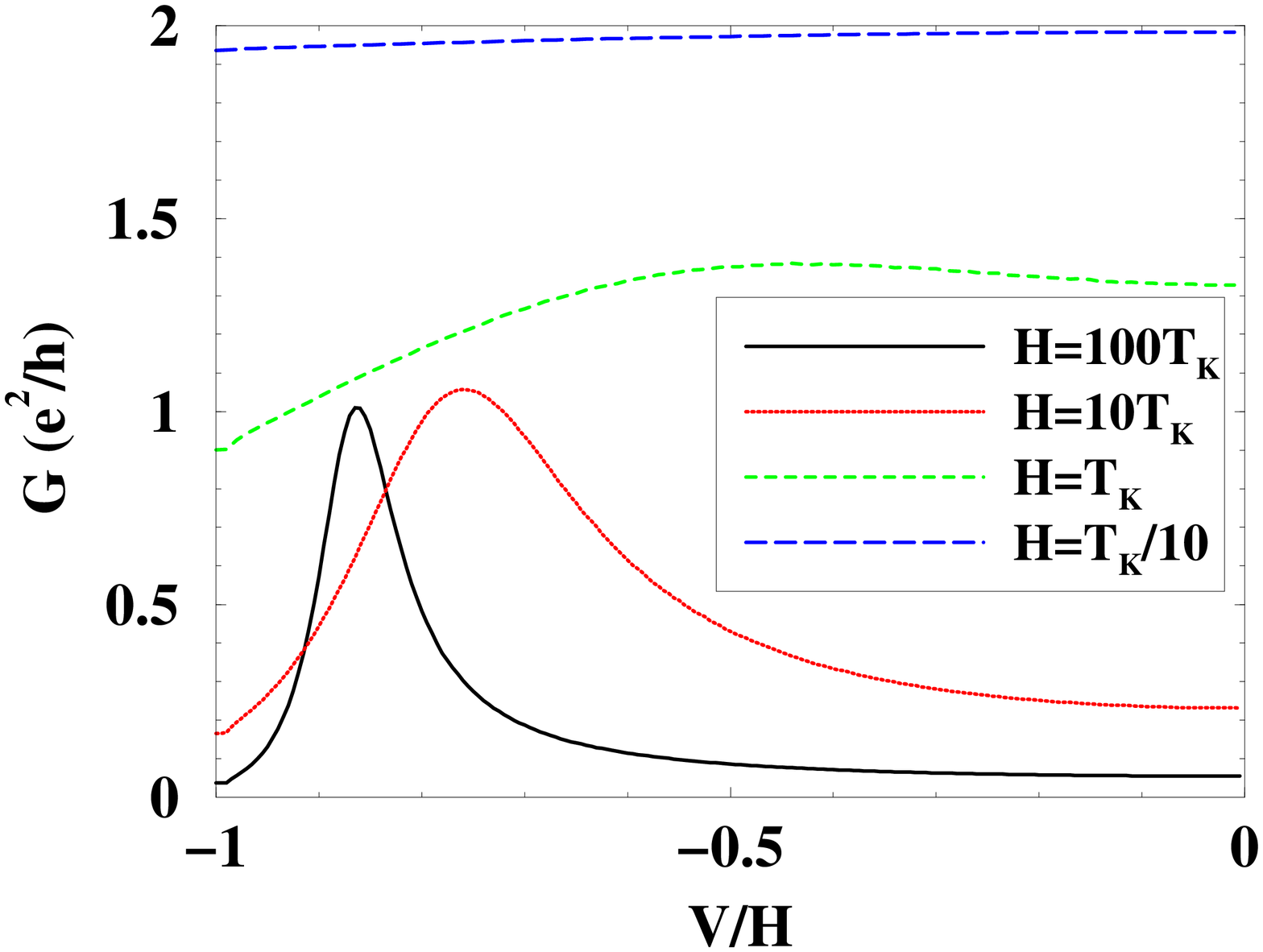,angle=-90,height=4in,width=4in}}
\vskip .15in
\indent\vbox{\noindent\hsize 5in Figure 15: A plot of the differential
conductance in a magnetic field at the symmetric point.  
The value of the parameters used in the plot
are $\Gamma = .05$ and $U = 10\Gamma$.}
\vskip .4in 

The results for the location and width of the differential 
conductance peak are similar to those
found by Moore and Wen \mwen~to characterize the location
and width of peaks appearing in the equilibrium Bethe ansatz
impurity density of states for the Kondo model.
In the large
H limit, the impurity density of states as found in the Bethe
ansatz then evidently shares certain properties with the non-equilibrium
spectral density
of states defined from the dot correlator,
${\rm Im}\langle dd^\dagger\rangle$.
However \mwen\ makes no prediction as to the height of the differential
conductance peak.  We in general do not expect the height of the
peak in the Bethe ansatz impurity density of states to be related
to the 
height of the
spectral density arising from ${\rm Im}\langle dd^\dagger\rangle$.
We already know that no such relationship exists at $H=0$ 
(see Section 5.2i) and there
is no reason to expect it to appear at finite $H$.

In Figure 16 we plot how peak characteristics evolve with
increasing $H$.  For comparison, we plot both the
asymptotic forms (\eVxxxiv -\eVxxxvi ) for the peak
characteristics against the exact
results.
We see that the location of the peak slowly approaches
$eV = -H$ with increasing H.  This approach will be logarithmic
in H as 
\eqn\eVxxxix{
{eV_{\rm max} \over H} + 1 \sim {1\over 2\pi} {1\over I^{-1} -b},
}
and $b \sim -\log(H)$.
Similarly the height of the peak approaches $e^2/h$
but again logarithmically in H.
We note that
the asymptotic forms reproduce the exact results with remarkable
accuracy even in the region, $H \sim T_k$, where
the assumptions of their derivation do not necessarily hold.

\vfill\eject

\vskip 2in
\centerline{\hskip -1in
\psfig{figure=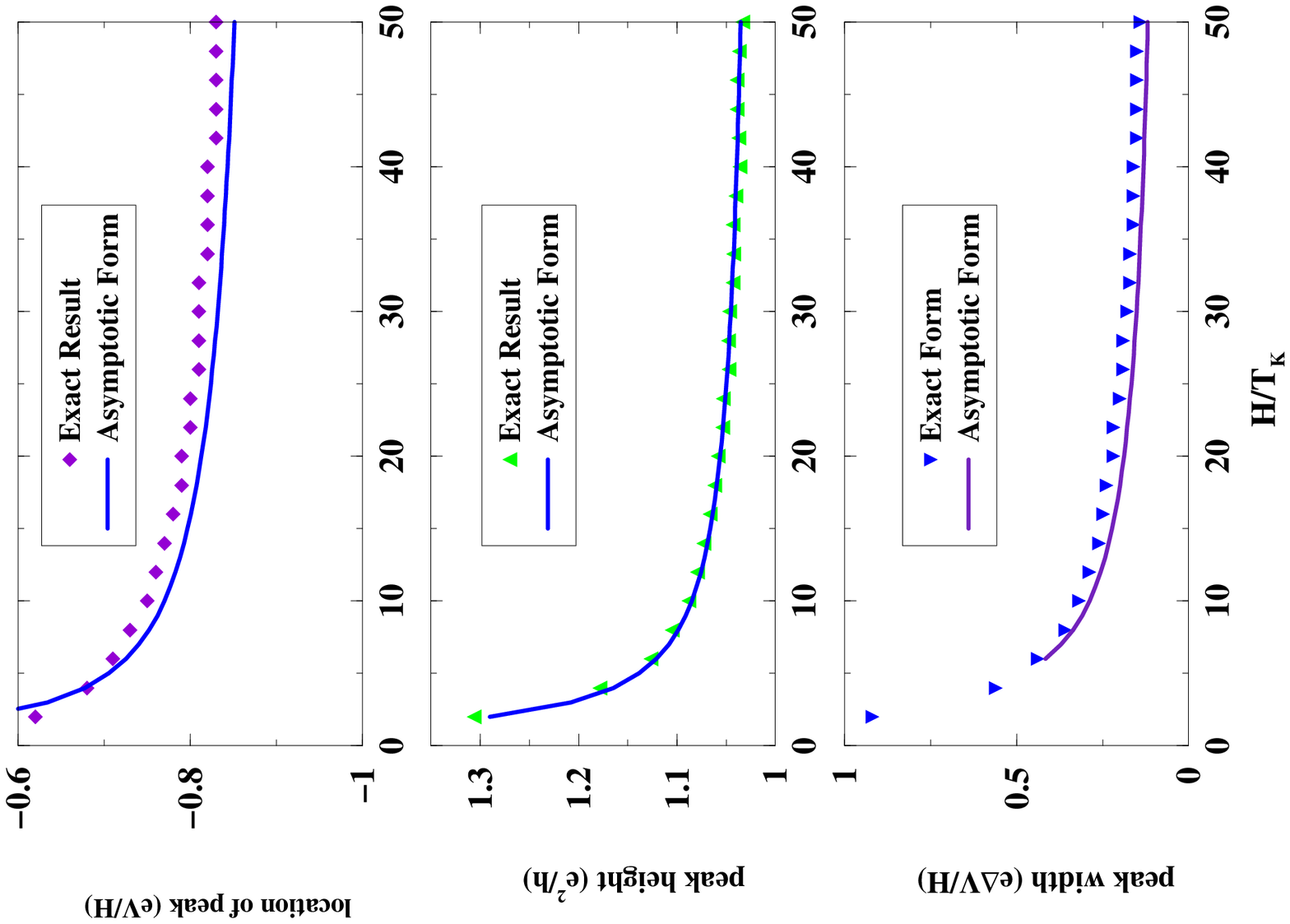,angle=-90,height=6in,width=2.5in}}
\vskip .15in
\indent\vbox{\noindent\hsize 5in Figure 16:
Plots describing the evolution
of the differential conductance peak with increasing magnetic field.
In the top panel is a plot of the location of the peak 
while the middle panel records the 
peak height and the
bottom panel gives the peak width.
The parameters used
are $U = .75/\pi D$ (D being the bandwidth) and $\Gamma$ = U/12.}

\newsec{Concluding Remarks}

Computing transport properties in 
a strongly interacting system is a difficult challenge. 
In this paper we have attacked the problem by combining
a Landauer-Buttiker approach together with data
from integrability.  
We have thus been able to provide a description
of the scattering states in the theory that have led to several
successes.  We have verified the
Friedel sum rule and provided a quantitative description of
the linear response conductance at T=0 both in and out of a magnetic field.
Our most striking result has, however, been the computation of
the finite temperature linear response conductance scaling curve.
This result is predicated upon an accurate description of the
scattering states away from the Fermi surface.  As such
we have also been able to compute the out-of-equilibrium current,
again both in and out of a magnetic field.  In particular,
we have provided a quantitative description of the differential 
magneto-conductance.  Given the nature of our construction of
the scattering states, we believe our computation of this latter quantity 
to become exact in the large field limit.

While our technique bears a degree of resemblance to
the successful, exact treatment of interacting quantum Hall edges \FLS,
the technical complexity of the two-lead Anderson 
model has prevented us from finding a definitive solution
of the problem in all regimes.  However, we have still been able to use
integrability to find extremely good approximations to the exact results in 
the cases of greatest experimental interest.  
The situation here is not altogether different from the 
use of form-factors in calculating correlation functions: although 
the techniques of integrability do not (yet) generically 
lead to closed form expressions,
they are nevertheless a breakthrough, providing 
excellent approximations which are valid from the lowest
energies through cross-over 
regimes. 
These approximations are far different from the standard, mean-field 
ones, for they contain all the crucial features of low dimensional, strong
interactions.  In particular the results presented here
represent an improvement on previous approximate methods found in 
the literature.
The quality of our approach can be gauged from the 
excellent reproduction (with no fitting parameter)
of Costi et al.'s NRG finite temperature
linear response curve.  It is unlikely other analytical techniques 
could do the same.

Our work raises two kinds of questions.  The first is whether 
higher order approximations can be devised in the present problem, 
akin to taking higher energy intermediate states into account 
in form-factors
calculations of correlation functions.  More precisely, is it possible
to develop further
the description of the fermions in terms of the integrable
states found in \eIIl .
The other, more practical question, is 
whether similar calculations can be performed in other models of interest,
and so obtain excellent approximations 
of out-of-equilibrium transport properties 
based upon the exact solution of the thermodynamics
and the proper identification of scattering states.

\newsec{Acknowledgments}
The authors would like to thank T. Costi for sharing
his NRG data.
R.K. would like to thank Y. Meir, N. Wingreen, L. Glazman,
and T. Costi for useful discussions.
H.S. thanks N. Andrei and A. Georges for many
useful discussions.
R.K. has been supported by
NSERC, the NSF through grant
number DMR-9802813 and through the Waterman Award under
grant number DMR-9528578, and the University of Virginia
Physics Department.  H.S. has been supported by
the Packard Foundation, the NYI
Program, and the DOE (H.S.). 
H.S. also acknowledges hospitality and support from 
the LPTHE (Jussieu) and LPTMS (Orsay). 
A.W.W.L. has been supported by the NSF through grant DMR-00-75064.
A.W.W.L. also acknowledges hospitality and partial support from NWO,
University of Amsterdam (the Netherlands).

\appendix{A}{Practical Computation of Density and Energy Functionals}

Here we present equations for the energy and density functionals that
are more amenable to numerical analysis.  The original density equations 
are given by
\eqn\eAi{\eqalign{
\rho_p (k) + \rho_h (k) &= {1\over 2\pi} + {\Delta (k) \over L} + 
g'(k) \int d\la a_1(g(k)-\la) \sigma_p (\la); \cr
\sigma_p (\la ) + \sigma_h (\la ) 
&= - {x'(\la)\over\pi} + {\tilde{\Delta}(\la)\over L}
- \int d\la' a_2(\la '-\la)\sigma_p (\la ') - 
\int dk a_1(\la -g(k))\rho_p (k).}}
Expressing these equations in terms of the Fourier transform of
$\sigma (\la )$ gives us
\eqn\eAii{\eqalign{
\rho_p (k) + \rho_h (k) &= {1\over 2\pi} + {\Delta (k) \over L} + 
g'(k) \int {d\om \over 2\pi} e^{-i\om g(k)}e^{-|\om |/2} \sigma_p (\om ); \cr
\sigma_p (\om ) + \sigma_h (\om ) 
&= - {x'(\om )\over\pi} + {\tilde{\Delta}(\om )\over L}
- e^{-|\om |}\sigma_p (\om ) - \int dk e^{i\om g(k)}e^{-|\om |/2}
\rho_p (k).}}
Solving for $\sigma_p(\om )$ and substituting into the r.h.s. of both
of the above equations gives
\eqn\eAiii{\eqalign{
\rho_p (k) + \rho_h (k) &= {1\over 2\pi} + {\Delta (k) \over L} -
g'(k) \int d\la \sigma_h (\la ) s(\la - g(k)) \cr
& - g'(k) \int {d\om\over 2\pi} e^{-i\om g(k)} 
\big( {x'(\om )\over \pi} - {\tilde{\Delta}(\om ) \over L}\big) 
{1\over 2 \cosh (\om /2)} \cr
& - g'(k) \int dk' \rho_p (k') R(g(k)-g(k'));\cr
\sigma_p (\la ) + \sigma_h (\la )
&= -{x'(\la )\over \pi} + \int d\la' R(\la '-\la ) 
{x'(\la ') \over \pi}
 + {\tilde{\Delta} (\la )\over L} - \int d\la' R(\la '-\la ) 
{\tilde{\Delta} (\la ')\over L} \cr
& + \int d\la ' \sigma_h (\la ' ) R(\la '-\la ) 
- \int dk \rho_p (k) s(\la - g(k))
,}}
where $R(\la )$ and $s(\la )$ are given by
\eqn\eAiv{\eqalign{
R(\la ) &= {1\over 2\pi} \int d\om
{e^{-i\om \la} \over 1+e^{|\om |} };\cr
s(\la ) &= {1\over 2 \cosh (\pi\la )} = {1\over 2\pi} \int d\om 
{e^{-i\om\lambda}\over 2\cosh(w/2)}.
}}
We can further simplify the first of the equations 
in \eAiii~by using the relations
\eqn\eAv{\eqalign{
-{x'(\la ) \over \pi} &= \int dk {d\om \over (2\pi)^2} e^{-i\om (\la - g(k))}
e^{-|\om | /2};\cr
\tilde{\Delta} (\la ) &= \int dk {d\om \over 2\pi} \Delta (k) 
e^{-i\om (\la - g(k))} e^{-|\om | /2}.\cr
}}
Then
\eqn\eAvi{\eqalign{
\rho_p (k) + \rho_h (k) 
&= \big({1\over 2\pi}+{\Delta (k)\over L}+g'(k)\int dk'
R(g(k)-g(k'))({1\over 2\pi}+{\Delta(k')\over L})\big)\cr
& - g'(k) \int d\la \sigma_h (\la ) s(\la - g(k)) 
- g'(k) \int dk' \rho_p (k') R(g(k)-g(k'));\cr
\sigma_p (\la ) + \sigma_h (\la )
&= \int dk ({\Delta (k) \over L}  + {1\over 2\pi})s(\la - g(k)) \cr
&+ \int d\la ' \sigma_h (\la ' ) R(\la '-\la ) 
- \int dk \rho_p (k) s(\la - g(k)).}}
These forms of the equations are better behaved numerically 
as
\eqn\eAvii{
\int d\la R(\la ) = {1\over 2},}
while
\eqn\eAviii{
\int d\la a_n (\la ) = 1.}
Hence when we go to solve these equations iteratively, successive iterations
grow increasingly small as $(1/2)^n$, whereas before we would not expect
convergence.

We can now derive new equations for the energy functional.  As in \eIIxxxi\
we find
\eqn\eAix{\eqalign{
\delta E &= L \ika \{\enp (k) \drp (k) - \enm (k) \drh (k)\}
+ L \ila \{\enp (\la )\dsp (\la ) - \enm (\la ) \dsh (\la) \}\cr
& = L \ika (k-{H\over 2}) \drp (k) +  2L \ila x(\la ) \dsp (\la).}}
But now from \eAvi\ we have
\eqn\eAx{\eqalign{
\delta\rho_p (k) + \delta\rho_h (k) &= 
 - g'(k) \int d\la~\delta\sigma_h (\la ) s(\la - g(k)) 
- g'(k) \int dk'~\delta\rho_p (k') R(g(k)-g(k')); \cr
\delta\sigma_p (\la ) + \delta\sigma_h (\la ) &= 
 \int d\la '~\delta\sigma_h (\la ' ) R(\la '-\la ) 
- \int dk~\delta\rho_p (k) s(\la - g(k)).}}
Solving for $\delta\sigma_p$/$\delta\rho_h$ and substituting into \eAix\
leaves us with
\eqn\eAxi{\eqalign{
\ep^+(k) + \ep^- (k) &= k - {H\over 2} - 2\int d\la x(\la ) s(\la - g(k))\cr
& + \int d\la \ep^+ (\la ) s(\la - g(k)) 
- \int dk' g'(k')\ep^- (k') R(g(k)-g(k'));\cr
\ep^+(\la ) + \ep^- (\la ) &= 2x(\la ) - 2\int d\la' R(\la -\la')x(\la')\cr
&+ \int d\la'\ep^+(\la' ) R(\la - \la') + \int dk g'(k) \ep^-(k) s(g(k)-\la).}}

\appendix{B}{Computing Scattering via the Impurity Energy}
In this appendix we compute the scattering phase of the electronic
excitations through examining the impurity energy.
To do so we will relate the impurity energy to the
impurity momentum and then use the already established
relations in Section 2.  In doing so, we will bring out
subtleties in defining the impurity energy for the purposes
of deriving scattering phases.

To determine the impurity energy of the
excitations, we play a game similar to that used previously
in deriving $\ep (k)/\ep (\la )$.  The total impurity
energy has the form (from \eIIvii )
\eqn\eBi{
E_{\rm imp} = -\ila \sp (\la ) (2 \Gamma (\la )) - \ika \rp (k) \delta (k),}
where $\Gamma (\la ) = {\rm Re \delta (x(\la ) + i y(\la ))}$.
Hence the bulk energy is given by
\eqn\eBii{\eqalign{
E_{\rm bulk} &= E - {1\over L}E_{\rm imp} \cr
&= L \ila \sp (\la) 2(x(\la ) + {\Gamma(\la) \over L})
+ L \ika \rp (k) (k - {H \over 2} + {\delta (k)\over L}) .}}
We can thus derive equations for the bulk energy of the excitations:
\eqn\eBiii{\eqalign{
\enph (k) + \enmh (k) &= k - {H\over 2} + {\delta(k)\over L} 
- \ila \enmh (\la ) a_1(\la - g(k));\cr
\enph (\la ) + \enmh (\la) &= 2(x(\la ) + {\Gamma (\la)\over L})\cr
& -\ilpa \enmh (\la ' )a_2(\la ' - \la)
+ \ika g'(k)\enmh(k) a_1(g(k)-\la) .}}
When we add particles to the system, we want to add them relative to the 
Fermi surface as determined by the host system (in other words we add them
far from the impurity).  In light of this, we have another constraint
determining the locations, $Q$ and $B$, of the two Fermi surfaces:
\eqn\eBiv{
\enh (k=B) = \enh (\la = Q) = 0.}
Of course, any bulk quantity will not distinguish between a Fermi surface set
as above or a Fermi surface determined by
\eqn\eBv{
\ep (k=B) = \ep (\la = Q) = 0.}
The difference between the two amounts to $1/L$ corrections.  However this
difference is important if one is looking at impurity quantities.
See in contrast \rev\ in
the context of the Kondo problem.

We are now in a position to relate the impurity energy to the impurity
density of states.  Differentiating the above leads to
\eqn\eBvi{\eqalign{
\del_k\enph (k) + \del_k\enmh (k) &= 1 +{\delta'(k)\over L} 
- g'(k) \ila \del_\la\enmh (\la ) a_1(\la - g(k));\cr
\del_\la\enph (\la ) + \del_\la\enmh (\la) &= 2(x'(\la ) + 
{\Gamma' (\la)\over L})\cr
& \hskip -.5in -\ilpa \del_{\la'}\enmh (\la ' )a_2(\la ' - \la)
+ \ika \del_k\enmh(k) a_1(g(k)-\la) .}}
Writing 
\eqn\eBvii{
\del\enh = \del\ep - {1\over L}\del\eni,}
where $\eni$ is the $1 \over L$ contribution to the energy of the
excitation, and comparing to \eIIxvi\ leads to the relations:
\eqn\eBviii{\eqalign{
\del_\la \eni (\la) &= -\del_\la p_{\rm imp} (\la) = 2\pi \si (\la);\cr
\del_k \eni (k) &= -\del_k p_{\rm imp}(k) = -2\pi \ri (k).}}
Hence for spin $\up$ electrons, the scattering phase
is given by 
\eqn\eBix{
\delta^\up_e = -\ep_{\rm imp}(k) - \ep_{\rm imp}(\la) .
}
Moreover we have,
\eqn\eBix{\eqalign{
\eni (\la) &= -2\pi \int^{\tilde{Q}}_\la \si (\la);\cr
\eni (k) &= -2\pi \int^k_{-D} \ri (k),}}
allowing us to prove the Friedel sum rule.  Note that
these relationships only hold due to our choice in defining
the Fermi surface as in \eBiv .
\vfill\eject

\appendix{C}{Direct Computation of the Scattering Phase}

It is possible to provide another derivation of the Friedel sum rule
that involves a direct computation of the scattering phase (as opposed
to working through the mediating agent of the impurity densities). 
For (purely technical) simplicity
we restrict ourselves to the case of a vanishing magnetic field
where there are no real $k's$ in
the ground state. 

As we discussed in Section 2, the computation of an electron
scattering phase involves the phases of a 
$k$-particle and a $\la$-hole.  As we work in the zero field limit,
the $k$-particle phase is zero and we can focus solely upon
the phase of the $\la$-hole.  To this end, we consider the
bulk density of the $\lambda_\alpha$'s, $\sigma_{\rm bulk} (\la )$.
In the ground state, $\sigma_{\rm bulk} (\la )$ obeys the equation
\eqn\eCi{\sigma_{\rm bulk}(\lambda)+
\int_Q^\infty d\la' a_2(\lambda-\lambda')\sigma_{\rm bulk}(\lambda')
=-{1\over\pi} x'(\lambda).}
Following the discussion in \ref\kor{V. Korepin, N. Bogoliubov, and
A. Izergin, {\bf Quantum Inverse Scattering Method and Correlation
Functions}, Cambridge, CUP (1993).}, the
key quantity in the following will be the shift of 
this distribution when a particle or a hole is created at 
rapidity $\Lambda$.
To study this
quantity, we go back to the discrete form of the 
Bethe ansatz equations, which read 
\eqn\eCii{2\pi J_\alpha
+\sum_{\beta=1}^M\theta_2\left(\lambda_\alpha-\lambda_\beta\right)
=-2L x(\lambda_\alpha).}
If a hole is made at $\Lambda$, the rapidities 
shift $\lambda_\alpha\to\lambda_\alpha^{(1)}$ and the above
equation becomes
\eqn\eCiii{2\pi J_\alpha
+\sum^M_{\beta=1}\theta_2\left(\lambda_\alpha^{(1)}-\lambda_\beta^{(1)}\right)
-\theta_2\left(\lambda_\alpha^{(1)}-\Lambda\right)
=-2L x(\lambda_\alpha^{(1)}).}
Setting 
\eqn\eCiv{\sigma_{\rm bulk}(\lambda_\alpha)
\left[\lambda_\alpha^{(1)}-\lambda_\alpha\right]
\equiv{1\over L} F(\lambda_\alpha|\Lambda),}
it easily follows using the equation for $\sigma_{\rm bulk} (\la )$ in 
\eCi\ that
\eqn\eCv{F(\lambda|\Lambda)+\int_Q^\infty d\la' a_2(\lambda-\lambda')
F(\lambda'|\Lambda) = -{1\over 2\pi}\theta_2(\lambda-\Lambda),}
where $\theta_2(x) = 2\tan^{-1}(x)-\pi.$
To proceed, it is convenient to introduce the integral 
operators $\hat{K}$ and $\hat{L}$ defined by
\eqn\eCvi{\hat{K}(f (\lambda)) 
\equiv -\int_Q^\infty d\la' a_2(\lambda-\lambda')f(\lambda') ,}
and 
\eqn\eCvii{(\hat{I}-\hat{K})(\hat{I}+\hat{L})=\hat{I}.}
From \eCvii, and the fact that 
$a_2(\lambda)={1\over 2\pi}{d\over d\lambda}\theta_2(\lambda)$, 
it follows that 
\eqn\eCviii{F(\lambda|\Lambda)
=\int_\Lambda^\infty d\la' L(\lambda,\lambda') .}
A similar formula with a minus sign would hold if a particle were created
in lieu of a hole at $\Lambda$.
One can represent $\hat{L}$ as a power series if one wishes:
\eqn\eCix{L(\lambda,\lambda')=-a_2(\lambda-\lambda')
+\int_Q^\infty d\la''a_2(\la-\la'')a_2(\la''-\la')+\ldots.}
Hence in particular, $L(\lambda,\lambda')=L(\lambda',\lambda)$. 

Now consider the impurity density of states, $\sigma_{imp}$.  It obeys
\eqn\eCx{\sigma_{imp}(\lambda)+\int_Q^\infty d\la'
a_2(\la-\la')\sigma_{imp}(\la')
=\tilde{\Delta}(\la ),}
from which it follows,
\eqn\eCxi{
\int_Q^\infty d\la \sigma_{imp}(\la )
=-{1\over 2\pi}\phi(Q)+\int_Q^\infty d\la F(\la |Q)\tilde{\Delta}(\la ) ,}
having set 
$\phi(\la )\equiv - 2\hbox{Re}~\delta\left[x(\la )+iy(\la )\right]$ 
and where we have used that $\phi(\infty)=0$. 
As
\eqn\eCxii{
n_d = 2\int^{\infty}_Q d\la \tilde{\sigma}_{imp}(\lambda),}
we find, 
\eqn\eCxiii{
n_d = -{1\over\pi}\phi(Q) 
+ 2\int_Q^\infty d\la \tilde{\Delta} (\la ) F(\lambda|Q) ,}
the key formula of this appendix.  The r.h.s. of the above 
equation is highly suggestive: the first term is proportional to
the bare scattering phase of the electron while the second term
represents the dressing of the bare phase that results from
the non-trivial ground state of the system.

To complete this section we now explicitly 
demonstrate the Friedel sum rule.  To do so we first imagine scattering the
unperturbed ground state through the impurity.  The entire scattering
phase is then
\eqn\eCxiv{
\sum_{\beta=1}^M \phi (\la_\beta ).
}
Now we imagine scattering the ground state plus hole through
the impurity with a resultant total phase
\eqn\eCxv{
-\phi (\Lambda ) + \sum_{\beta=1}^M \phi (\la^1_\beta ).
}
The difference of 
the two defines the scattering phase of the electron:
\eqn\eCxvi{
\delta_{el} = -\phi(\Lambda ) + 
2\pi \int^\infty_Q  d\la' F(\la'|\Lambda) \tilde{\Delta}(\la ').}
Comparing \eCxiii\ with \eCxvi\ shows that with $\Lambda = Q$ we arrive at
$\delta_{el} = \pi n_d = 2\pi n_{d\up /\do}$, the 
Friedel sum rule in the particular case when the magnetic field vanishes. 

\vfill\eject

\appendix{D}{Wiener-Hopf Analysis at the Symmetric Point}

\subsec{An alternative equation for $\ep^1(k)$}

We first solve the equation governing $\ep^1 (k)$, the energy
of excitations in lead 1 relative to the Fermi surface.  To do so we 
cast it in a different form than found in \eVvii .  For simplicity
we assume that $\mu_1$ is zero.  However finite $\mu_1$ does
not change the expression for $\ep^1$ provided $\mu_1 \ll D$.

Now $\ep^1 (k)$ is the energy
associated with adding or removing a k-excitation.  Thus imagine
removing a $k_o < B$.  This induces a change in the densities
$\rho (k)$ and $\sigma (\la )$.  At the symmetric point, the unperturbed
densities have the form (see \eAvi ):
\eqn\eDi{\eqalign{
\rho (k) &= \rho_o (k) - g'(k) \int^B_{-\infty} dk' 
\rho (k') R(g(k)-g(k')),\cr
\sigma (\la ) &= \sigma_o (\la ) - 
\int^B_{-\infty} dk \rho (k) s(\la - g(k)),}}
while the perturbed densities, $\rho_1 (k)$ and $\sigma_1 (\la )$,  
due to the hole at $k_o$, are
\eqn\eDii{\eqalign{
\rho_1 (k) &= \rho_o (k) - {1\over L} \delta(k-k_o)
- g'(k) \int^B_{-\infty} dk' \rho_1 (k') R(g(k)-g(k')),\cr
\sigma_1 (\la ) &= \sigma_o (\la ) - \int^B_{-\infty}dk\rho_1 (k) 
s(\la - g(k)),}}
where $L$ is the system size.  Rewriting $\rho_1$ by
$$
\rho_1 (k) \rightarrow \rho_1 (k) - {1\over L} \delta(k-k_o) ,
$$
yields
\eqn\eDiii{\eqalign{
\rho_1 (k) &= \rho_o (k) + {1\over L} g'(k) R(g(k_o)-g(k))
- g'(k) \int^B_{-\infty} dk' \rho_1 (k') R(g(k)-g(k'));\cr
\sigma_1 (\la ) &= \sigma_o (\la ) + {1\over L}s(\la-g(k_o))
- \int^B_{-\infty}dk\rho_1 (k) s(\la - g(k)).}}
And so changes in the density, apart from the $\delta (k-k_o)/L$ 
term already scaled out, are governed by
\eqn\eDiv{\eqalign{
\delta\rho (k) &\equiv L (\rho_1(k)-\rho (k)) = g'(k) R(g(k_o)-g(k))
- g'(k) \int^B_{-\infty} dk' \delta\rho (k') R(g(k)-g(k'));\cr
\delta\sigma (\la ) &\equiv L (\sigma_1 (\la )-\sigma (\la )) = 
s(\la-g(k_o))
- \int^B_{-\infty}dk~\delta\rho (k) s(\la - g(k)).}}
The energy of the excitation can be expressed in terms of
$\delta\rho$ and $\delta\sigma$:
\eqn\eDv{\eqalign{
-\ep^1 (k_o) &= - (k_o-{H\over 2}) + \int dk (k-H/2)\delta\rho (k) 
+ 2\int d\la x(\la)\delta\sigma (\la)\cr
&= -\big[ (k_o -H/2) - 2\int d\la x(\la ) s(\la - g(k_o))\big]\cr
&~~~~+\int dk ~\delta\rho (k) 
\big[ k-{H\over 2}-2\int d\la x(\la )s(\la - g(k))\big].}}
We see that $\ep^1 (k_o)$ depends now only upon $\delta\rho$.  That
this form for $\ep^1 (k_o)$ is equivalent to the equations in Section 2
or Appendix A can be shown using the technology found in Appendix C.

Provided $B<0$, we can introduce a change of variables that simplifies
matters:
\eqn\eDvi{
\rho (z) \equiv -{\rho (k)\over g'(k)}, ~~~ z\equiv g(k), ~~~ k<0.}
At energies not far in excess of the Kondo temperature we have
\eqn\eDvii{
(k -H/2) - 2\int d\la x(\la ) s(\la - g(k)) \approx {\sqrt{2U\Gamma}\over\pi}
e^{-\pi g(k)} - {H\over 2}.}
The above then simplifies to 
\eqn\eDviii{\eqalign{
\delta\rho (z) &= -R(z-g(k_o)) + \int^\infty_b dz' \delta\rho (z')R(z-z') , ~~ 
b={B^2\over 2U\Gamma};\cr
-\ep^1 (k_o) &= -\big[{\sqrt{2U\Gamma}\over\pi}e^{-\pi g(k_o)} - {H\over 2}\big]
+ \int dz \delta\rho (z)
\big[{\sqrt{2U\Gamma}\over\pi}e^{-\pi z}-{H\over 2}\big],\cr
&= -{\sqrt{2U\Gamma}\over\pi}(e^{-\pi g(k_o)}-\delta\rho(\om =i\pi)) 
+ {H\over 2}(1-\delta\rho(\om=0)),}}
where in the last line we have expressed $\ep^1 (k)$ in terms of the
Fourier transform of $\delta\rho$.
It is to the equation for $\delta\rho$
that we actually apply the Wiener-Hopf
analysis.

The expression for $\ep^1 (k)$ is valid provided we have removed
a particle, i.e. $g(k_o) > b$ or $k<B$.  
If we instead add a particle at $k> B$ or $g(k_o) <b$, we obtain 
in a similar fashion the following description of $\ep^1 (k)$:
\eqn\eDix{\eqalign{
\delta\rho (z) &= R(z-g(k_o)) + \int^\infty_b dz' \delta\rho (z')R(z-z') , ~~ 
b={B^2\over 2U\Gamma};\cr
\ep^1 (k_o) &=  
{\sqrt{2U\Gamma}\over\pi}(e^{-\pi g(k_o)}+\delta\rho(\om =i\pi)) 
- {H\over 2}(1+\delta\rho(\om=0)).}}
\vskip .2in

\subsec{Review of the Wiener-Hopf Analysis}

We so review the technique as presented in \rev
~on equations of the general form
\eqn\eDx{
f(z) = \int^\infty_A dz' f(z')h(z-z') + g(z).}
Writing $f^\pm (z) = f(z)\theta (\pm z \mp A)$, the Fourier transform
of the above equation yields
\eqn\eDxi{
f^+(\om ) + f^-(\om ) = f^+ (\om ) h(\om ) + g(\om ),}
where Fourier transforms are defined by
\eqn\eDxii{
a (\om ) = \int d\om e^{i\om z} a (z).}
The key step in the analysis is writing 
$1-h(\om)$ as a product of functions, $G^\pm$, which are analytic in the
upper/lower planes respectively:
\eqn\eDxiii{
1 - h(\om ) = {1\over G^+(\om ) G^- (\om )}.}
We can then write \eDxi\ as
\eqn\eDxiv{
e^{-i\om A} {f^+(\om )\over G^+(\om )} + 
e^{-i\om A} f^-(\om ) G^-(\om )  =  g(\om )G^-(\om ) e^{-i\om A}.}
Given $e^{-i\om A}f^\pm (\om)$ is analytic in the upper/lower half plane,
applying the operators
\eqn\eDxv{
\pm {1\over 2\pi i} \int d\om ' {1\over \om ' - (\om \pm i\delta)},}
to \eDxiv\ yields solutions for both $f^+$ and $f^-$:
\eqn\eDxvi{\eqalign{
f^+(\om ) &= G^+(\om ) {e^{i\om A} \over 2\pi i} \int d\om ' 
{1\over \om ' - (\om + i\delta)} g(\om ') G^-(\om ')e^{-i\om 'A};\cr
f^-(\om ) &= -{e^{i\om A} \over G^-(\om )} {1\over 2\pi i}\int d\om ' 
{1\over \om ' - (\om - i\delta)} g(\om ') G^-(\om ')e^{-i\om 'A}.}}
\vskip .2in

\subsec{Determination of $\delta\rho$}

Applying the above analysis to \eDviii ,
appropriate to the case of removing a particle, $z>b$ ($k<B$), 
we have
\eqn\eDxvii{\eqalign{
\delta\rho^+ (\om ) &= -e^{i\om b}{G^+(\om )\over 2\pi i}\int d\om '
{e^{i\om '(z_o - b)} \over \om ' - (\om + i\delta)} R(\om ') G^-(\om ');\cr
G^\pm (\om ) &= \sqrt{2\pi} 
{({\mp i \om + \delta \over 2\pi e})^{\mp {i \om \over 2\pi}} \over
\Gamma ({1\over 2}\mp {i\om \over 2\pi})};~~~1-R(\om ) 
= {1\over G_+(\om)G_-(\om )}\cr
b &= {1\over \pi} \log ({2\over H}\sqrt{U\Gamma\over \pi e}).}}
If $\om = 0$, we can continue the $\om'$-contour into the upper half
plane about the branch cut of $G^-(\om)$ while picking up the pole
at $\om ' = 0$:
\eqn\eDxvii{
\delta\rho (\om = 0 )  = -1 + {1\over \pi^{3/2}} 
\int^\infty_0 d\om '  {\sin (2\pi\om ' )\over \om '}  
\bigg({\om ' \over e}\bigg)^{-\om '} \Gamma ({1\over 2} + \om ') 
e^{-2\pi\om '(g(k_o)-b)}.}
With $\om = i\pi$, we find instead
\eqn\eDxviii{
\delta\rho (\om = i\pi )  = e^{-\pi g(k_o)} + 
e^{-\pi b}{1 \over\pi\sqrt{2e}}
\int^\infty_0 d\om '  {\sin (2\pi\om ' )\over \om ' - {1\over2}}  
\bigg({\om ' \over e}\bigg)^{-\om '} \Gamma ({1\over 2} + \om ') 
e^{-2\pi\om '(g(k_o)-b)}.}

If we now instead add a particle at $k> B$ or $g(k_o) <b$, we obtain 
from the Wiener-Hopf analysis of \eDix\
the following equations:
\eqn\eDxix{\eqalign{
\delta\rho (\om = 0 )  &= {1\over \sqrt{\pi}}
\int^\infty_0 {d\om '\over \om '}  
e^{2\pi\om '(g(k_o)-b)} 
{\tan (\pi\om ' ) \over \Gamma ({1\over 2} + \om ') }
\bigg({\om ' \over e}\bigg)^{\om '},\cr
\delta\rho (\om = i\pi )  &= {e^{-\pi b}\over \sqrt{2e}}
\int^\infty_0 {d\om '\over \om ' + {1\over 2}}  
e^{2\pi\om '(g(k_o)-b)} 
{\tan (\pi\om ' ) \over \Gamma ({1\over 2} + \om ') }
\bigg({\om ' \over e}\bigg)^{\om '}.}}
\vskip .2in

\subsec{Determination of $\delta (k) = 2\pi\int dk\rho_{\rm imp}$}

The impurity density of states, $\rho_{\rm imp}(k)$, obeys an
equation of the form
\eqn\eDxx{\eqalign{
\rho_{\rm imp} (z) &= \int^\infty_{b} dz' \rho_{\rm imp} (z') R(z-z')
+ {1\over 2} {1\over \cosh (\pi(z-I^{-1}))};\cr
\rho_{\rm imp} (z) &\equiv - {\rho_{\rm imp }(k) \over g'(k)}, ~~~ z=g(k),}}
provided $B < 0$, i.e $H \ll \sqrt{U\Gamma}$.
As we are interested in the scattering phase, we want to compute
\eqn\eDxxi{
\delta (k) = 2\pi\int^k_{-\infty} dk' \rho_{\rm imp}(k').}
If $z>b$, appropriate for when we are computing the phase
of a hole, $\delta (k)$
becomes
\eqn\eDxxii{
\delta (k) = 2\pi \int {d\om\over 2\pi} {-ie^{-i\om g(k)} \over \om - i\delta}
\rho^+(\om ).}
In this case Wiener-Hopf gives the solution of $\rho^+ (\om)$ as
\eqn\eDxxiii{\eqalign{
\rho^+_{\rm imp} (\om ) &= {e^{i\om b}G^+(\om ) \over 2\pi i}
\int d\om ' {1\over \om ' - (\om + i\delta)} \rho_o (\om ') G^-(\om ')
e^{-i\om ' b};\cr
\rho_o (\om ) &= {e^{i\om I^{-1}} \over 2 \cosh {\om \over 2}}.}}
Provided we assume $H > T_k$, or roughly
equivalently, $I^{-1} > b$, this simplifies to
\eqn\eDxxiv{\eqalign{
\delta (k) &= 2\tan^{-1} (2(I^{-1}-g(k))) + \pi\cr
&~~~ -{1\over\pi^2}\int^\infty_0 d\om {e^{-2\pi\om (g(k)-b)}\over \om}
\sin (2\pi\om ) \bigg({\om \over e}\bigg)^{-\om}\Gamma({1\over 2}+\om)\cr
&~~~~~~~~~~~~~~~~
\times \int^\infty_0 d\om ' {e^{-2\pi\om '(I^{-1}-b)}\over \om ' + \om}
\sin (\pi\om ') 
\bigg({\om '\over e}\bigg)^{-\om '}\Gamma({1\over 2}+\om ').}}

If on the other hand we are interested in the phase of an added 
particle (i.e. $z<b$), we
compute $\delta (k)$ via
\eqn\eDxxv{\eqalign{
\delta (k) &= 2\pi\int^k_{-\infty} dk' \rho_{\rm imp}(k')\cr
&= 2\pi - 2\pi\int^{g(k)}_{-\infty} dz \rho_{\rm imp}(z)\cr
& = 2\pi - 2\pi \int {d\om\over 2\pi} {e^{-i\om g(k)} \over -i\om + \delta}
\rho^-(\om ).}}
The Wiener-Hopf analysis then yields for $\rho^-(\om )$
\eqn\eDxxvi{
\rho^-_{\rm imp} (\om ) = -{e^{i\om b} \over 2\pi i G^-(\om )}
\int d\om ' {1\over \om ' - (\om - i\delta)} \rho_o (\om ') G^-(\om ')
e^{-i\om ' b}.}
This gives the scattering phase as 
\eqn\eDxxvii{\eqalign{
\delta (k) &= {3\pi \over 2} - \sin^{-1}(\tanh (\pi(g(k)-I^{-1}))) \cr
& ~~~ + {1\over 2\pi} {\bf P}\int d\om 
{e^{2\pi\om (g(k)-b)}\over \om}
{\tan (\pi\om ) \over \Gamma({1\over 2}+\om)}
\bigg({\om \over e}\bigg)^{\om}\cr
& ~~~~~~~~~~ \times {\bf P}\int d\om ' 
{e^{-2\pi\om '(I^{-1}-b)}\over \om ' - \om}
\sin (\pi\om ') 
\bigg({\om '\over e}\bigg)^{-\om '}\Gamma({1\over 2}+\om '),}}
where ${\bf P}$ indicates the principal value of the integral
is to be taken.

\listrefs
\bye